\documentclass[prb,amsmath,amssymb,showpacs,twocolumn]{revtex4}

\usepackage{epsfig}
\usepackage{graphics}
\usepackage{graphicx}
\usepackage{dcolumn} 
\usepackage{bm}

\newcommand{\sba}{\begin{subeqnarray}}
\newcommand{\sea}{\end{subeqnarray}}

\allowdisplaybreaks[1]

\def\cm-1{cm$^{-1}$}

\begin{document}

\title{Interaction of strongly correlated electrons
       and acoustical phonons}

\author{V.\ A.\ Moskalenko$^{1,2}$} \email{moskalen@thsun1.jinr.ru}
\author{P.\ Entel$^{3}$}
\author{D.\ F.\ Digor$^{1}$}

\affiliation{$^{1}$Institute of Applied Physics,  Moldova Academy
             of Sciences, Chisinau 2028, Moldova}
\affiliation{$^{2}$BLTP, Joint Institute for Nuclear Research,
             141980 Dubna, Russia}
\affiliation{$^{3}$University of Duisburg-Essen, 47048 Duisburg, Germany}

\date{\today}

\begin{abstract}
%
We investigate the interaction of correlated electrons with
acoustical phonons using the extended Hubbard-Holstein model in which
both, the electron-phonon interaction and the on-site Coulomb
repulsion are considered to be strong. The Lang-Firsov canonical
transformation allows to obtain mobile polarons for which a new
diagram technique and generalized Wick's theorem is used. This allows
to handle the Coulomb repulsion between the electrons emerged
into a sea of phonon fields (\textit{phonon clouds}). The physics
of emission and absorption of the collective phonon-field
mode by the polarons is discussed in detail. Moreover, we have
investigated the different behavior of optical and acoustical
phonon clouds when propagating through the lattice.
Initially the optical phonon fields are located at the lattice
sites and do not spread out through the crystal and their
evolution is limited by a time $\tau$, which means that the
exchange of polarons by the phonon clouds is localized and occurs
at one and the same lattice site. Hence, the renormalization of local
polaron propagators due to optical phonons conserves the hopping
matrix elements and the electronic band width. In the opposite case,
when the phonon fields are of acoustical nature, the propagation
of the fields is delocalized from the beginning and correlations
lead band narrowing effects. In addition, there is
the possibility of electron transfer without being accompanied by
phonon fields. In this case only the local electron propagators
are changed while the tunneling matrix elements remain unaffected.
In the strong-coupling limit of the electron-phonon interaction,
and in the normal as well as in the superconducting phase,
chronological thermodynamical averages of products of acoustical
phonon-cloud operators can be expressed by one-cloud operator
averages. While the normal one-cloud propagator has the form of a
Lorentzian, the anomalous one is of Gaussian form and considerably
smaller. Therefore, the anomalous electron Green's functions can
be considered to be more important than corresponding
polarons functions, i.e., pairing of electrons without
phonon-clouds is easier to achieve than pairing of polarons
with such clouds.
%
\end{abstract}

\maketitle

\section{Introduction}

Since the discovery of high-temperature superconductivity by
Bednorz and M\"uller \cite{Bednorz} the Hubbard model and related
models such as RVB and $t-J$ have widely been used to discuss the
physical properties of the normal and superconducting states
\cite{Dagotto,Kampf,Scalapino,Brenig,Shen}.
However, a unanimous explanation of the origin of the condensate
in high temperature superconductors has not emerged so far. One
of the unsolved questions is in how far can phonons be involved in
the formation of the superconducting state. The aim of the
present paper is to gain further insight into the mutual
influence of strong on-site Coulomb repulsion using the
single-band Hubbard-Holstein model \cite{Holstein,Hubbard}
and a recently developed diagram technique
\cite{Vladimir,Vakaru,Moskalenko,Bogoliubov,Moskalenko1}.
We consider now the most interesting case, namely superconductivity of
correlated electrons coupled to dispersive acoustical phonons. This
investigation differs from our previous studies
\cite{Moskalenko2,Dogotaru,Moskalenko3,Entel,Moskalenko4,Entel2}
of electrons coupled to dispersionless optical phonons, which was
addressed by most other authors
\cite{Freericks,Alexandrov,Alexandrov1,Mierzejewski}.

Because the interaction between electrons and phonons is
strong, we include the Coulomb repulsion in the zero-order
Hamiltonian and apply the canonical transformation of Lang and
Firsov \cite{Lang} to eliminate the linear electron-phonon interaction.
In the strong electron-phonon coupling limit, the resulting
Hamiltonian of hopping polarons (i.e., hopping electrons
surrounded by phonon clouds) can lead to an attractive
interaction among electrons meditated by the phonons. In this
limit, the chemical potential, the on-site and inter-site
Coulomb energies as well as  the frequency of the collective
phonon-cloud mode (which is much larger than the bare
acoustical phonon frequencies) are strongly renormalized
\cite{Entel,Moskalenko4,Entel2}.
This affects the dynamical properties of the polarons and the
character of the superconducting transition. We suggest that the
resulting superconducting state with polaronic Cooper pairs is
mediated  by the exchange of phonon clouds and their collective
mode during the hopping of the polarons.

\section{Theoretical approach}

\subsection{The Lang-Firsov transformation of the Hubbard-Holstein model}

The initial Hamiltonian of correlated electrons coupled to
longitudinal acoustical phonons (the polarization index is omitted)
has the form
%
\begin{equation} \label{1}
%
H = H_e + H_{ph}^0 + H_{e-ph},
%
\end{equation}
%
where
%
\begin{eqnarray}
H_e & = & \sum\limits_{ij,\sigma} \left \lbrace
     t(j - i) + \varepsilon_0 \delta_{ij} \right \rbrace
     a_{j\sigma}^{\dag} a_{i\sigma}
\nonumber \\
{} & + & U_0 \sum\limits_i n_{i,\uparrow} n_{i,\downarrow}
   + {\textstyle\frac{1}{2}} \sum_{ij}{}^{^{\prime }}
     V_{i,j}^c n_i n_j ,
     \label{2}
\\
H_{ph}^0 & = & \sum\limits_{\mathbf{k}}
      \omega_{\mathbf{k}} \left(
     b_{\mathbf{k}}^{\dag} b_{\mathbf{k}}
   + {\textstyle\frac{1}{2}} \right) ,
     \label{3}
\\
H_{e-ph} & = & \sum\limits_{ij}
     g(i - j)q_i n_j ,
     \label{4}
\\
&& {} n_i = \sum\limits_{\sigma} n_{i\sigma}, \quad
      n_{i\sigma} =a_{i\sigma}^{\dag} a_{i\sigma} .
\nonumber
%
\end{eqnarray}
%
%
Here $a_{i\sigma}$ $(a_{i\sigma}^{\dag})$ are annihilation (creation)
operators of electrons at lattice site $i$ with spin $\sigma$,
$b_{\mathbf{k}}$ $(b_{\mathbf{k}}^{\dag})$ are phonon operators with
wave vector $\mathbf{k}$; $q_i$ $(p_i)$  is the phonon
coordinate (momentum) at site $i$, which is related to
the phonon operators by
\[
q_i = \frac{1}{\sqrt{2}} \left(
     b_i + b_i^{\dag} \right) , \quad
p_i = \frac{i}{\sqrt{2}} \left(
    -b_i + b_i^{\dag} \right) .
\]
The Fourier representation of these quantities have the form
%
\begin{eqnarray} \begin{array}{ll}
%
{\displaystyle b_i = \frac{1}{\sqrt{N}} \sum\limits_{\mathbf{k}}
      b_{\mathbf{k}} e^{-i\mathbf{kR}_i}, } & \quad
{\displaystyle b_i^{\dag} = \frac{1}{\sqrt{N}} \sum\limits_{\mathbf{k}}
      b_{\mathbf{k}}^{\dag} e^{i\mathbf{kR}_i} },
\\*[0.2cm]
{\displaystyle q_i = \frac{1}{\sqrt{N}} \sum\limits_{\mathbf{k}}
      q_{\mathbf{k}} e^{-i\mathbf{kR}_i}, } &  \quad
{\displaystyle p_i = \frac{1}{\sqrt{N}} \sum\limits_{\mathbf{k}}
      p_{\mathbf{k}} e^{i\mathbf{kR}_i}, }
\\*[0.2cm]
{\displaystyle q_{\mathbf{k}} = \frac{1}{\sqrt{2}} \left(
      b_{\mathbf{k}} + b_{-\mathbf{k}}^{\dag} \right), } & \quad
{\displaystyle p_{\mathbf{k}} = \frac{i}{\sqrt{2}} \left(
      b_{\mathbf{k}}^{\dag} - b_{-\mathbf{k}} \right) . }
\label{(5)}
%
\end{array} \end{eqnarray}
%
In this Hamiltonian $U_0$ and $V_{ij}^c$ are the on-site and
inter-site Coulomb interactions, $t(i - j)$ is the nearest neigbor
two-center transfer integral (which may be extended to include
also next-nearest neighbor hopping of electrons), $g(i - j)$ is
the matrix element of the electron-phonon interaction,
$\varepsilon_0 = \overline{\varepsilon}_0 - \mu_0$, where
$\overline{\varepsilon}_0$ is the local electron energy and
$\mu_0$ is the chemical potential of the system. Here and in the
next part of the paper the Plank constant $\hbar$ is considered
equal to one. The Fourier representation of $t(j - i)$ is related
to the tight-binding dispersion $\varepsilon (\mathbf{k)}$ of the
bare electrons with band width $W$,
%
\begin{equation}
%
t(\mathbf{R}) = \frac{1}{N} \sum\limits_{\mathbf{k}}
   \varepsilon (\mathbf{k)} e^{-i\mathbf{kR}} ,
\nonumber
\label{t}
%
\end{equation}
%
with $\mathbf{R}$ as nearest neighbor distance. Apparently the
energy scale of the model Hamiltonian is fixed by the parameters
$W, \, U, \, g$ and $\omega _{\mathbf{k}}$. The band filling
$n$ is an additional parameter. After applying the displacement
transformation of Lang-Firsov \cite{Lang},
%
\begin{equation}
%
H_p = e^SHe^{-S},\quad
c_{\mathbf{i}\sigma} = e^Sa_{i\sigma} e^{-S},\quad
c_{i\sigma}^{\dag} = e^Sa_{i\sigma}^{\dag} e^{-S} ,
\label{(6)}
%
\end{equation}
%
with
%
\begin{eqnarray}
%
S & = & \frac{-i}{\sqrt{N}} \sum\limits_{\mathbf{k,i}}
      S(\mathbf{k}) p_{\mathbf{k}} n_i e^{i\mathbf{kR}_i} ,
\label{(7)}
\\[0.2cm]
S(\mathbf{k}) & = & \frac{g(\mathbf{k})}{\omega_{\mathbf{k}}}
   \equiv \overline{g}(\mathbf{k}),
\nonumber
%
\end{eqnarray}
%
we obtain the polaron Hamiltonian in the form:
%
\begin{eqnarray}
%
H_p & = & H_p^0
   + H_{ph}^0 + H_{int} ,
\label{(8)}
\\*[0.2cm]
H_p^0 & = & \sum\limits_i
     H_{ip}^0, \quad
     H_{ip}^0 = \epsilon \sum\limits_{\sigma} n_{i\sigma}
   + Un_{i\uparrow} n_{i\downarrow} ,
\label{(9)}
\\
H_{int} & = & \sum\limits_{ij,\sigma}
     t(j - i) c_{j\sigma}^{\dag} c_{i\sigma}
   + {\textstyle\frac{1}{2}} \sum_{ij}{}^{^{\prime}}
     V_{ij}^c n_i n_j ,
\label{(10)}
%
\end{eqnarray}
%
where
%
\begin{eqnarray}
%
\hspace*{-0.3cm}
c_{i\sigma}^{\dag} \! & = & \! a_{i\sigma}^{\dag} e^{-i\pi_i},
    \quad c_{i\sigma} = a_{i\sigma} e^{i\pi_i} ,
\label{(11)}
\\*[0.2cm]
\hspace*{-0.3cm}
\pi _i \! & = & \! \frac 1{\sqrt{N}} \sum\limits_{\mathbf{k}}
    \overline{g}(\mathbf{k}) p_{\mathbf{k}} e^{i\mathbf{kR}_i}
  = \sum\limits_{\mathbf{k}} p_j
    \overline{g}(\mathbf{R}_j - \mathbf{R}_i) ,
\label{(12)}
\\
\hspace*{-0.3cm}
\varepsilon \! & = & \! \overline{\varepsilon}_0 - \mu , \quad
    \mu = \mu_0 + {\textstyle\frac{1}{2}} V^{ph} , \quad
    U = U_0 - V^{ph},
\label{(13)}
%
\end{eqnarray}
%
and
%
\begin{eqnarray}
%
V^{ph}(i - j) & = & \frac{1}{N} \sum\limits_{\mathbf{k}}
    \frac{g(\mathbf{k}) g(-\mathbf{k})}
         {\omega_{\mathbf{k}}} \,
    e^{-\mathbf{k}(\mathbf{R}_i - \mathbf{R}_j)} .
\label{(14)}
%
\end{eqnarray}
%
Hence, the effective intersite interaction is
$V_{ij} = V_{ij}^c - V_{ij}^{ph}$ with $V_{i = j} = 0$.
The frequency $\omega_{\mathbf{k}}$ of acoustical phonons is linear
in $\mathbf{k}$ for sufficiently small wave vectors. In order to
have a reasonable expression for the parameter $S(\mathbf{k})$ of the
canonical transformation, it is necessary that the condition
$g(\mathbf{k} = \mathbf{0}) = 0$ is fulfilled. This condition means that
the movement of phonons with infinite wave length, which is equivalent to
the macroscopic displacement of the system, cannot influence its
properties and must be omitted. Therefore, the Fourier representation of
the direct attraction mediated by phonons must also vanish in this limit:
$V^{ph}(\mathbf{k} = \mathbf{0}) = 0$. It is important to note that the
Fourier representation of the Coulomb part of the inter-site interaction
must also vanish for vanishing wave vector:
$V^c(\mathbf{k} = \mathbf{0}) = 0$ as a consequence of
required charge neutrality of the system. Therefore, the resulting
direct interaction between electrons,
$V(\mathbf{R}) = V^c(\mathbf{R}) - V^{ph}(\mathbf{R})$,
fulfills $V(\mathbf{k} = \mathbf{0}) = 0$. This will be used when analyzing
the corresponding diagrammatic contribution.

When deriving the polaron Hamiltonian, it was necessary to include
the shift of the polaron coordinate $q_{\mathbf{k}}$ by
%
\begin{equation*}
%
e^S q_{\mathbf{k}} e^{-S} = q_{\mathbf{k}}
   - \frac{1}{\sqrt{N}} \sum\limits_{\mathbf{k}}
     \overline{g}(\mathbf{k}) n_ie^{i\mathbf{kR}_i},
%
\end{equation*}
%
which helps to eliminate the linear electron-phonon interaction.

The polaron Hamiltonian is by nature a polaron-phonon operator
because the new creation and annihilation operators
$c_{\mathbf{i}\sigma }^{\dag }$ and $c_{\mathbf{i}\sigma }$
entering $H_p$ must be interpreted as operators of polarons,
i.e., electrons dressed with displacements of ions that couple
dynamically to the momentum of acoustical phonons. In the
zero-order approximation (omitting $H_{int}$) polarons are
localized and phonons are free with a strongly renormalized
chemical potential $\mu $ and on-site
interaction $U$. This last quantity can become negative if
the phonon mediated attraction $V^{ph}$ is strong enough to
overcome the direct Coulomb repulsion. The first term of the
perturbation operator $H_{int}$ describes tunneling of polarons
between lattice sites, i.e., tunneling of electrons surrounded by
clouds of phonons. The second term of this operator describes the
renormalized polaron-polaron inter-site interactions.

\subsection{Averages of electron and phonon operators}

One problem is to deal properly with the impact of electronic
correlations on the polaron formation involving operators
like (11) for the electron and phonon subsystems. This can be done best by
using Green's functions provided one finds a way to deal
with the spin and charge degrees of freedom. In order to achieve this
in the limit of large $U$, the Hubbard term can be included in the
zero-order Hamiltonian. As a consequence, conventional
perturbation theory of quantum statical mechanics is not an
adequate tool because it relies on the expansion of the partition
function around the noninteracting state using
Wick's theorem and conventional Feynman diagrams.

In order to have a systematic description of correlated electrons,
Hubbard \cite{Hubbard} proposed a graphical expansion around the
atomic limit in powers of hopping integrals. This
diagrammatic approach was reformulated by Slobodyan and
Stasyuk \cite{Slobodyan} for the single-band Hubbard model and
independently by Zaitsev \cite{Zaitsev} and further developed by
Izyumov \cite{Izyumov}. In these approaches, the complicated
algebraic structure of the projection or Hubbard operators was used.

We have found an alternative way in the sense that
our diagram technique involves simpler creation and annihilation
operators for electrons at all intermediate  stages of the theory
and Hubbard operators only when evaluating final expressions
\cite{Vladimir,Vakaru,Moskalenko,Bogoliubov,Moskalenko1}. In this
approach, averages of chronological products of interactions
are reduced to \textit{n}-particle Matsubara Green's functions
of the atomic system. These functions can be factorized into
independent local averages using a generalization of Wick's theorem
(GWT) which takes strong local correlators into account, see Refs.\
\onlinecite{Vladimir,Vakaru} and \onlinecite{Entel} for details.
Application of the GWT yields
new irreducible on-site many-particle Green's functions or Kubo cumulants.
These new functions contain all local spin and charge fluctuations. A
similar linked-cluster expansion for the Hubbard model around the
atomic limit was recently formulated by Metzner \cite{Metzner}. But in
the latter work the Dyson equation for the renormalized
one-particle Green's function
was not derived, nor the correlation function which appears as main
element of this equation. It is the purpose of this paper to check
in how far we can use the GWT for the extended Hubbard-Holstein model
given by Eq.\ (1).

With respect to the transformed Hubbard-Holstein model, phonon
operators are averaged using ordinary Wick's theorem by taking into
account the factorization of the phonon partition function in
$\mathbf{k}$ space of phonon wave vectors. We define the
temperature Green's function for the polarons in the interaction
representation by
%
\begin{equation}
%
G_p({\bf {x}},\sigma,\tau | \mathbf{x}^{\prime},\sigma^{\prime},
     \tau ^{\prime })
   = - \left \langle \mathrm{T} \,
     c_{\mathbf{x}\sigma}(\tau) \overline{c}_{\mathbf{x}^{\prime}
     \sigma^{\prime}}(\tau^{\prime})
     U(\beta) \right \rangle_0^c ,
\label{(15)}
%
\end{equation}
%
with
%
\begin{eqnarray*}
%
c_{\mathbf{x}\sigma }(\tau) & = &
   e^{H^0\tau} c_{\mathbf{x}\sigma } e^{-H^0\tau} , \quad
\overline{c}_{\mathbf{x}\sigma}(\tau)
 = e^{H^0\tau}\overline{c}_{\mathbf{x}\sigma} e^{-H^0\tau} ,
\\
\pi _{\mathbf{x}}(\tau) & = & e^{H^0\tau} \pi_{\mathbf{x}} e^{-H^0\tau} ,
%
\end{eqnarray*}
%
for the polaron and phonon operators, respectively, with
$H^0 = H_p^0 + H_{ph}^0$. Instead of $i, \, j$ we now use
$\mathbf{x}, \, \mathbf{x}^{\prime}$  as site indices;
$\tau, \, \tau^{\prime}$ are imaginary time variables with
$0 < \tau < \beta$; $\mathrm{T}$ is the time ordering operator and
$\beta$ the inverse temperature. The evolution operator is given by
%
\begin{equation}
%
U(\beta) = \mathrm{T} \, \exp \left(
   - \int_0^\beta \! d\tau \, H_{int}(\tau) \right).
%
\end{equation}
%
The statistical averages $\langle ... \rangle_0^c$ are evaluated
with respect to the zero-order density matrix of the grand
canonical ensemble of localized polarons and free acoustical
phonons,
%
\begin{equation}
%
\frac{\displaystyle e^{-\beta H^0}}{\displaystyle
    \mathrm{Tr} \, e^{-\beta H^0}}
  = \prod\limits_{i}
\frac{\displaystyle e^{-\beta H_{ip}^0}}{\displaystyle \mathrm{Tr} \,
    e^{-\beta H_{ip}^0}} \prod\limits_{\mathbf{k}}
\frac{\displaystyle e^{-\beta \omega_{\mathbf{k}}
    b_{\mathbf{k}}^{\dag} b_{\mathbf{k}}}}
    {\displaystyle \mathrm{Tr} \,
    e^{-\beta \omega_{\mathbf{k}}
    b_{\mathbf{k}}^{\dag} b_{\mathbf{k}}}} \, .
\label{(17)}
%
\end{equation}
%
The subscript $c$ in Eq.\ (15) indicates that only connected diagrams
have to be taken into account. The polaron part of the density matrix
(17) is factorized with respect to the lattice sites. The on-site
polaron Hamiltonian contains the polaron-polaron interaction which
is proportional to the renormalized parameter $U$.
Therefore, this Hamiltonian can be diagonalized only by using
Hubbard operators \cite{Hubbard}. At this stage no special assumption is
made about the value of the quantity $U$ and its sign. So we can
set up the equations of motion for the dynamical
quantities in this general case. Wick's theorem of weakly coupled
quantum field theory can be used to evaluate statistical
averages of phonon operators, including, the propagator of
phonon clouds.

\subsection{Phonon-cloud propagators}

The zero-order one-phonon Matsubara Green's function has the form
%
\begin{eqnarray}
%
\sigma (x,x^{\prime}) & = &
   \sigma(\mathbf{x} - \mathbf{x}^{\prime}|\tau -\tau^{\prime})
 = \left \langle \mathrm{T} \,
   \pi_{\mathbf{x}}(\tau) \pi_{\mathbf{x}^{\prime}}(\tau^{\prime})
   \right \rangle_0
\nonumber \\ & = &
   \frac{1}{2N} \sum\limits_{\mathbf{k}}
   |\overline{g}(\mathbf{k})|^2
   \cos \mathbf{k(x-x^{\prime}})
\nonumber \\ & & \times
   \frac{\cosh \omega_{\mathbf{k}}(\beta/2 -
         |\tau -\tau^{\prime}|)}
        {\sinh \omega_{\mathbf{k}}\beta/2} ,
\label{(18)}
%
\end{eqnarray}
%
with
%
\begin{equation}
%
\pi_{\mathbf{x}}(\tau ) = \sum_j p_j(\tau)
     \overline{g}(\mathbf{R}_j - \mathbf{R}_x).
\nonumber
%
\end{equation}
%
Here $\mathbf{x}$ is again the position and $\tau$ the imaginary
time while $x$ in Eq.\ (18) stands for $(\mathbf{x},\tau)$

This function makes an essential contribution for small values of
distances $|\mathbf{x} - \mathbf{x}^{\prime}|$ and
$|\tau - \tau^{\prime}|$ close to zero or $\beta$. For
$\mathbf{x} = \mathbf{x}^{\prime}$ the minimum value of this function
is obtained for $|\tau - \tau^{\prime}| = \beta/2$. Since all approximations
in this paper concern the strong-coupling limit of the electron-phonon
interaction, we will use the series expansion of $\sigma(x,x^{\prime})$
near $\tau = 0$ and $\tau = \beta$:
%
\begin{equation}
%
\sigma (0|\tau ) = \left\{
\begin{array}{lc}
\sigma (0|0) - \omega_c\tau,  & \quad \tau \gtrsim 0 \\
\sigma (0|0) + \omega_c(\tau  - \beta), & \quad \tau \lesssim \beta
\end{array}
\right.
\label{(19)}
%
\end{equation}
%
with
%
\begin{equation}
%
\omega_c = \frac{1}{2N} \sum\limits_{\mathbf{k}}
     |\overline{g}(\mathbf{k})|^2\omega_{\mathbf{k}}
\label{(20)}
%
\end{equation}
%
as collective phonon cloud frequency \cite{Entel,Moskalenko4}. Besides
the one-phonon propagator we have also many-phonon cloud propagators.
There are two kind of one-cloud propagators, of which
$\phi(x|x^{\prime})$ is the normal-state one and
$\varphi(x|x^{\prime})$ the anomalous one of the superconducting state,
given by
%
\begin{eqnarray}
%
\phi(x|x^{\prime}) & = &
   \phi(\mathbf{x-x}^{\prime}|\tau -\tau^{\prime})
 = \langle \mathrm{T} \,
   e^{i\pi_{\mathbf{x}}(\tau) -
   i\pi_{\mathbf{x}^{\prime}}(\tau^{\prime})}
   \rangle_0
\nonumber \\ & = &
   \exp \left( -{\textstyle\frac{1}{2}}
   \langle \mathrm{T} \,
   [\pi_{\mathbf{x}}(\tau) -
   \pi_{\mathbf{x}^{\prime}}(\tau^{\prime})]^2
   \rangle_0\right)
\nonumber \\ & = &
   \exp [ -\sigma (0|0) + \sigma(\mathbf{x} - \mathbf{x}^{\prime}|
   \tau - \tau^{\prime}) ] ,
\label{(21)}
\\*[0.2cm]
\varphi(x|x^{\prime}) & = &
   \varphi(\mathbf{x-x}^{\prime}|\tau - \tau^{\prime})
 = \langle \mathrm{T} \,
   e^{i\pi_{\mathbf{x}}(\tau) + i\pi_{\mathbf{x}^{\prime}}
   (\tau^{\prime})}
   \rangle_0
\nonumber \\ & = &
   \exp \left( - {\textstyle\frac{1}{2}}
   \langle \mathrm{T} \,
   [\pi_{\mathbf{x}}(\tau) +
   \pi_{\mathbf{x}^{\prime}}(\tau^{\prime})]^2
   \rangle_0 \right)
\nonumber \\ & = &
   \exp [ -\sigma (0|0) - \sigma(\mathbf{x} - \mathbf{x}^{\prime}|
   \tau - \tau^{\prime}) ] .
\label{(22)}
%
\end{eqnarray}
%
For the first function the maximum value of the one-phonon propagator
$\sigma(\mathbf{x}|\tau)$ is favored while for the second one the
corresponding minimum value is preferred. The Fourier representations in
$\tau$-space have the form
%
\begin{subequations}
%
\begin{align}
\phi(0|\tau) & = \frac{1}{\beta} \sum\limits_\Omega e^{-i\Omega_n\tau}
   \widetilde{\phi}(i\Omega_n) ,
\\
\varphi(0|\tau) & = \frac{1}{\beta} \sum\limits_\Omega e^{-i\Omega_n\tau}
   \widetilde{\varphi}(i\Omega_n) ,
\end{align}
\label{(23a,b)}
%
\end{subequations}
%
where
%
\begin{subequations}
%
\begin{align}
\widetilde{\phi }(i\Omega_n) & = \int\nolimits_o^\beta \!
   d\tau \, e^{i\Omega \tau} e^{-\sigma(0|0) + \sigma(0|\tau)},
\\
\widetilde{\varphi}(i\Omega_n) & = \int\nolimits_o^\beta \!
   d\tau \, e^{i\Omega\tau} e^{-\sigma(0|0) - \sigma(0|\tau)} .
\end{align}
\label{(24a,b)}
%
\end{subequations}
%
Here is $\Omega _n$ the even Matsubara frequency
$\Omega_n = 2\pi n/\beta$. In order to find the Fourier
representations of these functions we have used the peculiarities of the
$\sigma$-propagator in the strong-coupling limit of the
electron-phonon interaction. As proven in Appendix A, the first
propagator can be written as
$$
  \phi(\mathbf{x}|\tau^{\prime}) \simeq \phi(\mathbf{x})\phi(\tau),
  \quad \phi(\mathbf{x})\approx \delta_{\mathbf{x},0}
$$
with
%
\begin{equation}
%
\widetilde{\phi}(i\Omega_n) \approx \frac{2\omega_c}
  {(i\Omega_n)^2 - (\omega_c)^2}, \quad
  \widetilde{\phi}(\mathbf{q}) \approx 1.
\label{(25)}
%
\end{equation}
%
A more realistic value for $\widetilde{\phi}(\mathbf{q})$ is obtained
by using the dependence of $\sigma(\mathbf{x}|\tau)$ on small
values of $\mathbf{x}$. In this more precise approximation we find
%
\begin{equation}
%
\widetilde{\phi}(\mathbf{q}) = \left( \frac{2\pi}{\sigma_1} \right)^{3/2}
   \! e^{- \mathbf{q}^2/(2\sigma_1)}, \quad
   \phi(\mathbf{x}) \simeq e^{- \sigma_1\mathbf{x}^2/2},
 \label{(26)}
%
\end{equation}
%
where
%
\begin{equation}
%
\sigma_1 = \frac{1}{6N} \sum\limits_{\mathbf{k}}
   |\overline{g}(\mathbf{k})|^2 \, \mathbf{k}^2
   \coth {\textstyle\frac{1}{2}} \omega_{\mathbf{k}}\beta .
\label{(27)}
%
\end{equation}
%
This result has been obtained by an expansion of $\cos \mathbf{kx}$
in terms of $\mathbf{x}$. We also assume that
$\overline{g}(\mathbf{k})$ depends on $\mathbf{k}$ only through its
modulo $|\mathbf{k}|$. Then the Fourier representation of the normal
phonon cloud propagator is a Lorentzian and therefore the
time dependence of this phonon cloud corresponds to that of an
oscillator with the large collective frequency $\omega_c$. For the
anomalous one-cloud propagator $\varphi(x|x^{\prime})$ we obtain
in this approximation a Gaussian representation, see Appendix A:
%
\begin{eqnarray}
%
\widetilde{\varphi}(i\Omega _n) & = & \sqrt{2\pi/\sigma_2} \,
   \exp \left[ {\textstyle\frac{1}{2}} i\beta \Omega_n \right.
\nonumber \\ & - & {} \left.
   \sigma(0|0) - \sigma(0|\beta/2) - (\Omega_n)^2/(2\sigma_2) \right] ,
\label{(28)}
%
\end{eqnarray}
%
where
%
\begin{equation}
%
\sigma_2 = \sigma^{\prime \prime} (0|\beta/2) .
\label{(29)}
%
\end{equation}
%
The space dependence of $\varphi(\mathbf{x}|i\Omega_n)$
is more complicated compared to the space dependence of
$\phi(\mathbf{x}|i\Omega_n)$ because we cannot restrict the discussion
to small values of $|\mathbf{x}|$. In the following we will
discuss many-cloud propagators, both in the normal and superconducting
states. We start with the two-cloud propagators [as before,
$x = (\mathbf{x},\tau)$]:
%
\begin{widetext}
%
\begin{eqnarray}
%
\phi_2(x_1,x_2|x_3,x_4) & = &
   \langle \mathrm{T} \, \exp \left[
   i\left[ \pi_{\mathbf{x}_1}(\tau_1) \right]
 + \pi_{\mathbf{x}_2}(\tau_2) - \pi_{\mathbf{x}_3}(\tau_3)
 - \pi_{\mathbf{x}_4}(\tau_4) \right] \rangle_0
\nonumber \\ & = &
   \exp \left( -{\textstyle\frac{1}{2}}
   \langle \mathrm{T} \, \left[
   \pi_{\mathbf{x}_1}(\tau_1) + \pi_{\mathbf{x}_2}(\tau_2)
 - \pi_{\mathbf{x}_3}(\tau_3) - \pi_{\mathbf{x}_4}(\tau_4)
   \right]^2 \rangle_0 \right)
\nonumber \\ & = &
   \exp \big( \Sigma(\mathbf{x}_1,\tau_1;\mathbf{x}_2,\tau_2|
   \mathbf{x}_3,\tau_3;\mathbf{x}_4,\tau_4) \big) ,
\label{(30)}
%
\end{eqnarray}
%
\begin{eqnarray}
%
\varphi_2(x_1,x_2,x_3|x_4) & = &
   \langle \mathrm{T} \, \exp \left[
   i\left[ \pi_{\mathbf{x}_1}(\tau_1) \right]
 + \pi_{\mathbf{x}_2}(\tau_2) + \pi_{\mathbf{x}_3}(\tau_3)
 - \pi_{\mathbf{x}_4}(\tau_4)\right] \rangle_0
\nonumber \\ & = &
   \exp \left( - {\textstyle\frac{1}{2}}
   \langle \mathrm{T} \, \left[
   \pi_{\mathbf{x}_1}(\tau_1) + \pi_{\mathbf{x}_2}(\tau_2)
 + \pi_{\mathbf{x}_3}(\tau_3) - \pi_{\mathbf{x}_4}(\tau_4)
   \right]^2 \rangle_0 \right)
\nonumber \\ & = &
   \exp \big( \Sigma(\mathbf{x}_1,\tau_1;\mathbf{x}_2,\tau_2;
   \mathbf{x}_3,\tau_3|\mathbf{x}_4,\tau_4) \big) ,
\label{(31)}
%
\end{eqnarray}
%
where
%
\begin{eqnarray}
%
\lefteqn{
\Sigma(\mathbf{x}_1,\tau_1;\mathbf{x}_2,\tau_2|
   \mathbf{x}_3,\tau_3;\mathbf{x}_4,\tau_4)
 = \sigma(\mathbf{x}_1 - \mathbf{x}_3||\tau_1 - \tau_3|)
 + \sigma(\mathbf{x}_1 - \mathbf{x}_4||\tau_1 - \tau_4|)
}
\nonumber \\
& & {}
 + \sigma(\mathbf{x}_2 - \mathbf{x}_4||\tau_2 - \tau_4|)
 + \sigma(\mathbf{x}_2 - \mathbf{x}_3||\tau_2 - \tau_3|)
 - \sigma(\mathbf{x}_1 - \mathbf{x}_2||\tau_1 - \tau_2|)
 - \sigma(\mathbf{x}_3 - \mathbf{x}_4||\tau_3 - \tau_4|)
 - 2\sigma(0|0),
\label{(32)}
%
\end{eqnarray}
%
\begin{eqnarray}
%
\lefteqn{
\Sigma(\mathbf{x}_1,\tau_1;\mathbf{x}_2,\tau_2;\mathbf{x}_3,\tau_3|
   \mathbf{x}_4,\tau_4)
 = \sigma(\mathbf{x}_1 - \mathbf{x}_4||\tau_1 - \tau_4|)
 + \sigma(\mathbf{x}_2 - \mathbf{x}_4||\tau_2 - \tau_4|)
}
\nonumber \\
& & {}
 + \sigma(\mathbf{x}_3 - \mathbf{x}_4||\tau_3 - \tau_4|)
 - \sigma(\mathbf{x}_1 - \mathbf{x}_2||\tau_1 - \tau_2|)
 - \sigma(\mathbf{x}_1 - \mathbf{x}_3||\tau_1 - \tau_3|)
 - \sigma(\mathbf{x}_2 - \mathbf{x}_3||\tau_2 - \tau_3|)
 - 2\sigma(0|0) .
\label{(33)}
%
\end{eqnarray}
%
The following relations exist between two- and one-cloud
Green's functions:
%
\begin{subequations} \begin{align}
%
\phi_2(x_1,x_2|x_3,x_4) & = \phi(x_1|x_3) \phi(x_2|x_4)
   \exp \left[ \sigma(x_1|x_4) + \sigma(x_2|x_3)
 - \sigma(x_1|x_2) - \sigma(x_3|x_4) \right]
\\
&
 = \phi(x_1|x_4) \phi(x_2|x_3)
   \exp \left[ \sigma(x_1|x_3) + \sigma(x_2|x_4)
 - \sigma(x_1|x_2) - \sigma(x_3|x_4) \right] ,
\label{(34a,b)}
%
\end{align} \end{subequations}
%
\begin{subequations} \begin{align}
%
\varphi_2(x_1,x_2,x_3|x_4) & = \varphi(x_1|x_2) \phi(x_3|x_4)
   \exp \left[ \sigma(x_1|x_4) + \sigma(x_2|x_4) - \sigma(x_1|x_3)
  - \sigma(x_2|x_3) \right]
\\ & =
   \varphi(x_1|x_3) \phi(x_2|x_4)
   \exp \left[ \sigma(x_1|x_4) + \sigma(x_3|x_4) - \sigma(x_1|x_2)
 - \sigma(x_2|x_3) \right]
\\ & =
   \varphi(x_2|x_3) \phi(x_1|x_4)
   \exp \left[ \sigma(x_2|x_4) + \sigma(x_3|x_4)
 - \sigma(x_1|x_2) - \sigma(x_1|x_3) \right] .
\label{(35a,b,c)}
%
\end{align} \end{subequations}
%
\end{widetext}
%
Many-cloud phonon propagators will be present in all diagrams of
the thermodynamical perturbation theory to be formulated here.
As above equations show, all sites of the diagrams are joint and appear to be
connected in the presence of acoustical phonons. In order to
classify the diagrams as connected and disconnected ones, it is
necessary to have the analogy of Wick's theorem for
many-cloud propagators similar to the theorem we had formulated
for correlated electrons \cite{Vladimir,Vakaru,Entel}. In the absence
of such a theorem we cannot prove the existence of a
linked-cluster theorem for the thermodynamical potential and for
other extensive quantities.

This problem has been discussed in detail in Ref.\ \cite{Moskalenko5},
however, only now we are able to present a solution. In order to
obtain this solution, we observe that the two-cloud functions
determined by Eqs.\ (34) and (35) have their maximum values when the
arguments of the normal one-cloud functions $\phi (x|x^{\prime})$
coincide ($x = x^{\prime}$) and the corresponding exponential factors
close to these arguments approach one. There are several possibilities
to achieve this and all of them have to be taken into account. We
assume that as main approximation the following expressions
for the two-cloud propagators will result,
%
\begin{eqnarray}
%
\phi_2(x_1,x_2|x_3,x_4) & = & \phi(x_1|x_3) \phi(x_2|x_4)
\nonumber \\ & + &
   \phi(x_1|x_4) \phi(x_2|x_3)
\nonumber \\ & + &
   \phi_2^{ir}(x_1,x_2|x_3,x_4) ,
\label{(36)}
%
\end{eqnarray}
%
\begin{eqnarray}
%
\varphi_2(x_1,x_2,x_3|x_4) & = & \varphi(x_1|x_2) \phi(x_3|x_4)
\nonumber \\ & + &
   \varphi(x_1|x_3) \phi(x_2|x_4)
\nonumber \\ & + &
   \varphi(x_2|x_3) \phi(x_1|x_4)
\nonumber \\ & + &
   \phi_2^{ir}(x_1,x_2,x_3|x_4) .
\label{(37)}
%
\end{eqnarray}
%
These last equations also define the irreducible parts of the
two-cloud propagators or phonon-cloud cumulants. In the
strong-coupling limit the irreducible functions are small and can
be omitted as shown below. The validity of this statement is discussed
in Appendix A, in which the Fourier representation of the normal
two-cloud propagator,
%
\begin{eqnarray}
%
\lefteqn{\hspace*{-1.0cm}
   \phi_2(\mathbf{x}_1,i\Omega_1;\mathbf{x}_2,i\Omega_2|
   \mathbf{x}_3,i\Omega_3;x_4,i\Omega_4)
 = \int_0^\beta \! ... \! \int_0^\beta \!
   d\tau _1...d\tau_4 \,
}
\nonumber \\*[0.1cm] && \times
   \exp \left(
   i\Omega_1\tau_1 + i\Omega_2\tau_2 - i\Omega_3\tau_3
 - i\Omega_4\tau_4 \right)
\nonumber \\ &&
   \times \phi_2(\mathbf{x}_1,\tau_1;\mathbf{x}_2,\tau_2|
   \mathbf{x}_3,\tau_3;\mathbf{x}_4,\tau_4) ,
\label{(38)}
%
\end{eqnarray}
%
has been calculated in the strong-coupling limit leading to
%
\begin{eqnarray}
%
\lefteqn{\hspace*{-0.5cm}
   \phi_2(\mathbf{x}_1,i\Omega_1;\mathbf{x}_2,i\Omega_2|
   \mathbf{x}_3,i\Omega_3;x_4,i\Omega_4)
}
\nonumber \\ && {} \simeq
   \phi(\mathbf{x}_1 - \mathbf{x}_2|i\Omega_1) \,
   \delta_{\Omega_1\Omega_3} \,
   \phi(\mathbf{x}_2 - \mathbf{x}_4|i\Omega_2) \,
   \delta_{\Omega_2\Omega_4}
\nonumber \\ && {} +
   \phi(\mathbf{x}_1 - \mathbf{x}_4|i\Omega_1) \,
   \delta_{\Omega_1\Omega_4} \,
   \phi (\mathbf{x}_2 - \mathbf{x}_3|i\Omega_2) \,
   \delta_{\Omega_2\Omega_3}.
\label{(39)}
%
\end{eqnarray}
%
The last equation shows that in this limit the
irreducible function is not relevant and Wick's theorem has a
simple form, which does no contain significant irreducible
contributions. Similarly we obtain for $\varphi_2$ a form without
irreducible contributions,
%
\begin{eqnarray}
%
\lefteqn{
\varphi _2(\mathbf{x}_1,i\Omega _1;\mathbf{x}_2,i\Omega _2;\mathbf{x}%
_3,i\Omega _3|x_4,i\Omega _4)
}
\nonumber \\ &&  {} =
\int_0^\beta \! ... \! \int_0^\beta \! d\tau_1 ... d\tau_4 \,
   e^{-\left( i\Omega_1\tau_1 + i\Omega_2\tau_2 + i\Omega_3\tau_3
 - i\Omega_4\tau_4\right) }
\nonumber \\*[0.1cm] &&
   \times \varphi_2(\mathbf{x}_1,\tau_1;\mathbf{x}_2,\tau_2;
   \mathbf{x}_3,\tau_3|\mathbf{x}_4,\tau_4)
\label{(40)}
\\*[0.2cm] && {} \hspace*{-0.5cm} \simeq
   \varphi(\mathbf{x}_1 - \mathbf{x}_2|i\Omega_1) \,
   \delta_{\Omega_2,-\Omega_1}
   \phi(\mathbf{x}_3 - \mathbf{x}_4|i\Omega_3) \,
   \delta_{\Omega_3,\Omega_4}
\nonumber \\ && {} \hspace*{-0.5cm} +
   \varphi(\mathbf{x}_1 - \mathbf{x}_3|i\Omega_1) \,
   \delta_{\Omega_3,-\Omega_1} \,
   \phi(\mathbf{x}_2 - \mathbf{x}_4|i\Omega_2) \,
   \delta_{\Omega_2,\Omega_4}
\nonumber \\ && {} \hspace*{-0.5cm} +
   \varphi(\mathbf{x}_2 - \mathbf{x}_3|i\Omega_2) \,
   \delta_{\Omega_3,-\Omega_2} \,
   \phi(\mathbf{x}_1 - \mathbf{x}_4|i\Omega_4) \,
   \delta_{\Omega_1,\Omega_4} .
\label{(41)}
%
\end{eqnarray}
%
These results correspond to our preliminary estimates that the
irreducible parts in  Eqs.\ (36) and (37) can be omitted
because they are not important in the strong-coupling limit, see
Appendix A. Hence, without the irreducible parts the equations
assume a form corresponding to Wick's theorem applied to two-cloud
propagators. This can easily be generalized to the case of a
larger number of clouds.
Thus, there is an analogy of having a generalized Wicks's theorem
for the case of correlated electrons \cite{Vladimir,Vakaru} and a
corresponding theorem for correlated phonon clouds. This allows
us now to develop a thermodynamical perturbation theory for
correlated electrons interacting strongly with phonons.

As is shown below the tunneling of polarons between lattice sites
can be accompanied by either preserving or by exchanging phonon
clouds. In the strong-coupling limit these clouds are heavy,
therefore, in case of preserving the cloud, the effective matrix
transfer matrix element is considerably diminished leading to
band narrowing effects. In the other case, when clouds are exchanged,
the transfer matrix element and the electronic band width remain
unchanged.

\section{Polaron Green's functions}

\subsection{Local approximation}

The zero-order one-polaron Green's function is given by
%
\begin{align}
%
G_p^0(x,x^{\prime}) & =
   - \langle \mathrm{T} \, c_{\mathbf{x}\sigma}(\tau)
     \overline{c}_{\mathbf{x}^{\prime}\sigma^{\prime}}(\tau)
     \rangle_0
\nonumber \\ & =
   - \langle \mathrm{T} \, a_{\mathbf{x}\sigma}(\tau)
     \overline{a}_{\mathbf{x}^{\prime}\sigma^{\prime}}(\tau)
     \rangle_0 \,
     \phi(\mathbf{x},\tau|\mathbf{x}^{\prime},\tau^{\prime})
\nonumber \\ & =
     \delta_{\mathbf{x},\mathbf{x}^{\prime}}
     \delta_{\sigma,\sigma^{\prime}} \,
     G_\sigma^0(\tau - \tau^{\prime}) \,
     \phi(\tau - \tau^{\prime}) ,
\label{(42)}
%
\end{align}
%
where $x$ stands now for $x=(\mathbf{x},\sigma,\tau)$. In order
to discuss the influence of the collective mode on
$G_p^0(x,x^{\prime})$, we write down its Fourier transformation
by making use of Eqs.\ (25) (see Ref.\ \onlinecite{Entel2}):
%
\begin{align}
%
\widetilde{G}_{p\sigma}^0(i\omega_n) & =
   \int_0^\beta \! d\tau \,
   e^{i\omega_n\tau} G_{p\sigma}^0(\tau)  ,
\label{(43)}
\\*[0.2cm]
\widetilde{G}_{p\sigma}^0(i\omega _n) & =
   \frac{1}{Z_0} \left[
   \frac{e^{-\beta E_0} + \overline{N}(\omega_c)
   \left( e^{-\beta E_0} + e^{-\beta E_\sigma} \right)}
   {i\omega_n + E_0 - E_\sigma - \omega_c} \right.
\nonumber \\ & + {}
   \frac{e^{-\beta E_\sigma} + \overline{N}(\omega_c)
   \left( e^{-\beta E_0} + e^{-\beta E_\sigma } \right)}
   {i\omega_n + E_0 - E_\sigma + \omega_c}
\nonumber \\ & + {}
   \frac{e^{-\beta E_{-\sigma}} + \overline{N}(\omega_c)
   \left( e^{-\beta E_\sigma} + e^{-\beta E_2} \right)}
   {i\omega_n + E_{-\sigma} - E_2 - \omega_c}
\nonumber \\ & \left. {} + {}
   \frac{e^{-\beta E_2} + \overline{N}(\omega_c)
   \left( e^{-\beta E_\sigma} + e^{-\beta E_2} \right)}
   {i\omega_n + E_{-\sigma} - E_2 + \omega_c} \right] ,
\label{(44)}
%
\end{align}
%
where $\omega_n$ is the odd Matsubara frequency and
%
\begin{subequations} \begin{align}
%
Z_0 & = 1 + e^{-\beta E_\sigma} + e^{-\beta E_{-\sigma}}
   + e^{-\beta E_2} ,
\label{(45a)}
\\
E_{0} & = 0,\quad E_{\pm \sigma} = \varepsilon,\quad E_2 =U + 2\varepsilon,
\label{(45b)}
\\
\overline{n}(\varepsilon) & = (e^{\beta \varepsilon} + 1)^{-1} , \quad
\overline{N}(\omega_c) = (e^{\beta \omega_c} - 1)^{-1} .
\label{(45c)}
%
\end{align} \end{subequations}
%
Equation (44) shows that the on-site transition energies of
polarons are changed by the energy $\omega_c$ of the collective
mode. The delocalization of polarons due to hopping and intersite
Coulomb interaction leads to broadening of the polaronic
energy levels. The polaron propagator has the following antisymmetry
property:
%
\begin{equation}
%
\widetilde{G}_{p\sigma }^0(i\omega_n;\varepsilon;\omega_c)
   = -\widetilde{G}_{p\sigma }^0(-i\omega_n;-\varepsilon - U;\omega_c)
\label{(46)}
%
\end{equation}

\subsection{Expansion around the atomic limit}

We will now investigate polaron delocalization under the influence
of $H_{int}$ in Eq.\ (10) by making use of thermodynamical
perturbation theory in the interaction representation. The averages of
chronological products of interactions are reduced to $n$-particle
Green's functions of the atomic system, which can be factorized into
independent local averages of electron operators and chronological
products of phonon operators. The procedure relies on a
generalized Wick's theorem for electron operators, which takes into
account the strong local electronic correlations, and Wick's theorem
for phonon cloud operators. In addition to the normal one-polaron
propagator in Eq.\ (15), we will also investigate the anomalous
propagators defined by
%
\begin{subequations} \begin{align}
%
F_p(x|x^{\prime}) & = - \langle \mathrm{T} \,
   c_x c_{x^{\prime}} U(\beta) \rangle_0^c ,
\label{(47b)}
\\
\overline{F}_p(x|x^{\prime}) & = - \langle \mathrm{T} \,
   \overline{c}_x \overline{c}_{x^{\prime}} U(\beta) .
   \rangle_0^c
\label{(47b)}
%
\end{align} \end{subequations}
%
As before, $x$ stands for $(\mathbf{x},\sigma,\tau)$.
The easiest way to establish (47) is to make use of a local source
term of Cooper pairs,
%
\begin{equation*}
%
H_\Delta^0 = \Delta \sum\limits_{i} \left(
   a_{i\uparrow}^{\dag} a_{i\downarrow}^{\dag }
 + a_{i\downarrow} a_{i\uparrow} \right) ,
%
\end{equation*}
%
which is added to the local Hamiltonian (2) and switched off at the
end of the calculation.

In first order perturbation theory the contributions to the normal
polaron Green's function (15) and anomalous Green's function (47a)
are shown in Fig.\ 1(a) and Fig.\ 1(b), respectively.

\begin{figure*}[t]
%
\centering
\includegraphics[width=0.65\textwidth,clip]{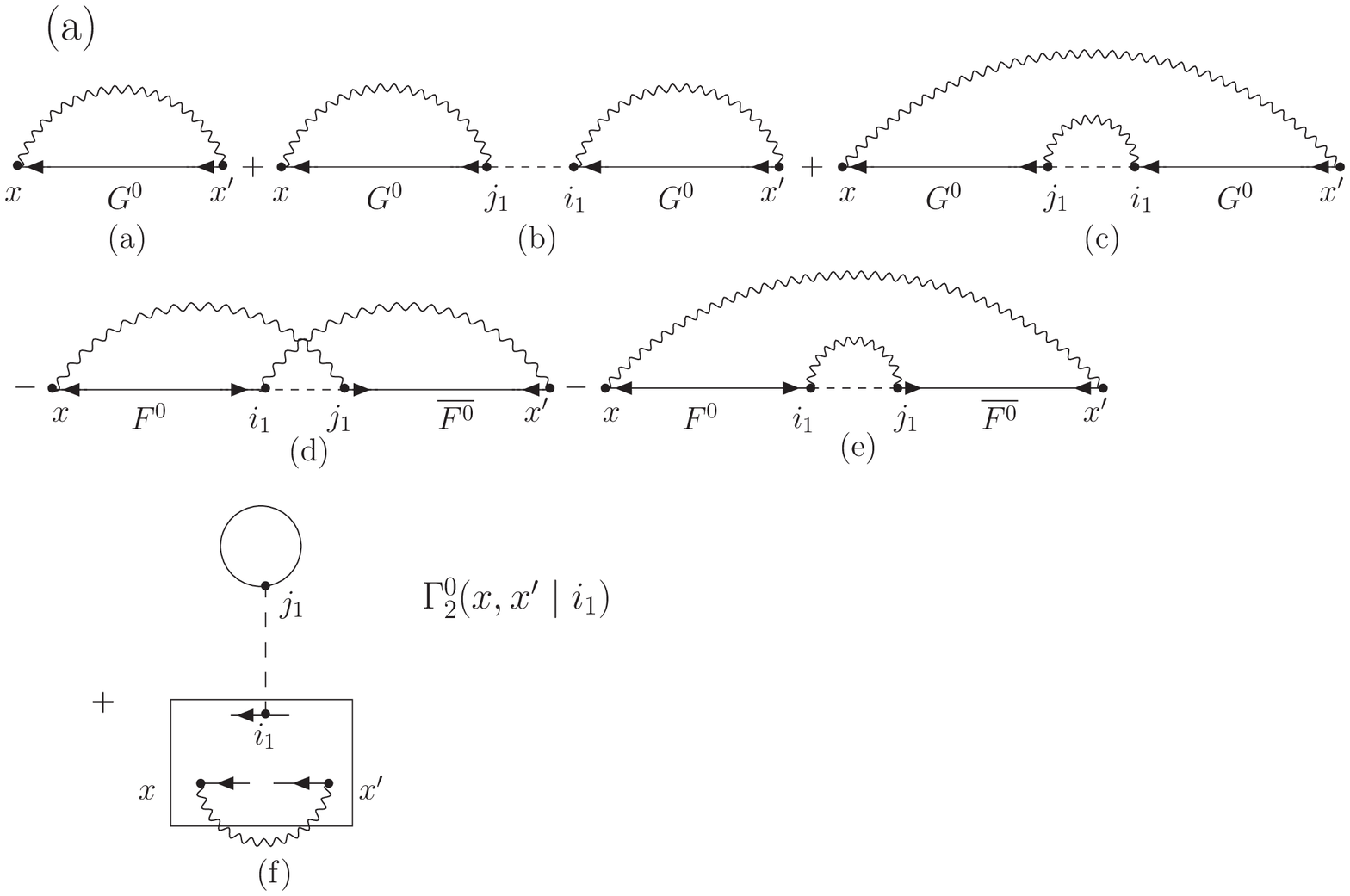}
\includegraphics[width=0.65\textwidth,clip]{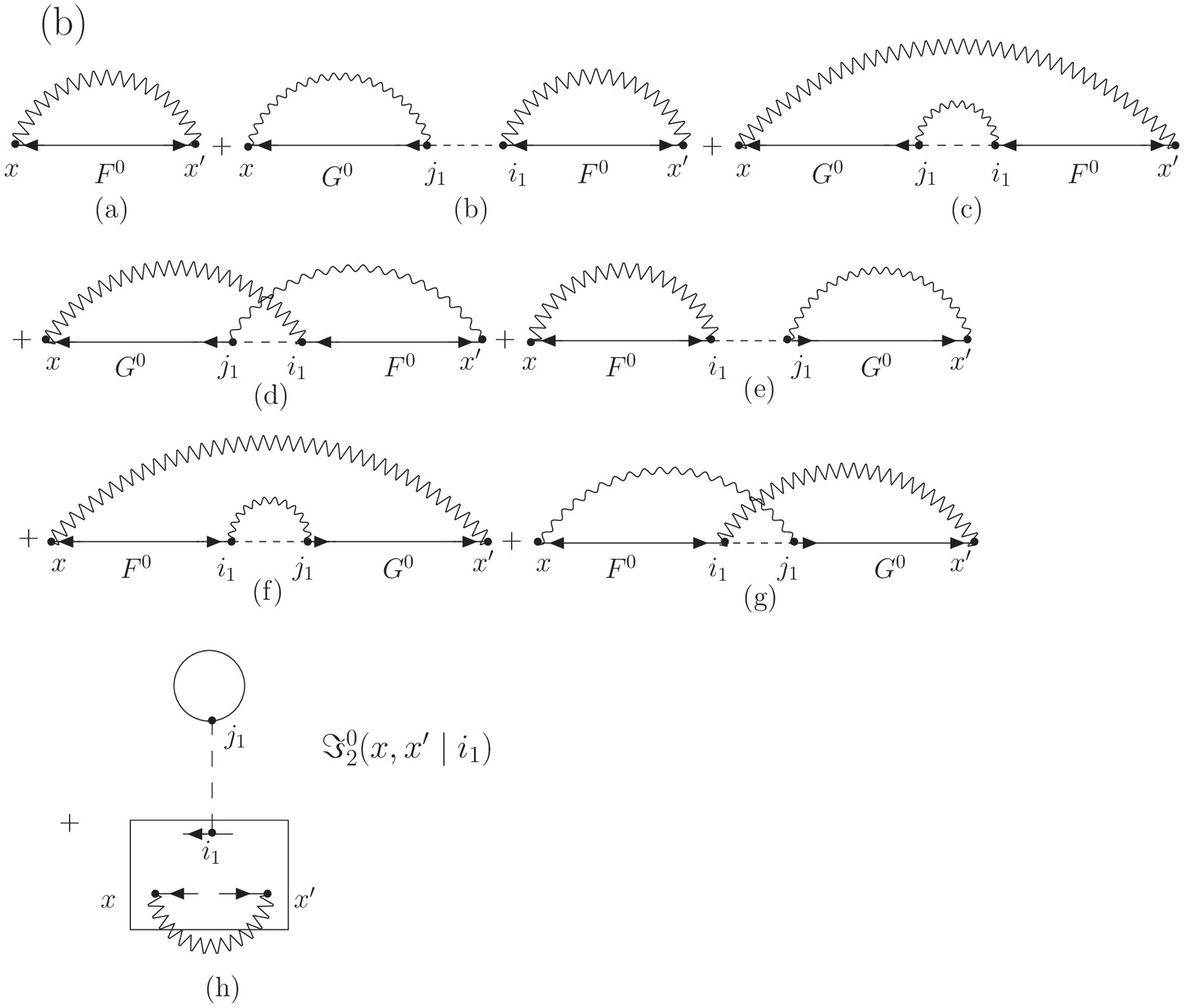}
\caption{
The simplest diagrams contributing to the normal (a) and
anomalous (b) one-polaron Green's functions. Solid lines with
arrows in same direction represent normal ($G^0$) and lines with arrows in
opposite directions anomalous ($\overline{F}^0$, $F^0$) propagators,
respectively. Short-dashed lines are for the hopping matrix elements
$t(i - j)$, long-dashed lines represent the
direct polaron-polaron interactions $V(i - j)$, the wiggly lines
stand for the normal phonon (cloud) propagators $\phi(x|x^{\prime})$,
and the zigzag lines are the anomalous phonon (cloud)
propagators $\varphi(x|x^{\prime})$. Rectangles depict the
correlation functions $\Gamma_2^0$ and $\Im_2^0$.
}
\label{fig-1}
%
\end{figure*}

The diagrammatic elements are self-explanatory (see caption of Fig.\
1). Since the correlation functions $\Gamma_2^0(x,x^{\prime}|i)$
and $\Im_2^0(x,x^{\prime}|i)$ and the two-particle irreducible
function $G_2^{0\ ir}[x_1,x_2|x_{3,}x_4]$ are local quantities, all
site indices are equal:
%
\begin{subequations} \begin{align}
%
\Gamma _2^0(x,x^{\prime}|i) & =
   \delta_{\mathbf{x},\mathbf{x}^{\prime}} \delta_{\mathbf{x},i} \,
   \Gamma_2^0(\sigma,\tau;\sigma^{\prime},\tau^{\prime}|\tau_1) ,
\label{(48a)}
\\
\Im_2^0(x,x^{\prime}|i) & =
   \delta_{\mathbf{x},\mathbf{x}^{\prime}} \delta_{\mathbf{x},i} \,
   \Im_2^0(\sigma,\tau;\sigma^{\prime},\tau^{\prime}|\tau_1) ,
 \label{(48b)}
%
\end{align} \end{subequations}
%
where
%
\begin{subequations} \begin{align}
%
\Gamma_2^0(\sigma,\tau;\sigma^{\prime},\tau^{\prime}|\tau_1) & =
   \langle \mathrm{T} \,
   a_\sigma(\tau) \overline{a}_{\sigma^{\prime}}(\tau^{\prime})
   n(\tau_1) \rangle_0
\nonumber \\ & - {}
   \langle \mathrm{T} \,
   a_\sigma(\tau) \overline{a}_{\sigma^{\prime}}(\tau^{\prime})
   \rangle_0 \langle n(\tau_1) \rangle_0  ,
\label{(49a)}
\\
\Im_2^0(\sigma,\tau;\sigma^{\prime},\tau^{\prime}|\tau_1) & =
   \langle \mathrm{T} \,
   a_\sigma(\tau) a_{\sigma^{\prime}}(\tau^{\prime})
   n(\tau_1) \rangle_0
\nonumber \\ & - {}
   \langle \mathrm{T}
   a_\sigma(\tau) a_{\sigma^{\prime}}(\tau^{\prime}) \rangle_0
   \langle n(\tau_1)\rangle_0 .
\label{(49b)}
%
\end{align} \end{subequations}
%
These functions can be compared with the two-particle irreducible
quantities of Ref.\ \onlinecite{Vladimir,Vakaru} defined by
%
\begin{subequations} \begin{align}
%
G_2^{0 \, ir}[x_1,x_2|x_{3,}x_4] & =
   \delta_{\mathbf{x}_1,\mathbf{x}_2}
   \delta_{\mathbf{x}_1,\mathbf{x}3}
   \delta_{\mathbf{x}_1,\mathbf{x}_4}
\nonumber \\ & \times
   G_2^{0 \, ir}[\sigma_1\tau_1;\sigma_2\tau_2|
                 \sigma_3\tau_3;\sigma_4\tau_4] ,
\label{(50a)}
\\
G_2^{0 \, ir}[x_1,x_2,x_3|x_4] & =
   \delta_{\mathbf{x}_1,\mathbf{x}_2}
   \delta_{\mathbf{x}_1,\mathbf{x}_3}
   \delta_{\mathbf{x}_1,\mathbf{x}_4}
\nonumber \\ & \times
   G_2^{0 \, ir}[\sigma_1\tau_1;\sigma_2\tau_2;
                 \sigma_3\tau_3|\sigma_4\tau_4] ,
\label{(50b)}
%
\end{align} \end{subequations}
%
where
%
\begin{eqnarray}
%
\lefteqn{\hspace*{-0.5cm}
G_2^{0 \, ir}[\sigma_1\tau_1;\sigma_2\tau_2|
   \sigma_3\tau_3;\sigma_4\tau_4]
}
\nonumber \\ && {} =
   \langle \mathrm{T}
   a_{\sigma_1}(\tau_1)a_{\sigma_2}(\tau_2)
   \overline{a}_{\sigma_3}(\tau_3) \overline{a}_{\sigma_4}(\tau_4)
   \rangle_0
\nonumber \\ && {} -
   \langle \mathrm{T} \,
   a_{\sigma_1}(\tau_1) \overline{a}_{\sigma_4}(\tau_4) \rangle_0 \,
   \langle \mathrm{T} \,
   a_{\sigma_2}(\tau_2) \overline{a}_{\sigma 3}(\tau_3) \rangle_0
\nonumber \\ && {} +
   \langle \mathrm{T} \,
   a_{\sigma_1}(\tau_1) \overline{a}_{\sigma_3}(\tau 3)\rangle_0 \,
   \langle \mathrm{T} \,
   a_{\sigma_2}(\tau_2) \overline{a}_{\sigma_4}(\tau_4) \rangle_0 .
\label{(51)}
%
\end{eqnarray}
%
The second function, $G_2^{0 \, ir}[x_1,x_2,x_3|x_4]$, can be obtained
from Eq.\ (51) by replacing the operator
$\overline{a}_{\sigma_3}(\tau_3)$ by $a_{\sigma_3}(\tau_3)$
and corresponds to an anomalous superconducting contribution.
Between the non-full cumulant
$\Gamma_2^0(\sigma,\tau;\sigma^{\prime},\tau^{\prime}|\tau_1)$
and the full one $G_2^{0 \, ir}$ exists the following relation
%
\begin{eqnarray}
%
\hspace*{-0.5cm}
\Gamma_2^0(\sigma,\tau;\sigma^{\prime},\tau^{\prime}|\tau_1) & = &
   -\sum\limits_{\sigma_1}
   G_2^{0 \, ir}[\sigma\tau;\sigma_1\tau_1|\sigma_1\tau_1^{+};
   \sigma^{\prime}\tau^{\prime}]
\nonumber \\ & + &
   \sum\limits_{\sigma 1}
   G^0(\sigma\tau|\sigma_1\tau_1) \, G^0(\sigma_1\tau_1|
   \sigma^{\prime}\tau^{\prime}) .
\label{(52)}
%
\end{eqnarray}
%
In second order perturbation theory we have to deal with
more complicated functions like
%
\begin{subequations} \begin{align}
%
\Gamma _3(x,x^{\prime}|i_1,i_2) & =
   \delta_{\mathbf{x},\mathbf{x}^{\prime}} \,
   \delta_{\mathbf{x},i_1} \, \delta _{\mathbf{x},i_2} \,
   \Gamma_3(\sigma\tau;\sigma^{\prime}\tau^{\prime}|\tau_1,\tau_2)) ,
\label{(53a)}
\\
\Gamma_3(\sigma\tau;\sigma^{\prime}\tau^{\prime}|
   \tau_1,\tau_2) & =
   \langle \mathrm{T} \,
   a_\sigma(\tau) \overline{a}_{\sigma^{\prime}}(\tau^{\prime})
   n(\tau_1) n(\tau_2) \rangle_0
\nonumber \\ & -
   \langle \mathrm{T}
   a_\sigma(\tau) \overline{a}_{\sigma^{\prime}}(\tau^{\prime})
   n(\tau_1) \rangle_0 \, \langle n(\tau_2) \rangle_0
\nonumber \\ & -
   \langle \mathrm{T}
   a_\sigma (\tau) \overline{a}_{\sigma^{\prime}}(\tau^{\prime})
   n(\tau_2) \rangle_0 \, \langle n(\tau_1) \rangle_0
\nonumber \\ & -
   \langle a_\sigma(\tau)
   \overline{a}_{\sigma^{\prime}}(\tau^{\prime})
   \rangle_0 \, \langle n(\tau_1) n(\tau_2) \rangle_0
\nonumber \\ & +
   2 \langle a_\sigma(\tau) \overline{a}_{\sigma^{\prime}}
   (\tau^{\prime}) \rangle_0 \langle n(\tau_1)
   \rangle_0 \langle \, n(\tau_2) \rangle_0 ,
\label{(53b)}
%
\end{align} \end{subequations}
%
with the following relation between the non-full and full
three-particle cumulants,
%
\begin{eqnarray}
%
\lefteqn{\hspace*{-0.5cm}
\Gamma_3(\sigma \tau ;\sigma ^{\prime }\tau ^{\prime }|
   \tau _1,\tau_2)
}
\nonumber \\ && {} \hspace*{-0.4cm} =
 - \sum_{\sigma_1\sigma_2}
   G_3^{0 \, ir}[\sigma\tau \sigma_1\tau_1;\sigma_2\tau_2|
   \sigma_1\tau_1^{+};\sigma_2\tau_2^{+};\sigma^{\prime}\tau^{\prime}]
\nonumber \\ && {} \hspace*{-0.4cm} -
   \sum_{\sigma_1\sigma_2}
   G^0(\sigma\tau|\sigma_1\tau_1)
   G^0(\sigma_1\tau_1|\sigma_2\tau_2)
   G^0(\sigma_2\tau_2|\sigma^{\prime}\tau^{\prime})
\nonumber \\ && {} \hspace*{-0.4cm} -
   \sum_{\sigma_1\sigma_2}
   G^0(\sigma\tau|\sigma_2\tau_2)
   G^0(\sigma_2\tau_2|\sigma_1\tau_1)
   G^0(\sigma_1\tau_1|\sigma^{\prime}\tau^{\prime}) ,
\label{(54)}
%
\end{eqnarray}
%
where $\tau _1^{+}=\tau _1+0$ and $\tau _2^{+}=\tau _2+0$.
The diagrams of Fig.\ 1 contain also this element which is
marked as a circle at the end of the long-dashed lines.
This element is the average of the electron number operator
$\widehat{n}$. The contribution of such diagrams is proportional to
$V(\mathbf{k} = 0)$ which is equal to zero. Hence, they can be omitted.
In still higher order perturbation theory these circles appear with
$m$ long-dashed lines ending on them, which represents the
$m$-order Kubo cumulant for the $m$th degree of the number operator
$\hat{n}$:
%
\begin{eqnarray}
%
n^{2c} & = & \langle (\hat{n} - \langle \hat{n} \rangle)^2 \rangle,
   \quad
   n^{3c} = \langle (\hat{n} - \langle \hat{n} \rangle)^3 \rangle,
\nonumber \\
n^{4c} & = & \langle (\hat{n} - \langle \hat{n} \rangle)^4 \rangle
   - 3[ \langle (\hat{n} - \langle \hat{n} \rangle)^2 \rangle]^2 .
\label{(55)}
%
\end{eqnarray}

The first diagram of the right-hand part of Fig.\ 1(a) is a local
normal electron propagator $G^0(x|x^{\prime})$ represented by a
solid line, which is renormalized by the phonon-cloud propagator
$\phi(x|x^{\prime})$. The corresponding first diagram in Fig.\ 1(b)
is the anomalous electron Green's function renormalized by the
phonon-cloud. In this case the renormalization is determined by the
anomalous function $\varphi(x|x^{\prime})$, which is smaller than
the corresponding contribution from the propagator $\phi(x|x^{\prime})$.
The diagrams (b)-(e) of Fig.\ 1(a) and (b)-(g) of Fig.\ 1(b)
take into account intersite tunneling, of which the corresponding last
diagrams describe direct polaron-polaron intersite interaction.

Among these diagrams we have also weakly connected ones, e.g.,
diagram (b) in Fig.\ 1(a) and diagrams (b) and (e) in Fig.\ 1(b),
which can be divided into two parts by cutting one tunneling line.
All other diagrams are strongly connected. All contributions of
Fig.\ 1(b) for the anomalous Green's functions contain the
phonon-cloud propagator $\varphi(x|x^{\prime})$ and are therefore
less important than the corresponding diagrams of Fig.\ 1(a).

In second order perturbation theory there are both groups of
weakly and strongly-connected diagrams. The weakly connected ones are
the simple repetition of the first-order diagrams. Therefore, in
Figs.\ 2-4 only the strongly-connected diagrams for the normal and
anomalous polaron Green's functions are shown.

\begin{figure*}[t]
%
\centering
\includegraphics[width=0.50\textwidth,clip]{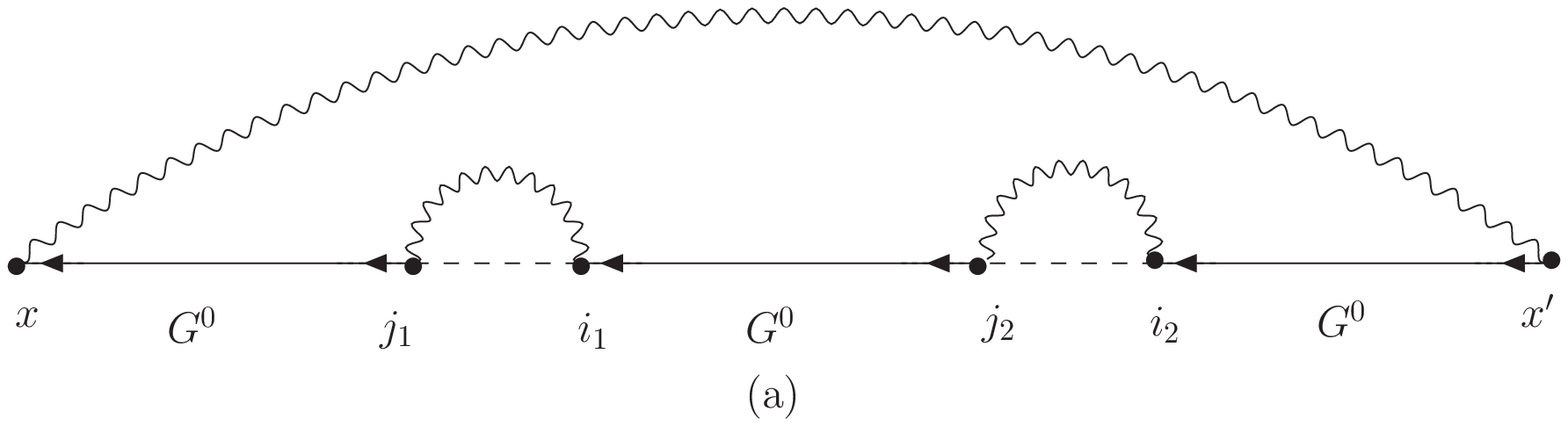}
\includegraphics[width=0.50\textwidth,clip]{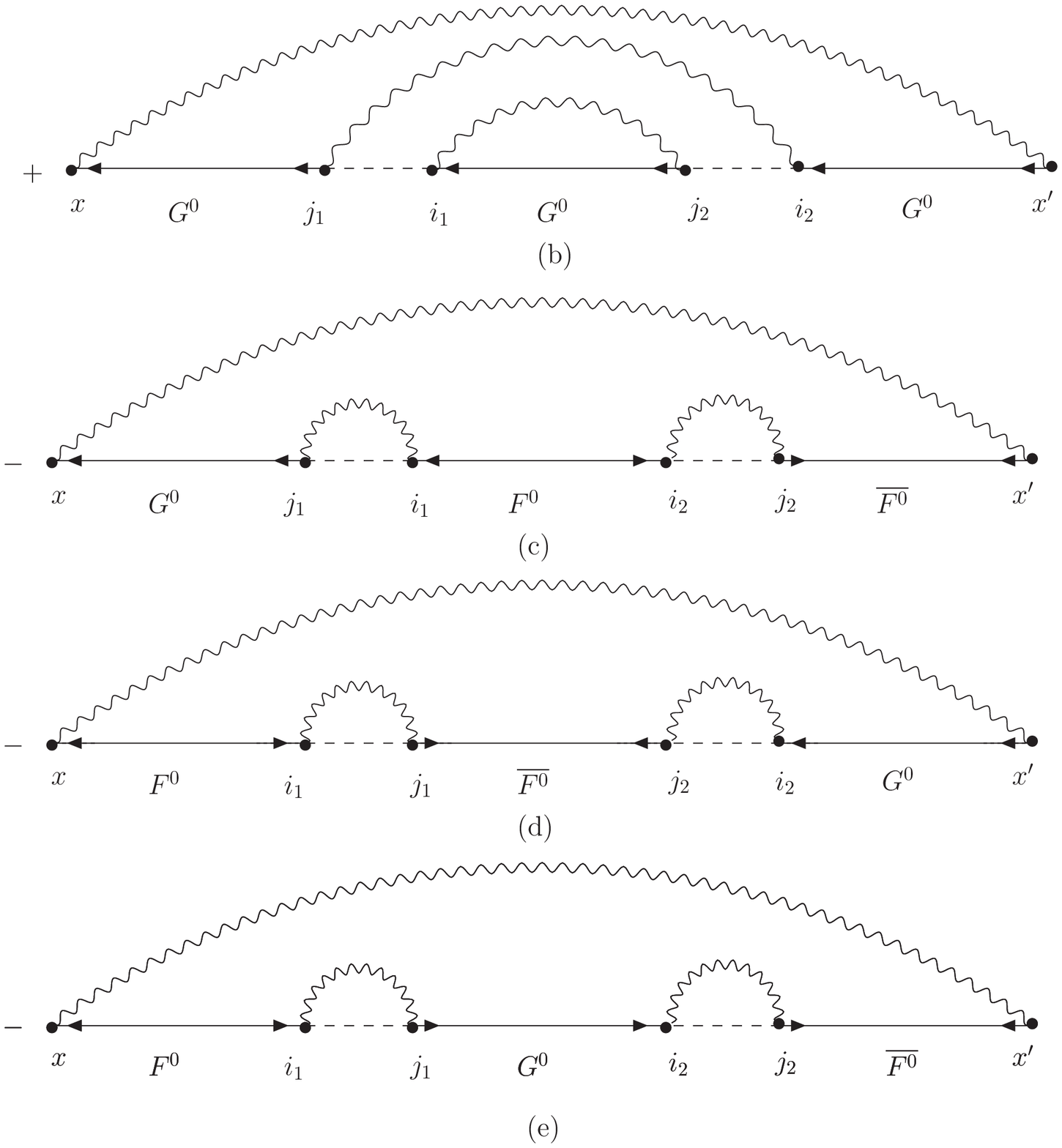}
\caption{
Strongly-connected diagrams of second order perturbation
for the normal polaron Green's function.
}
\label{fig-2}
%
\end{figure*}

\begin{figure*}[t]
%
\centering
\includegraphics[width=0.75\textwidth,clip]{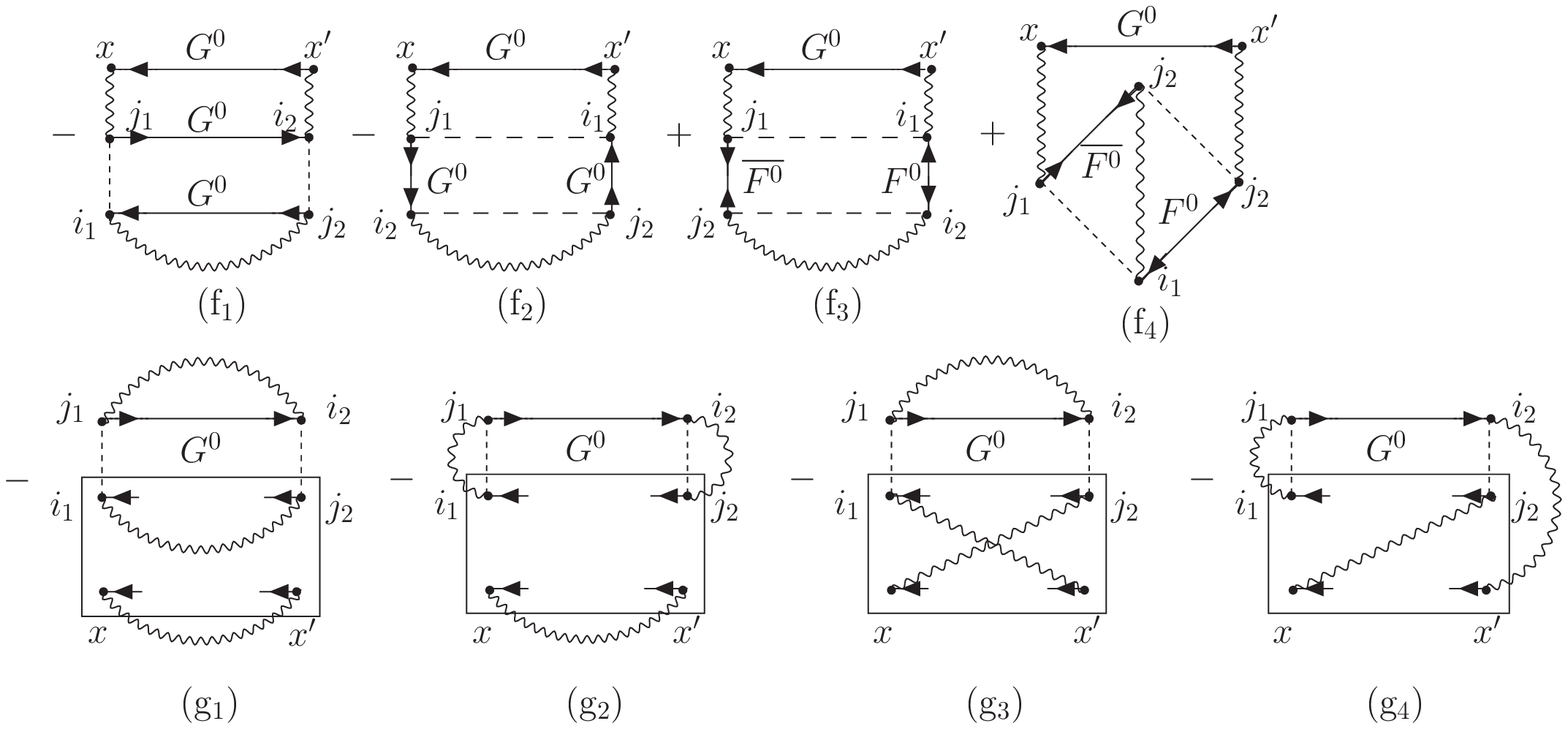}
\includegraphics[width=0.75\textwidth,clip]{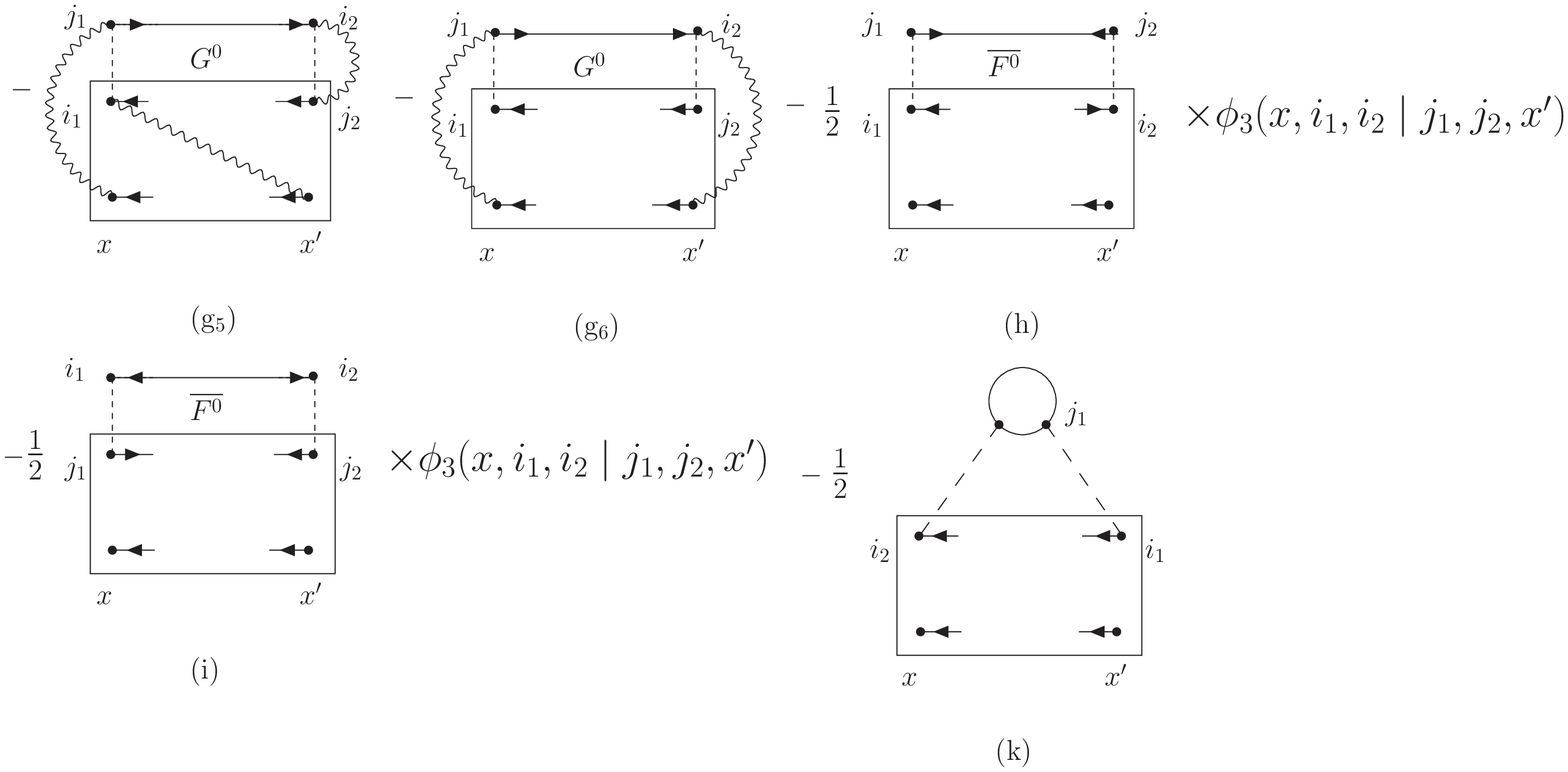}
\caption{
More complex diagrams which appear in second order perturbation theory
for the normal polaron Green's function.
}
\label{fig-3}
%
\end{figure*}

\begin{figure*}[t]
%
\centering
\includegraphics[width=0.75\textwidth,clip]{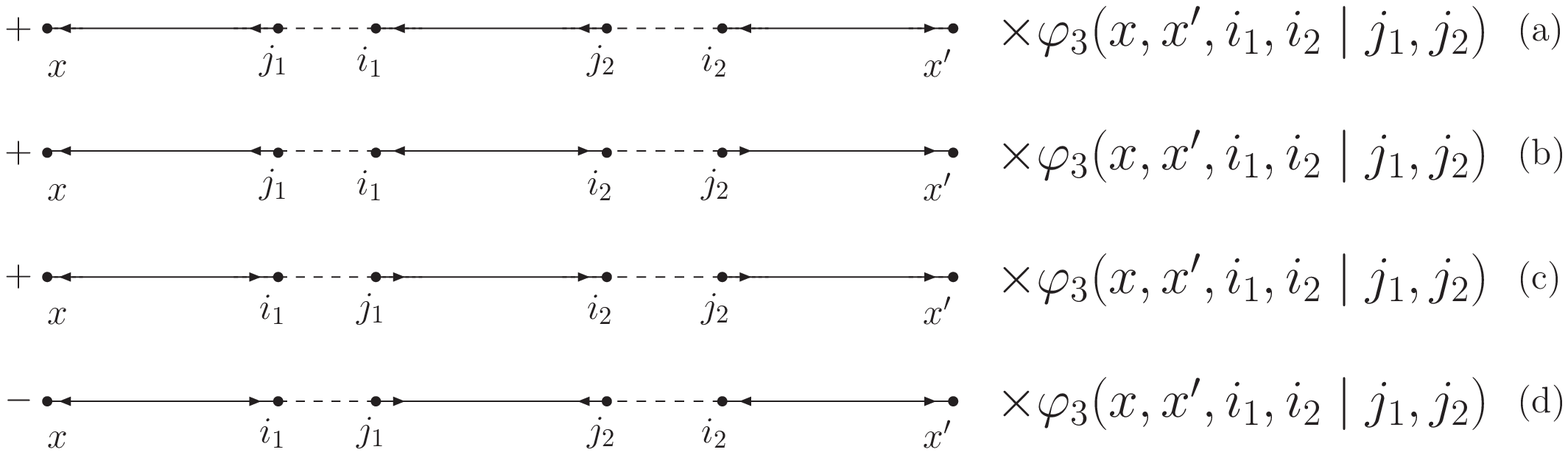}
\includegraphics[width=0.75\textwidth,clip]{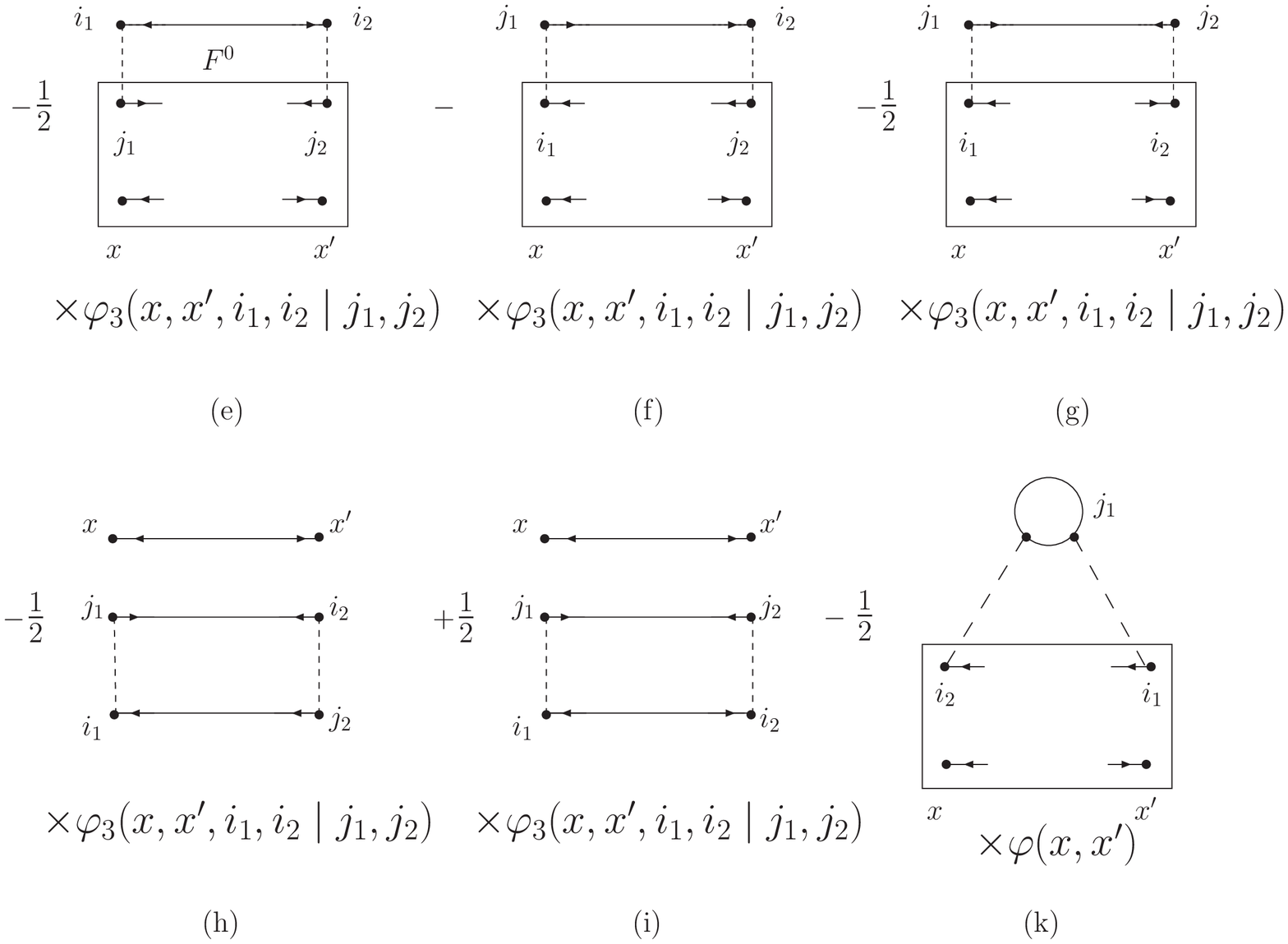}
\caption{ Diagrams of second order perturbation theory for the
anomalous polaron Green's function. The factors $\varphi_n$
account for the phonon-cloud contributions. } \label{fig-4}
%
\end{figure*}

The diagrams (a)-(i) in Figs.\ 2 and 3 originate from polaron
tunneling processes. In the group of more complex diagrams the
contributions from (f) appear only because of the electron-phonon
interaction. Diagrams (g)-(k), each with a rectangle, are special
contributions characteristic of strongly correlated electron
systems. The diagrams (h) and (i) contain the phonon (cloud)
correlation function $\phi_3(xi_1i_2|j_1j_2x^{\prime})$, each
consisting of six terms with different order of phonon propagation.
To illustrate this, the corresponding diagrammatical representation
for one of the anomalous Green's functions is shown in Fig.\ 4
(without resolving the structure of the less important
polaron-cloud propagator $\varphi_n$).

With respect to the full normal Green's function $G_p$ we now introduce
the correlation function $Z_p(x|x^{\prime})$ in analogy with previous works
\cite{Vladimir,Vakaru,Moskalenko,Bogoliubov,Moskalenko1}.
In diagrammatical form typical contributions to this function
are given by (c)-(f) in Fig.\ 1(a) and by all contributions from
Figs.\ 2 and 3 and by corresponding strongly-connected contributions from
higher order perturbation theory. If we then connect the
strongly-connected diagrams in all possible ways by the weakly connected
diagrams with non-renormalized tunneling matrix elements, we are able
to formulate a Dyson-like equation,
%
\begin{align}
%
\hspace*{-0.3cm}
G_{p\sigma\sigma^{\prime}}(\mathbf{x},\tau|
   \mathbf{x}^{\prime},\tau^{\prime}) & =
   \Lambda_{p\sigma\sigma^{\prime}}(\mathbf{x},\tau|
   \mathbf{x}^{\prime},\tau^{\prime})
\nonumber \\*[0.1cm] & {} +
   \int \! d\tau_1 \sum_{\mathbf{x}_1\mathbf{x}_2} \sum_{\sigma_1}
   \Lambda_{p\sigma\sigma_1}(\mathbf{x},\tau|\mathbf{x}_1,\tau_1)
\nonumber \\ & \times
   t(\mathbf{x}_1 - \mathbf{x}_2) \,
   G_{p\sigma_1\sigma^{\prime}}(\mathbf{x}_2,\tau_1|
   \mathbf{x}^{\prime},\tau^{\prime}) ,
\label{(56)}
%
\end{align}
%
where
%
\begin{equation}
%
\Lambda_p(x|x^{\prime}) = G_p^0(x|x^{\prime}) + Z_p(x|x^{\prime}) ,
\label{(57)}
%
\end{equation}
%
which in Fourier representation,
%
\begin{eqnarray}
%
\lefteqn{\hspace*{-0.5cm}
G_{p\sigma\sigma^{\prime}}(x|x^{\prime})
}
\nonumber \\ && \hspace*{-0.7cm} =
   \frac{1}{\beta N} \sum_{\omega_n} \sum_{\mathbf{k}}
   G_{p\sigma\sigma^{\prime}}(k|i\omega_n) \,
   e^{-i\mathbf{k}(\mathbf{x} - \mathbf{x}^{\prime})
 - i\omega_n(\tau - \tau^{\prime})} ,
\label{(58)}
%
\end{eqnarray}
%
has a simple algebraical form:
%
\begin{equation}
%
G_{p\sigma \sigma }(k|i\omega_n)
  = \frac{\Lambda_{p\sigma\sigma}(k|i\omega_n)}
   {1 - \varepsilon(\mathbf{k}) \,
   \Lambda_{p\sigma\sigma}(k|i\omega_n)} ,
\label{(59)}
%
\end{equation}
%
with the tight-binding dispersion $\varepsilon(\mathbf{k})$
of the bare electrons defined before.

Here we have assumed a paramagnetic ground state and spin conservation
$\sigma^{\prime} = \sigma$. It is important to note that the
form of the Dyson equation is the same for both the superconducting and
normal states of the system. The states differ with respect to the
correlation function $Z_p$. This situation is somewhat different from
the Hubbard-Holstein model for optical phonons \cite{Entel}, where
in the superconducting state normal and anomalous Green's functions
are interrelated. However, here two anomalous polaron Green's
functions are proportional to the anomalous one-phonon-cloud
propagator $\varphi(x|x^{\prime})$. Nevertheless, Dyson's equation
for the anomalous polaron Green's functions requires the knowledge of
the normal polaron Green's function. Then, by summing of
corresponding diagrams we obtain:
%
\begin{eqnarray}
%
&&
F_{p\sigma\overline{\sigma}}(\mathbf{x},\tau|
   \mathbf{x}^{\prime},\tau^{\prime }) =
   \Omega_{p\sigma\overline{\sigma}}(\mathbf{x},\tau|
   \mathbf{x}^{\prime},\tau^{\prime})
\nonumber  \\ && {} +
\sum_{\mathbf{x}_1,\mathbf{x}_2} \int \! d\tau _1 \,
   \Lambda_{p\sigma\sigma}(\mathbf{x},\tau|\mathbf{x}_1,\tau_1 )
\nonumber  \\ && \times
   t(\mathbf{x}_1 - \mathbf{x}_2)F_{p\sigma\overline{\sigma}}
   (\mathbf{x}_2,\tau_1|\mathbf{x}^{\prime},\tau^{\prime})
\nonumber \\ && {} +
\sum\limits_{\mathbf{x}_1,\mathbf{x}_2} \int \! d\tau_1 \,
   \Omega_{p\sigma,\overline{\sigma}}
   (\mathbf{x},\tau|\mathbf{x}_{1},\tau_1) \,
   t(\mathbf{x}_2 - \mathbf{x}_1)
\nonumber \\ && \times
   G_{p\overline{\sigma},\overline{\sigma}}
   (\mathbf{x}^{\prime},\tau^{\prime}|\mathbf{x}_2,\tau_1)
\label{(60)}
%
\end{eqnarray}
%
with
%
\begin{equation}
%
\Omega_p(x|x^{\prime}) = F_p^0(x|x^{\prime}) + Y_p(x|x^{\prime}) ,
\label{(61)}
%
\end{equation}
%
where $Y_p(x|x^{\prime})$ is the sum of all strongly-connected
diagrams for the anomalous Green's functions. This quantity is analogous
to the function $Z_p(x|x^{\prime})$ but differs from it by the direction
of external electron lines.
The Fourier representation of Eq.\ (60) has the form:
%
\begin{eqnarray}
%
\lefteqn{\hspace*{-1.5cm}
F_{p\sigma\overline{\sigma}}(\mathbf{k}|i\omega_n) \left[
   1 - \varepsilon(\mathbf{k}) \, \Lambda_{p\sigma\sigma}
   (\mathbf{k}|i\omega_n) \right]
 = \Omega_{p\sigma\overline{\sigma}}(\mathbf{k}|i\omega_n)}
\nonumber \\ && \times
   \left[ 1 + \varepsilon(-\mathbf{k}) \,
   G_{p\overline{\sigma}\overline{\sigma}}(-\mathbf{k}|-i\omega_n)
   \right] .
\label{(62)}
%
\end{eqnarray}
%
We obtain by making use of Eq.\ (59):
%
\begin{align}
%
F_{p\sigma\overline{\sigma}}(k|i\omega_n) & =
   \frac{\Omega_{p\sigma\overline{\sigma}}(\mathbf{k}|i\omega_n)}
   {\left[ 1 - \varepsilon(\mathbf{k}) \,
   \Lambda_{p\sigma\sigma}(\mathbf{k}|i\omega_n) \right]}
\nonumber \\ & \times
   \frac{1}{\left[ 1 - \varepsilon(-\mathbf{k}) \,
   \Lambda_{p\overline{\sigma}\overline{\sigma}}
   (-\mathbf{k}|-i\omega_n) \right]}
\label{(63)}
%
\end{align}
%

The anomalous quantities $\Omega_p$, $\overline{\Omega}_p$, $F_p$ and
$\overline{F}_p$ are proportional to the anomalous one-phonon-cloud
propagator $\varphi$, which in the strong-coupling limit is an
exponentially small quantity. Therefore, it is more important to
consider the propagators $F_e(x|x^{\prime})$ and
$\overline{F}_e(x|x^{\prime})$ defined in terms of the electron
operators $a_\sigma(\tau)$ and $\overline{a}_\sigma (\tau)$
and not in terms of the polarons operators $c(\tau)$ and
$\overline{c}_\sigma(\tau)$. In addition we have to discuss
the normal electron propagator $G_e(x|x^{\prime})$. These functions
are defined by
%
\begin{subequations} \begin{align}
%
G_e(x|x^{\prime }) & =
   - \langle \mathrm{T} \, a_{\mathbf{x}\sigma}(\tau)
     \overline{a}_{\mathbf{x}^{\prime}\sigma^{\prime}}
     (\tau^{\prime})U(\beta) \rangle_0^c ,
\label{(64a)} \\
F_e(x|x^{\prime}) & =
   - \langle \mathrm{T} \, a_{\mathbf{x}\sigma}(\tau)
     a_{\mathbf{x}^{\prime}\sigma^{\prime}}(\tau^{\prime})
     U(\beta) \rangle_0^c .
\label{64b} \\
\overline{F}_e(x|x^{\prime}) & =
   - \langle \mathrm{T} \overline{a}_{\mathbf{x}\sigma}(\tau)
     \overline{a}_{\mathbf{x}^{\prime}\sigma^{\prime}}(\tau^{\prime})
     U(\beta) \rangle_0^c .
\label{(64c)}
%
\end{align} \end{subequations}
%
Diagrammatical contributions to the first two functions are
shown in Fig.\ 5 and 6, respectively.

\begin{figure*}[t]
%
\centering
\includegraphics[width=0.65\textwidth,clip]{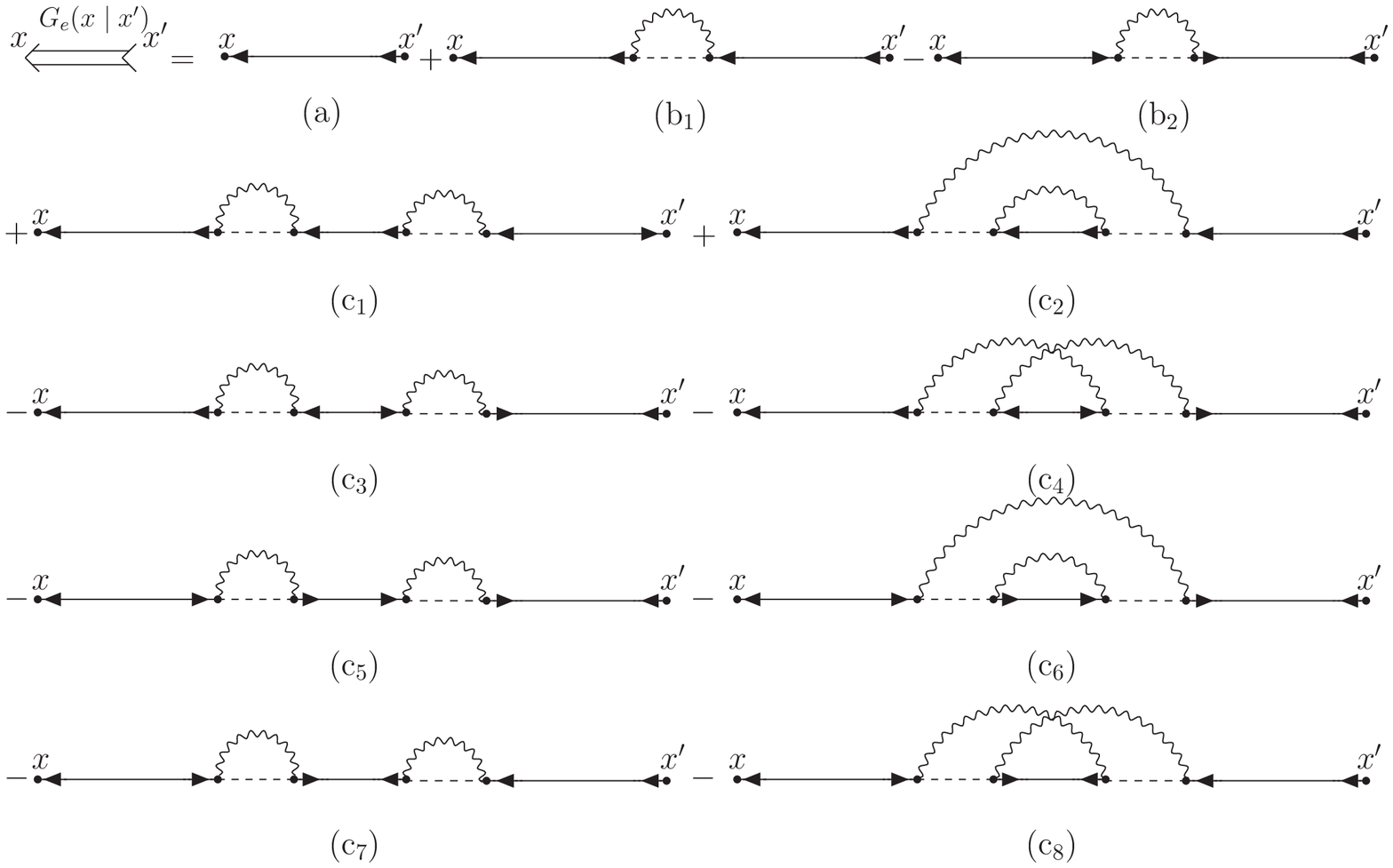}
\includegraphics[width=0.65\textwidth,clip]{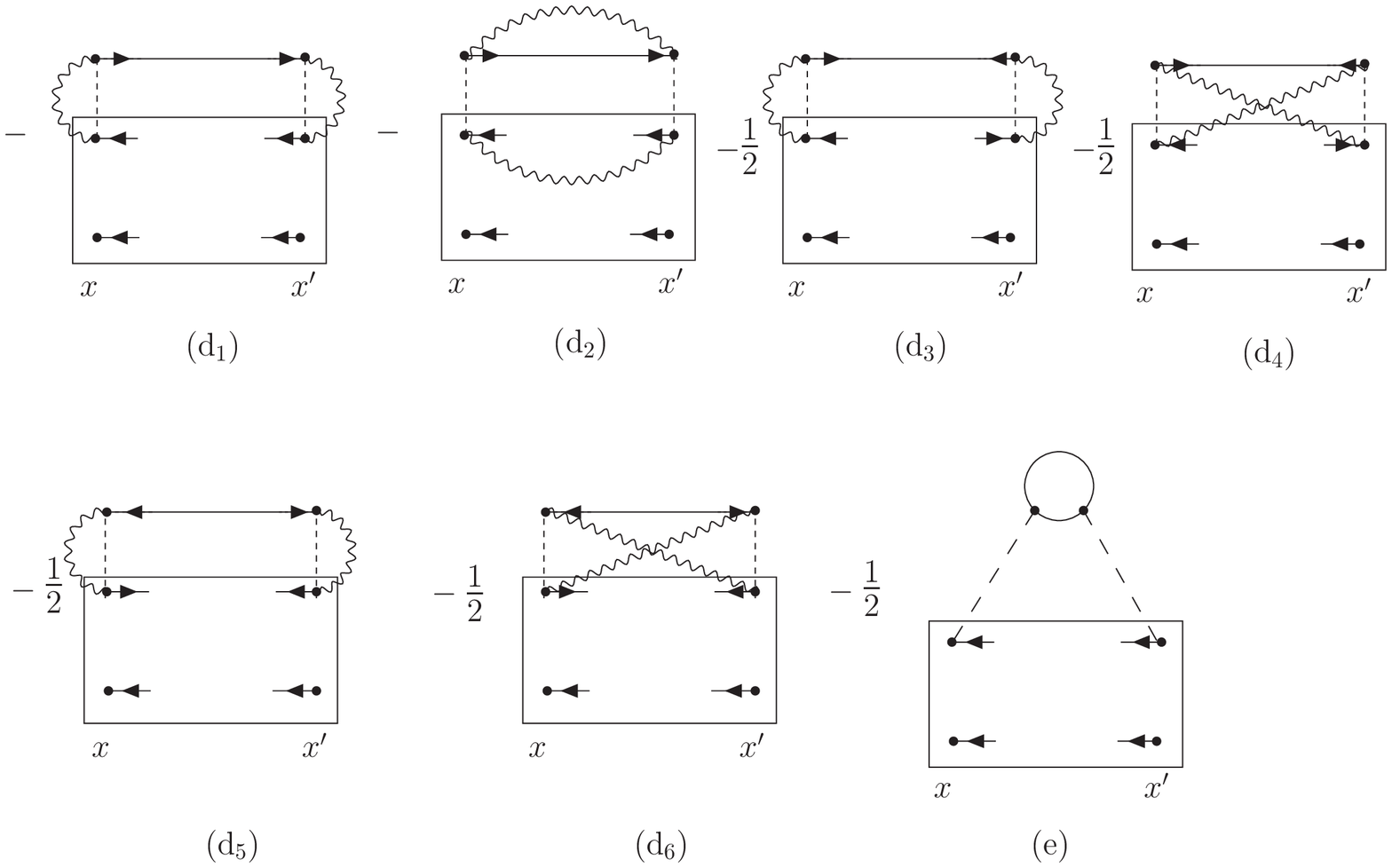}
\caption{
Diagrammatic contributions to the normal one-electron propagator in
the presence of phonon clouds.}
\label{fig-5}
%
\end{figure*}

\begin{figure*}[t]
%
\centering
\includegraphics[width=0.75\textwidth,clip]{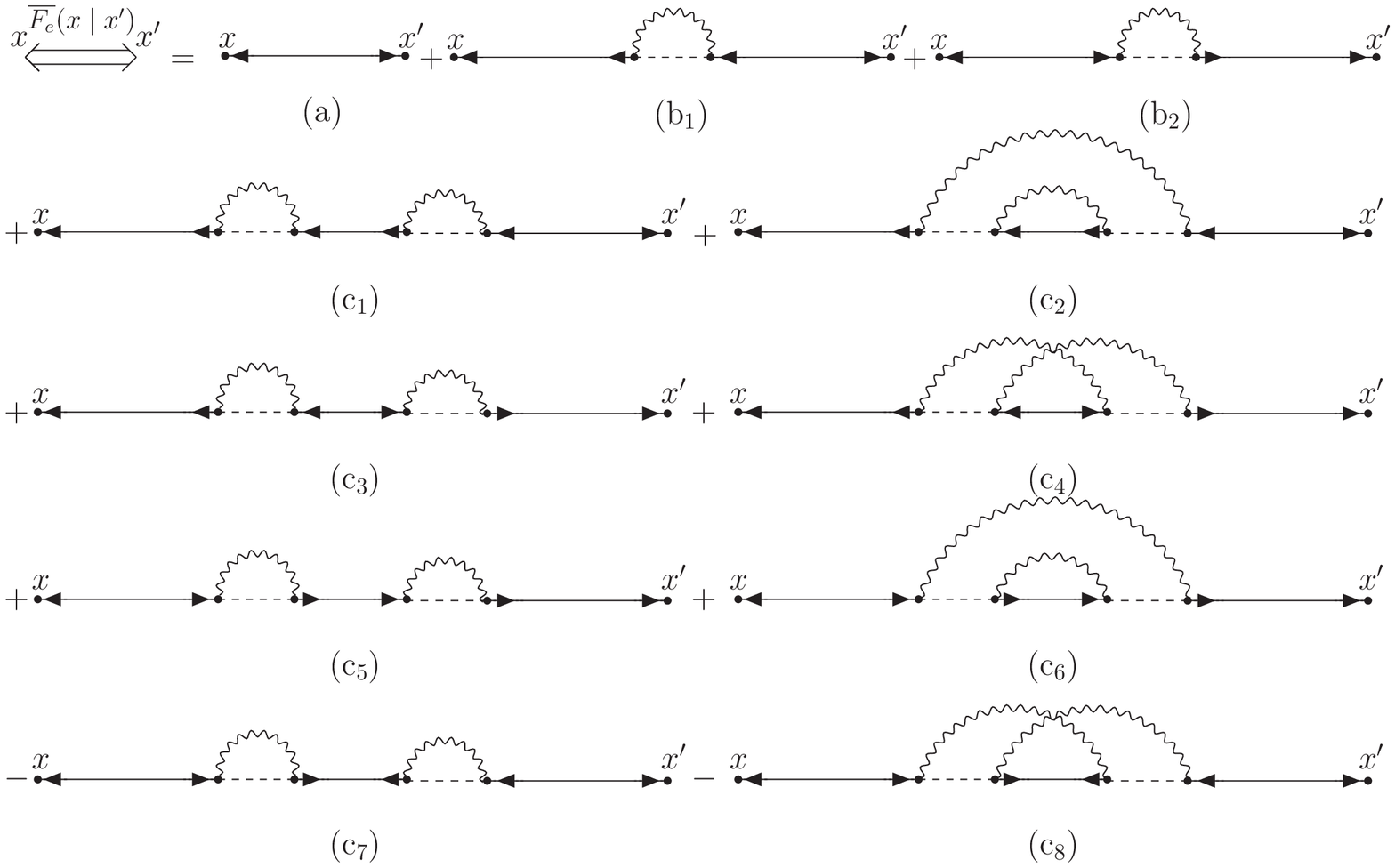}
\includegraphics[width=0.75\textwidth,clip]{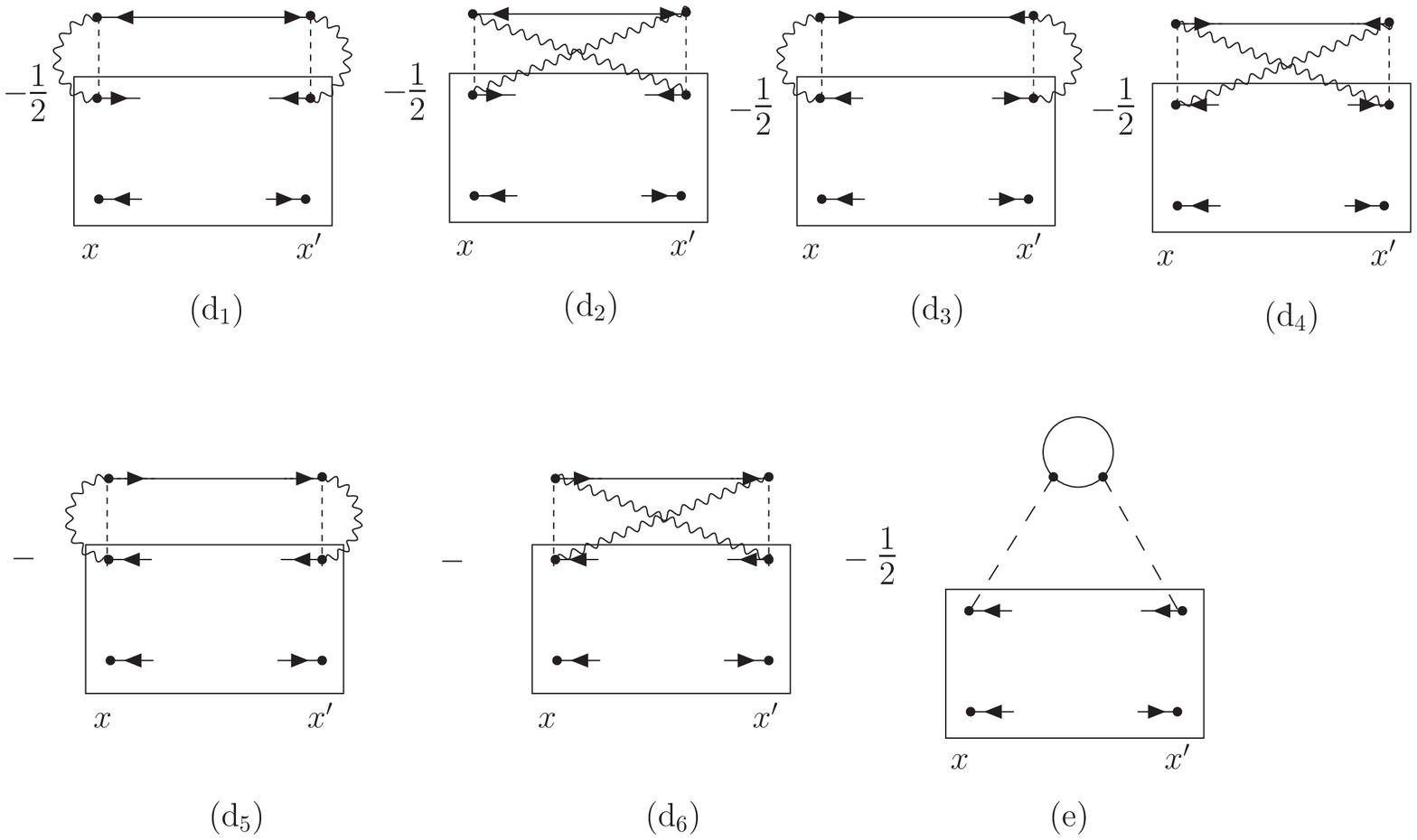}
\caption{
Diagrammatic contributions to the anomalous one-electron propagator in
the presence of phonon clouds.}
\label{fig-6}
%
\end{figure*}

Figure 5 shows the diagrams contributing to the normal one-electron
propagator in the presence of acoustical phonons (clouds).
We again find weakly connected diagrams which can be divided into two parts by cutting
one electron line like c$_1$, c$_3$, c$_5$, and c$_7$ . Furthermore, we introduce
normal, $\Sigma(x|x^{\prime})$, and anomalous, $\Xi(x|x^{\prime})$
and $\overline{\Xi}(x|x^{\prime})$, mass operators, of which the
simplest contributions are shown in Fig.\ 7. For example, diagram
a$_1$ is the renormalized tunneling matrix element, whereas (b) and
(c) are the simplest contributions to the anomalous mass operators.
In analogy to the rectangles representing irreducible Green's
functions, $G_n^{0 \, ir}$, or non-full cumulants, $\Gamma_n^0$,
$\Im_n^0$, and $\overline{\Im}_n^0$ in Figs.\ 5 and 6, we introduce
here the correlation functions $Z_e(x|x^{\prime})$ for the normal
state and $Y_e(x|x^{\prime})$ and $\overline{Y}_e(x|x^{\prime})$
for the superconducting state. For example, to (d) and (e) in
Fig.\ 5 corresponding diagrams contribute here to
$Z_e(x|x^{\prime})$, while to (d) and (e) in Fig.\ 6 corresponding
diagrams contribute to the correlation function $Y_e(x|x^{\prime})$,
which leads to

\begin{figure*}[t]
%
\centering
\includegraphics[width=0.65\textwidth,clip]{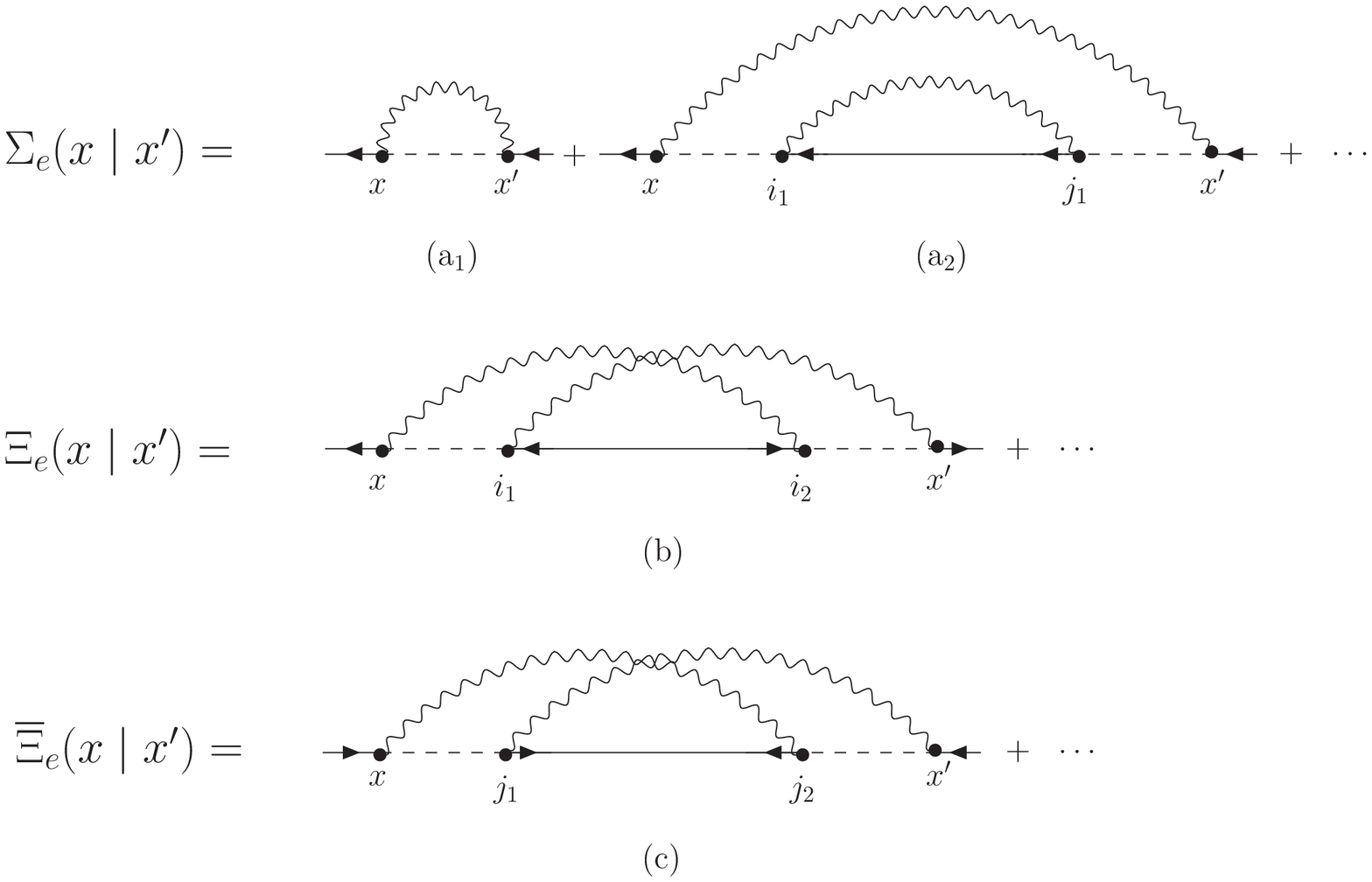}
\caption{
The simplest diagrams contributing to the normal,
$\Sigma_e(x|x^{\prime})$, and anomalous, $\Xi_e(x|x^{\prime})$
and $\overline{\Xi}_e(x|x^{\prime})$, mass operators.
}
\label{fig-7}
%
\end{figure*}

\begin{subequations} \begin{align}
%
\Lambda_e(x|x^{\prime}) & = G^0(x|x^{\prime}) + Z_e(x|x^{\prime}),
\label{(65a)} \\
\Omega_e(x|x^{\prime}) & = F^0(x|x^{\prime}) + Y_e(x|x^{\prime}),
\label{(65b)} \\
\overline{\Omega}_e(x|x^{\prime}) & = \overline{F}^0(x|x^{\prime})
 + \overline{Y}_e(x|x^{\prime}) .
\label{(65c)}
%
\end{align} \end{subequations}

Thus we have introduced the main dynamical quantities which determine
the to Fig.\ 5 and 6 corresponding diagrammatical structure. This
allows us to derive Dyson equations for the electron Green's
function system. The full electron Green's functions can be
expressed as
%
\begin{subequations} \begin{align}
%
G_e(x|x^{\prime}) & =
   \Lambda_e(x|x^{\prime}) + \Lambda_e(x|x_1) \,
   \Sigma_e(x_1|x_2) \, G_e(x_2|x^{\prime})
\nonumber \\ & -
   \Lambda_e(x|x_1) \, \Xi _e(x_1|x_2) \,
   \overline{F}_e(x_2|x^{\prime})
\nonumber \\ & -
   \Omega_e(x|x_1)\Sigma_e(x_2|x_1) \, \overline{F}_e(x_2|x^{\prime})
\nonumber \\ & -
   \Omega_e(x|x_1) \, \Xi _e(x_1|x_2) \, G_e(x_2|x^{\prime}),
\label{(66a)}
\\*[0.2cm]
F_e(x|x^{\prime}) & =
   \Omega_e(x|x^{\prime}) + \Omega_e(x|x_1) \,
   \Sigma_e(x_2|x_1) \, G_e(x^{\prime}|x_2)
\nonumber \\ & -
   \Omega_e(x|x_1) \, \overline{\Xi}_e(x_1|x_2) \,
   F_e(x_2|x^{\prime})
\nonumber \\ & -
   \Lambda_e(x|x_1) \, \Sigma_e(x_1|x_2) \, F_e(x_2|x^{\prime})
\nonumber \\ & +
   \Lambda_e(x|x_1) \, \Xi_e(x_1|x_2) \, G_e(x^{\prime}|x_2),
\label{66b}
\\*[0.2cm]
\overline{F}_e(x|x^{\prime }) & =
   \overline{\Omega}_e(x|x^{\prime}) +
   \overline{\Omega}_e(x|x_1) \, \Sigma_e(x_1|x_2) \,
   G_e(x^{\prime}|x_2)
\nonumber \\ & -
   \overline{\Omega}_e(x|x_1) \, \Xi_e(x_1|x_2) \,
   \overline{F}_e(x_2|x^{\prime})
\nonumber \\ & +
   \Lambda_e(x|x_1) \, \Sigma_e(x_2|x_1) \,
   \overline{F}_e(x_2|x^{\prime})
\nonumber \\ & +
   \Lambda_e(x_1|x) \, \overline{\Xi}_e(x_1|x_2) \, G_e(x_2|x^{\prime})
\label{(66c)}
%
\end{align} \end{subequations}
%
Here $x$ stands for $(\mathbf{x},\sigma,\tau)$. Double repeated indices
imply summation over $\mathbf{x}$ and $\sigma$ and integration over
$\tau$. All quantities in these equations are renormalized functions
containing all diagrammatical contributions to the normal and anomalous
one-electron Green's functions. The Fourier representation of the
first two equations can be written as [$k =\mathbf{k},i\omega_n$]:
%
\begin{subequations} \begin{eqnarray}
%
\lefteqn{
\Lambda_{\overline{\sigma}}^e(-k)
}
\nonumber \\ && {} =
G_{\overline{\sigma}}^e(-k) \big [
   1 - \Lambda_{\overline{\sigma}}^e(-k) \,
   \Sigma_{\overline{\sigma}}^e(-k)
 + \overline{\Omega}_{\overline{\sigma}\sigma}^e(k) \,
   \Xi_{\sigma\overline{\sigma}}^e(k) \big ]
\nonumber \\ && {} +
   F_{\sigma \overline{\sigma}}^e(k) \big [
   \Lambda_{\overline{\sigma}}^e(-k) \,
   \overline{\Xi}_{\overline{\sigma}\sigma}(k)
 + \overline{\Omega}_{\overline{\sigma}\sigma}^e(k) \,
   \Sigma_\sigma^e(k) \big ] ,
\label{(67a)}
\\*[0.3cm]
\lefteqn{
 \Omega_{\sigma\overline{\sigma}}^e(k)
}
\nonumber \\ && {} =
   F_{\sigma\overline{\sigma}}^e(k) \big [
   1 - \Lambda_\sigma^e(k) \, \Sigma_\sigma^e(k)
 + \Omega_{\sigma\overline{\sigma}}^e(k) \,
   \overline{\Xi}_{\overline{\sigma }\sigma }^e(k) \big ]
\nonumber \\ && {} -
   G_{\overline{\sigma}}^e(-k) \big [
   \Omega_{\sigma\overline{\sigma}}^e(k) \,
   \Sigma_{\overline{\sigma}}^e(-k)
 + \Lambda_\sigma^e(k) \, \Xi_{\overline{\sigma}\sigma}^e(k)
   \big ] .
\label{(67b)}
%
\end{eqnarray} \end{subequations}
%
The solutions of these equations are
%
\begin{subequations} \begin{align}
%
G_{\overline{\sigma}}^e(-k) & =
   \frac{1}{d_{\sigma}(k)} \big \lbrace
   \Lambda_{\overline{\sigma}}^e(-k)
\nonumber \\ & -
   \Sigma_{\sigma}^e(k) \big [
   \Lambda_{\sigma}^e(k) \, \Lambda_{\overline{\sigma}}^e(-k)
 + \Omega_{\sigma\overline{\sigma}}^e(k) \,
   \overline{\Omega}_{\overline{\sigma}\sigma}^e(k) \big ] \big
   \rbrace ,
\label{68a}
\\
F_{\sigma\overline{\sigma}}^e(k) & =
   \frac{1}{d_{\sigma}(k)} \big \lbrace
   \Omega_{\sigma\overline{\sigma}}^e(k)
\nonumber \\ & +
   \Xi_{\sigma\overline{\sigma}}^e(k) \big [
   \Lambda_{\sigma}^e(k) \, \Lambda_{\overline{\sigma}}^e(-k)
 + \Omega_{\sigma\overline{\sigma}}^e(k) \,
   \overline{\Omega}_{\overline{\sigma}\sigma}^e(k) \big] \big \rbrace,
\label{68b} \\
\overline{F}_{\overline{\sigma}\sigma}^e(k) & =
   \frac{1}{d_{\sigma}(k)} \big \lbrace
   \overline{\Omega}_{\overline{\sigma}\sigma}^e(k)
\nonumber \\ & +
   \overline{\Xi}_{\overline{\sigma}\sigma}^e(k) \big [
   \Lambda_{\sigma}^e(k) \, \Lambda_{\overline{\sigma}}^e(-k)
 + \Omega_{\sigma\overline{\sigma}}^e(k) \,
   \overline{\Omega}_{\overline{\sigma}\sigma}^e(k) \big ] \rbrace ,
\label{(68c)}
%
\end{align} \end{subequations}
%
where
%
\begin{align}
%
d_{\sigma}(k) & =
   1 -\Lambda_{\sigma}^e(k) \, \Sigma_{\sigma}^e(k)
 - \Lambda_{\overline{\sigma}}^e(-k) \,
   \Sigma_{\overline{\sigma}}^e(-k)
\nonumber \\ & +
   \Omega_{\sigma\overline{\sigma}}^e(k) \,
   \overline{\Xi}_{\overline{\sigma}\sigma}^e(k)
 + \overline{\Omega}_{\overline{\sigma}\sigma}^e(k) \,
   \Xi_{\sigma\overline{\sigma}}^e(k)
\nonumber \\ & +
   \left [ \Lambda_{\sigma}^e(k) \, \Lambda_{\overline{\sigma}}^e(-k)
 + \Omega_{\sigma\overline{\sigma}}^e(k) \,
   \overline{\Omega}_{\overline{\sigma}\sigma}^e(k) \right ]
\nonumber \\ & \times
   \left [ \Sigma_{\sigma}^e(k) \, \Sigma_{\overline{\sigma}}^e(-k)
 + \Xi_{\sigma\overline{\sigma}}^e(k) \,
   \overline{\Xi}_{\overline{\sigma}\sigma}^e(k) \right ] .
\label{(69)}
%
\end{align}

The functions $F$, $\Omega$ and $\Xi$ obey the following
symmetry relations:
%
\begin{align}
%
F_{\sigma\overline{\sigma}}^e(k) & =
  - F_{\overline{\sigma }\sigma}^e(-k), \quad
    \overline{F}_{\overline{\sigma}\sigma}^e(k) =
  - \overline{F}_{\sigma\overline{\sigma}}^e(-k),
\nonumber \\
\Omega_{\sigma\overline{\sigma}}^e(k) & =
  - \Omega_{\overline{\sigma}\sigma}^e(-k), \quad
    \overline{\Omega}_{\overline{\sigma}\sigma}^e(k) =
  - \overline{\Omega}_{\sigma\overline{\sigma}}^e(-k),
\nonumber \\
\Xi_{\sigma\overline{\sigma}}^e(k) & =
  - \Xi_{\overline{\sigma}\sigma}^e(-k), \quad
    \overline{\Xi}_{\overline{\sigma}\sigma}^e(k) =
  - \overline{\Xi}_{\sigma\overline{\sigma}}^e(-k) ,
\label{(70)}
%
\end{align}
%
and hence,
%
\begin{equation}
%
d_\sigma (k)=d_{\overline{\sigma }}(-k) .
\label{(71)}
%
\end{equation}
%

These equations for the renormalized electron Green's functions are
exact. Since they do not contain the exponentially small anomalous
phonon-cloud propagator $\varphi(x|x^{\prime})$, superconducting
pairing is easier to achieve by electrons without phonon clouds
but moving in the environment of the clouds belonging to other polarons,
than by polarons moving in the same environment. We can now switch off
the superconducting source term, which means that $F^0$ and
$\overline{F}^0$ are identically zero. However, the functions
$\Omega_{\sigma,\overline{\sigma}}^e$ and
$\overline{\Omega}_{\overline{\sigma},\sigma}^e$
survive in this limit and are equal to the order parameters of the
superconducting state, $Y_{\sigma,\overline{\sigma}}^e$ and
$\overline{ Y}_{\overline{\sigma},\sigma}^e$, respectively.

\section{Solvable limits}

The three correlation functions $Z_{e\sigma}$,
$Y_{e\sigma,\overline{\sigma}}$ and
$\overline{Y}_{e\overline{\sigma},\sigma}$
are the infinite sums of diagrams which contain both partially
and completely irreducible many-particle Green's functions.
In order to obtain a closed set of equations which can be solved
(at least numerically), we restrict ourselves to a class of rather
simple diagrams which, however, contain the most important spin,
charge and pairing correlations.

One way to do this, is to check how the individual diagrams are
influenced by the phonon fields, for example by distinguishing the case
of moderate coupling, when Eq.\ (26) can be used, from the
strong-coupling case where Eq.\ (25) holds. This helps to eliminate
from the diagrams the less important ones.
For example, in the strong-coupling limit
$\phi(x|x^{\prime}) \propto \delta_{\mathbf{x},\mathbf{x}^{\prime}}$
holds, which allows to discard all renormalized tunneling matrix
elements of the form
$t(\mathbf{x}^{\prime}\mathbf{-x}) \, \phi(\mathbf{x-x}^{\prime}|0)$.
In this limiting case the narrowing of the electronic energy band is
maximum, i.e., its width is equal to zero. Since this extreme case
is a bit unrealistic, we will in the following consider the case of
moderate electron-phonon coupling when also the band narrowing is
moderate and Eq.\ (26) must be used.
After summing the infinite series of the most important contributions
we obtain the result which is shown graphically in Fig.\ 8.

\begin{figure*}[t]
%
\centering
\includegraphics[width=0.65\textwidth,clip]{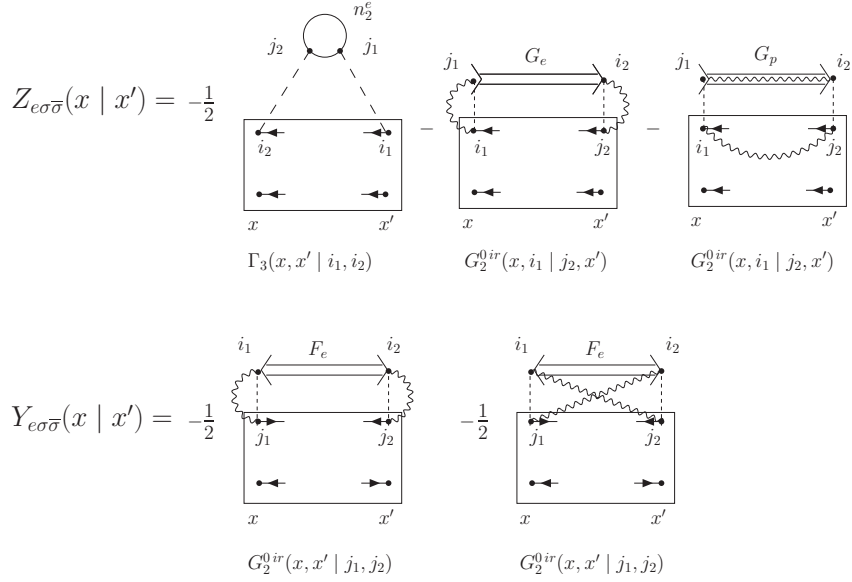}
\caption{
Schematical representation of the renormalization function
$Z_{e\sigma}$ and the superconducting order parameter
$Y_{e\sigma,-\sigma}$. Rectangles with four inner points
depict the two-particle completely irreducible Green's functions
$\widetilde{G}_2^{0\,ir}$ or partial cumulant $\Gamma_3^0$.
Double lines depict the renormalized one-particle Green's
functions of electron and polaron kind, short-dashed lines stand
for the hopping matrix elements and long-dashed lines represent
the direct electron-electron interaction $V\left(i-j\right)$.
}
\label{fig-8}
%
\end{figure*}

It is evident that in this approximation for the correlation functions
no closed set of equations is obtained because of the complicated
nature of mass operators. So we have to simplify the latter
quantities of which the simplest diagrams are depicted in Fig.\ 7.
For the normal mass operator we will use the contribution (a$_1$)
in Fig.\ 7 which is given by
%
\begin{align}
%
\Sigma_e(x|x^{\prime}) & \simeq
   t(\mathbf{x} - \mathbf{x}^{\prime}) \,
   \phi(\mathbf{x} - \mathbf{x}^{\prime}|0)
\nonumber \\ & \simeq
   t(\mathbf{x} - \mathbf{x}^{\prime})
   \exp (-\sigma_{1}(\mathbf{x} - \mathbf{x}^{\prime})^2/2) .
\nonumber
%
\end{align}
%
For simplicity we replace in the exponential function the
distance $|\mathbf{x} - \mathbf{x}^{\prime}|$ by the lattice
constant $a$ being a characteristic length over which the
electrons tunnel:
%
\begin{subequations} \begin{align}
%
\Sigma_e(x|x^{\prime}) & =
   \tilde{t}(\mathbf{x} - \mathbf{x}^{\prime}) \,
   \delta(\tau - \tau^{\prime})
\nonumber \\ & =
   t(\mathbf{x} - \mathbf{x}^{\prime }) \,
   \exp(-W_p) \, \delta(\tau - \tau^{\prime}) ,
\label{(72a)}
\\
W_p & = {\textstyle\frac{1}{2}} \sigma_1 a^2 .
\label{(72b)}
%
\end{align} \end{subequations}

This result means that tunneling of phonon-fields leads to
electronic band-narrowing effects by which the bare
energy $\varepsilon (\mathbf{k})$ is replaced by
$\widetilde{\varepsilon}(\mathbf{k})
   = \varepsilon(\mathbf{k})e^{-W_p}$.
For moderate electron-phonon interaction the quantity $W_p$ is
about unity. With respect to the anomalous mass operators, $\Xi_e$ and
$\overline{\Xi}_e$, we observe that they are smaller than
the normal one and can therefore safely be neglected. This will be
used when expanding the equations close to the superconducting
transition temperature.

Another approximation is related to a simplification of the
exact Dyson Eqs.\ (68) by omitting all anomalous mass operators,
which yields [$k = (\mathbf{k},i\omega)]$
%
\begin{subequations} \begin{align}
%
G_{e\,\sigma}(k) & =
   \frac{i}{D_\sigma^e(k)} \big \lbrace
   \Lambda_\sigma^e(k) \left[ 1 - \tilde{\varepsilon}(-\mathbf{k}) \,
   \Lambda _{\overline{\sigma}}^e(-k) \right ] ,
\nonumber \\ & -
   \tilde{\varepsilon}(-\mathbf{k}) \, Y_{\sigma\overline{\sigma}}^e(k) \,
  \overline{Y}_{\overline{\sigma}\sigma}^e(-k) \big \rbrace ,
\label{(73a)}
\\
F_{\sigma\overline{\sigma}}^e(k) & =
   \frac{Y_{\sigma\overline{\sigma}}^e(k)}{D_\sigma^e(k)} ,
\label{(73b)}
\\
D_\sigma^e(k) & =
   \left [ 1 - \tilde{\varepsilon}(\mathbf{k}) \,
   \Lambda_\sigma^e(k) \right ]
   \left [ 1 - \tilde{\varepsilon}(-\mathbf{k}) \,
   \Lambda_{\overline{\sigma}}^e(-k) \right ] ,
\nonumber \\ & +
   \tilde{\varepsilon}(\mathbf{k)} \, \tilde{\varepsilon}(-\mathbf{k)}
   \, Y_{\sigma\overline{\sigma}}^e(k) \,
   \overline{Y}_{\overline{\sigma}\sigma}^e(-k) ,
\label{(73c)}
\\
\Lambda_\sigma^e(k) & = G_{\sigma}^0(k) + Z_{e\,\sigma}(k) .
\label{(73d)}
%
\end{align} \end{subequations}
%

These equations are identical in form with the Dyson equations for
polaron superconductivity mediated by optical phonons \cite{Entel}.
The difference to the previous work is related to the
appearance of the renormalized energy $\tilde{\varepsilon}(\mathbf{k})$
and new correlation functions shown in Fig.\ 8. These irreducible
functions depicted by rectangles are on-site quantities with equal
site indices. Hence, all right-hand parts in Fig.\ 8 are proportional
to $\delta_{\mathbf{x},\mathbf{x}^{\prime}}$ meaning that
$Z_{e \, \sigma\overline{\sigma}}$ and
$Y_{e \, \sigma\overline{\sigma}}$ are also local functions and
corresponding Fourier representations to be independent of the polaron
momentum $\mathbf{k}$. In the diagrams $x$ and $i_1$ stand for
$(\mathbf{x},\sigma,\tau)$, and
$(\mathbf{i}_1,\sigma_1.\tau_1)$, respectively, whereby
summation over
$i_1, \, j_1, \, i_2, \, j_2$ and $\sigma_1, \, \sigma_2$
and integration over $\tau _1$ and $\tau _2$ is assumed.
These quantities have the following analytical structure:
%
\begin{align}
%
&
Z_e(\mathbf{x},\sigma,\tau|\mathbf{x}^{\prime},\sigma^{\prime},\tau^{\prime})
 = -{\textstyle\frac{1}{2}}
    \delta_{\mathbf{x},\mathbf{x}^{\prime}}
    \sum_j V^2(\mathbf{x} - j)
\nonumber \\
& \times \int_0^\beta \! d\tau_1 d\tau_2 \,
    \Gamma_3(\sigma,\tau;\sigma^{\prime},\tau^{\prime}|
    \tau_1,\tau_2) \, n^{2c} - \delta_{\mathbf{x},\mathbf{x}^{\prime}}
\nonumber \\
& \times \sum_{\sigma_1,\sigma_2} \sum_{ij} \int_0^\beta \!
    d\tau_1 d\tau_2 \, G_2^{0 \,ir}[\sigma,\tau;\sigma_1,\tau_1|
    \sigma_2,\tau_2;\sigma^{\prime},\tau^{\prime}]
\nonumber \\
& \times \tilde{t}(j - \mathbf{x}) \, \tilde{t}(\mathbf{x} - i) \,
    G_e(i,\sigma_2,\tau_2|j,\sigma_1,\tau_1)
  - \delta_{\mathbf{x},\mathbf{x}^{\prime}}
\nonumber \\
& \times \sum_{\sigma_1,\sigma_2} \sum_{ij} \int_0^\beta \!
    d\tau_1 d\tau_2 \, G_2^{0 \,ir}[\sigma,\tau;\sigma_1,\tau_1|
    \sigma_2,\tau_2;\sigma^{\prime},\tau^{\prime}]
\nonumber \\
& \times t(j - \mathbf{x}) \, t(\mathbf{x} - i) \,
     \phi (0|\tau_1 - \tau_2) \,
     G_p(i,\sigma_2,\tau_2|j,\sigma_1,\tau_1) ,
\label{(74)}
%
\end{align}
%
%
\begin{align}
%
&
Y^e(\mathbf{x},\sigma,\tau|\mathbf{x}^{\prime},-\sigma^{\prime},\tau^{\prime})
 = -{\textstyle\frac{1}{2}}
    \delta_{\mathbf{x},\mathbf{x}^{\prime}}
    \sum_{\sigma_1,\sigma_2} \sum_{\mathbf{i}_1,\mathbf{i}_2}
    \int_0^\beta \! d\tau_1 d\tau_2
\nonumber \\
& \times G_2^{0 \,ir}[\sigma,\tau;-\sigma^{\prime},\tau^{\prime}|
    \sigma_1,\tau_1;\sigma_2,\tau_2] \, \tilde{t}(\mathbf{x} - i_1) \,
    \tilde{t}(\mathbf{x} - i_2)
\nonumber \\
& \times F^e(i_1,\sigma_1,\tau_1|i_2,\sigma_2,\tau_2)
   -{\textstyle\frac{1}{2}} \delta_{\mathbf{x},\mathbf{x}^{\prime}}
\nonumber\\
& \times \sum_{\sigma_1,\sigma_2}\sum_{i_1 i_2}
    G_2^{0 \,ir}[\sigma,\tau;-\sigma^{\prime},\tau^{\prime}|
    \sigma_1,\tau_1;\sigma_2,\tau_2]  \, \tilde{t}(\mathbf{x} - i_1) \,
\nonumber \\
& \times \tilde{t}(\mathbf{x} - i_2) \, F^e(i_1,\sigma_1,\tau_1|
    i_2,\sigma_2,\tau_2)\phi(0|\tau_1 - \tau_2)
\nonumber \\
& \times \phi(0|\tau_2 - \tau _1) ,
\label{(75)}
%
\end{align}
%
and corresponding Eq.\ for $\overline{Y}_{-\sigma,\sigma}$.
Since Eqs.\ (74) and (75) are the result of summing an infinite series
of diagrams, the thin lined representing one-particle propagators
are replaced by full normal electron ($G_e$) and polaron ($G_p$)
and anomalous ($F_e$) functions. Fourier transformation of these
quantities leads in case of spin-singlet channel of superconductivity
to
%
\begin{align}
%
&
Z_{e\,\sigma}(i\omega) =
 - {\textstyle\frac{1}{2}} V_2n^{2c} \, \Gamma_{\sigma\sigma}(i\omega)
 - \frac{1}{\beta N} \sum_{\omega_1,\mathbf{k},\sigma_1}
   [\tilde{\varepsilon}(\mathbf{k})]^2
\nonumber \\
&  \times G_{e\sigma_1,\sigma_1}(\mathbf{k|}i\omega_1) \,
   \widetilde{G}_2^{0 \,ir}[\sigma,i\omega;\sigma_1,i\omega_1|
   \sigma_1,i\omega_1;\sigma,i\omega]
\nonumber \\
& - \frac{1}{N\beta^2} \sum_{\mathbf{k},\omega_1,\Omega_1} \! \!
   \varepsilon^2(\mathbf{k}) \,
   G_{p\sigma_1,\sigma_1}(\mathbf{k|}i(\omega_1 + \Omega_1)) \,
   \phi(i\Omega_1)
\nonumber \\
& \times \widetilde{G}_2^{0 \,ir}[\sigma,i\omega;\sigma_1,i\omega_1|
  \sigma_1,i\omega_1;\sigma,i\omega]  ,
\label{(76)}
%
\\*[0.5cm]
&
Y_{e\sigma,-\sigma }(i\omega ) =
 - \frac{1}{2\beta N} \sum_{\mathbf{k},\omega_1,\sigma_1} \! \!
   \tilde{\varepsilon}(\mathbf{k}) \, \tilde{\varepsilon}(-\mathbf{k})
   \, F_{e\sigma_1,-\sigma_1}(\mathbf{k|}i\omega_1)
\nonumber \\
& \times \widetilde{G}_2^{0 \,ir}[\sigma,i\omega;-\sigma,-i\omega|
   \sigma_1,i\omega_1;-\sigma_1,-i\omega_1]
\nonumber \\
& - \frac{1}{2N} \sum_{\sigma_1,\mathbf{k}} \frac{1}{\beta^3} \!
   \sum_{\omega_1\Omega_1\Omega_2} \! \!
   \tilde{\varepsilon}(\mathbf{k}) \, \tilde{\varepsilon}(-\mathbf{k})
\nonumber \\
& \times F_{e\sigma_1,-\sigma_1}(\mathbf{k|}i(\omega_{1-}\Omega_1 +
   \Omega_2)) \, \phi(i\Omega_1) \, \phi(i\Omega_2)
\nonumber \\
& \times \widetilde{G}_2^{0 \,ir}[\sigma,i\omega;-\sigma,-i\omega|
   \sigma_1,i\omega_1;-\sigma_1,-i\omega_1] ,
\label{(77)}
%
\end{align}
%
where
%
\begin{subequations} \begin{align}
%
&
V_2 = \frac{1}{N} \sum_{\mathbf{k}} |V(\mathbf{k})|^2,
\label{(78a)}
\\
&\Gamma_{\sigma,\sigma^{\prime}}(\tau - \tau^{\prime})
   \equiv \int_0^\beta \! d\tau_1d\tau_2 \,
   \Gamma_3(\sigma,\tau;\sigma^{\prime}\tau^{\prime}|
   \tau_1,\tau_2),
\label{(78b)}
\\*[0.2cm]
& \widetilde{G}_2^{0 \,ir}[\sigma_1,i\omega_1;\sigma_2,i\omega_2|
   \sigma_3,i\omega_3;\sigma_4,i\omega_4]
\nonumber \\
& = \beta \, \delta (\omega_1 + \omega_2 - \omega_3 - \omega_4)
\nonumber \\
& \times \widetilde{G}_2^{0 \,ir}[\sigma_1,i\omega_1;\sigma_2,
   i\omega_2|\sigma_3,i\omega_3;\sigma_4,i(\omega_1 + \omega_2
 - \omega_3)] .
\label{(78c)}
%
\end{align} \end{subequations}
%
From Eq.\ (76) for $Z_e(i\omega)$ we obtain the following expression
for $\Lambda_e(i\omega )$,
%
\begin{align}
%
&
\Lambda_{e\sigma}(i\omega)
 = G_\sigma^0(i\omega) - {\textstyle\frac{1}{2}} V_2n^{2c} \,
   \Gamma_{\sigma,\sigma}(i\omega) - \frac{1}{\beta N}
   \sum_{\omega_1,\mathbf{k},\sigma_1} \!
   [\tilde{\varepsilon}(\mathbf{k})]^2
\nonumber \\
& \times G_{e\sigma}(\mathbf{k|}i\omega_1) \,
   \widetilde{G}_2^{0 \,ir}[\sigma,i\omega;\sigma_1,i\omega_1|
   \sigma_1,i\omega_1;\sigma,i\omega]
\nonumber \\
& - \frac{1}{N} \sum_{\mathbf{k},\sigma_1} \frac{1}{\beta^2} \!
   \sum_{\omega_1\Omega_1} \! \varepsilon^2(\mathbf{k}) \,
   G_{p\sigma}(\mathbf{k|}i(\omega_1 + \Omega_1)) \,
   \phi(i\Omega_1)
\nonumber \\
& \times \widetilde{G}_2^{0 \,ir}[\sigma,i\omega;\sigma_1,i\omega_1|
   \sigma_1,i\omega_1;\sigma,i\omega] .
\label{(79)}
%
\end{align}
%
Here, the renormalized and unrenormalized tunneling matrix elements
accompany the electron and polaron propagation, respectively.
Equations (77) and (79) together with the expressions
for the one-particle Green's functions and the definitions of
the irreducible Green's functions in Eq.\ (50) and
Kubo cumulants in (55) determine completely the properties of the
superconducting phase and allow to discuss the influence of strong
electron-phonon interaction.

The three irreducible two-particle Green's functions already
calculated (see Eqs.\ (5-7) in Ref.\ \onlinecite{Moskalenko1})
are given by
%
\begin{widetext} \begin{subequations} \begin{align}
%
&
\widetilde{G}_2^{0 \,ir}[\sigma,i\omega;\sigma,i\omega_1|
   \sigma,i\omega_1;\sigma,i\omega]
 = \frac{\beta U^2 [1 - \delta_{\omega,\omega_1}]
   (1 + e^{\beta\mu}) (e^{\beta\mu} + e^{\beta(2\mu - U)})}
   {Z_0^2 \, \lambda(i\omega) \, \overline{\lambda}(i\omega) \,
   \lambda(i\omega_1) \, \overline{\lambda}(i\omega_1)} ,
\label{(80a)}
\\*[0.3cm]
& \widetilde{G}_2^{0 \,ir}[\sigma,i\omega;\overline{\sigma},i\omega_1|
   \overline{\sigma},i\omega_1;\sigma,i\omega]
 = \frac{U}{Z_0} \bigg \lbrace
   \frac{\beta U e^{2\beta\mu}(e^{-\beta U} - 1)}
   {Z_0\, \lambda(i\omega) \, \overline{\lambda}(i\omega) \,
   \lambda(i\omega_1) \, \overline{\lambda}(i\omega_1)}
 - \frac{\beta U e^{\beta\mu} \delta_{\omega,\omega_1}}
   {\lambda(i\omega) \, \overline{\lambda }(i\omega) \,
   \lambda(i\omega_1) \, \overline{\lambda}(i\omega_1)}
\nonumber \\
& - \frac{(e^{\beta(2\mu - U)} - 1) [\overline{\lambda}(i\omega)
 + \overline{\lambda}(i\omega_1)]}
   {[\lambda(i\omega) + \overline{\lambda}(i\omega_1)] \,
   \overline{\lambda}^2(i\omega) \, \overline{\lambda}^2(i\omega_1)}
 - (1 + e^{\beta\mu}) \bigg [
   \frac{1}{\lambda(i\omega) \, \overline{\lambda}(i\omega)}
   \times \bigg ( \frac{1}{\lambda^2(i\omega_1)} +
   \frac{1}{\overline{\lambda}^2(i\omega_1)} \bigg )
\nonumber \\
& + \frac{1}{\lambda(i\omega_1) \, \overline{\lambda}(i\omega_1)}
   \times \bigg ( \frac{1}{\lambda^2(i\omega)}
 + \frac{1}{\overline{\lambda}^2(i\omega)} \bigg )
 - \frac{2}{\lambda(i\omega) \, \overline{\lambda}(i\omega) \,
   \lambda(i\omega_1) \, \overline{\lambda}(i\omega_1)} \bigg ]
   \bigg \rbrace ,
\label{(80b)}
\\*[0.3cm]
& \widetilde{G}_2^{0 \,ir}[\sigma,i\omega;\overline{\sigma},-i\omega|
   \sigma,i\omega_1;\overline{\sigma},-i\omega_1]
 = \frac{U}{Z_0} \bigg \lbrace
   \frac{\beta U}{[\mu^2 - (i\omega)^2]
   [(\mu - U)^2 - (i\omega)^2]}
\nonumber \\
& \times \bigg [ \delta_{\omega + \omega_1,0} \,
   e^{\beta\mu} + \delta_{\omega - \omega_1,0} \,
   \frac{e^{2\beta\mu}}{Z_0} (1 - e^{-\beta U} ) \bigg ]
 + \bigg ( 1 + \frac{U}{2\mu - U} \bigg)
   \frac{(1 + e^{\beta\mu})}{[\mu^2 - (i\omega)^2]
   [\mu^2 - (i\omega_1)^2]}
\nonumber \\
& + \bigg ( 1 - \frac{U}{2\mu - U} \bigg )
   \frac{e^{\beta\mu} + e^{-\beta(U - 2\mu)}}
   {[(\mu - U)^2 - (i\omega)^2] [(\mu - U)^2 - (i\omega_1)^2]}
   \bigg \rbrace ,
\label{(80c)}
%
\end{align} \end{subequations} \end{widetext}
%
where $\delta_{\omega, \omega_1}$ is the Kronecker symbol for Matsubara
frequencies and
%
\begin{align}
%
Z_0 & = 1 + 2e^{\beta\mu} + e^{\beta (2\mu - U)},
\nonumber \\
\lambda (i\omega) & = i\omega + \mu, \quad
   \overline{\lambda}(i\omega) = i\omega+ \mu - U, \quad
   \overline{\sigma} = -\sigma .
\label{(81)}
%
\end{align}
%

For the present study $\mu$ and $U$ in Eqs.\ (80a-c) and (81) are
the renormalized quantities of Eq.\ (13). The
fore-standing equations are generalized Eliashberg equations
of strong-coupling superconductivity for the case that
strong electron correlations have been taken into account in a
self-consistent way. In spite of the approximations involved the
equations are rather complicated. In order to gain further insight
into the physics behind Eqs.\ (77) and (79), we will linearize the
equations in terms of the order parameter
$Y_{\sigma,\overline{\sigma}}$, but not in terms of
$\Lambda_{\sigma}$. Then the critical temperature
$T_c$ of the superconducting transition can be obtained from
%
\begin{widetext} \begin{align}
%
Y_{e\sigma,\overline{\sigma}}(i\omega)
  & = -\frac{1}{\beta N} \sum \limits_{\mathbf{k},\omega_1}
    \frac{
    \tilde{\varepsilon}(\mathbf{k}) \,
    \tilde{\varepsilon}(-\mathbf{k}) \,
    Y_{e\sigma,\overline{\sigma}}(i\omega_1)
    \widetilde{G}_2^{0 \,ir}[\sigma,i\omega;\overline{\sigma},-i\omega|
    \sigma,i\omega_1;\overline{\sigma},-i\omega_1]
    }{
    [1 - \tilde{\varepsilon}(\mathbf{k}) \,
    \Lambda_{e\sigma}(i\omega_1)]
    [1 - \tilde{\varepsilon}(-\mathbf{k}) \,
    \Lambda_{e\overline{\sigma}}(-i\omega_1)]
    }
\nonumber \\
& - \frac{1}{\beta N} \sum \limits_{\mathbf{k,}\omega_1}
   \frac{1}{\beta^2} \! \sum_{\Omega_1,\Omega_2}
   \frac{
   \tilde{\varepsilon}(\mathbf{k}) \,
   \tilde{\varepsilon}(-\mathbf{k}) \,
   Y_{e\sigma,\overline{\sigma}}(i(\omega_1 - \Omega_1 + \Omega_2)) \,
   \phi(i\Omega_1) \, \phi(i\Omega _2) \,
   \widetilde{G}_2^{0 \,ir}[\sigma,i\omega;\overline{\sigma},-i\omega|
   \sigma,i\omega_1;\overline{\sigma},-i\omega_1]
   }{
   [1 - \tilde{\varepsilon}(\mathbf{k}) \,
   \Lambda_{e\sigma}(i\omega_1 - i\Omega_1 + i\Omega_2)]
   [1 - \tilde{\varepsilon}(-\mathbf{k}) \,
   \Lambda_{e\overline{\sigma}}(-i\omega_1 + i\Omega_1 - i\Omega_2)]
   } ,
\label{(82)}
\\
\Lambda_{e\sigma}(i\omega) & =
   G_\sigma^0(i\omega) - {\textstyle\frac{1}{2}} V_2n^{2c} \,
   \Gamma_{\sigma,\sigma}(i\omega)
 - \frac{1}{\beta N}\sum \limits_{\mathbf{k,}\omega_1,\sigma_1}
   \frac{
   [\tilde{\varepsilon}(\mathbf{k})]^2
   \Lambda_{e\sigma_1}(i\omega_1) \,
   \widetilde{G}_2^{0 \,ir}[\sigma,i\omega;\sigma_1,i\omega_1|
   \sigma_1,i\omega _1;\sigma,i\omega]
   }{
   1 - \tilde{\varepsilon}(\mathbf{k}) \,
   \Lambda_{e\sigma_1}(i\omega)
   }
\nonumber \\
& - \frac{1}{\beta^2N} \sum \limits_{\mathbf{k},\omega_1,
   \Omega_1,\sigma_1} \! \! \!
   \frac{
   \varepsilon^2(\mathbf{k}) \,
   \Lambda_{p\sigma_1}(i\omega_1 + i\Omega_1) \,
   \phi (i\Omega_1)
   }{
   1 - \tilde{\varepsilon}(\mathbf{k}) \,
   \Lambda_{p\sigma_1}(i\omega)
   } \,
   \widetilde{G}_2^{0 \,ir}[\sigma,i\omega;\sigma_1,i\omega_1|
   \sigma_1,i\omega_1;\sigma,i\omega] .
\label{(83)}
%
\end{align} \end{widetext}
%
In order to determine $T_c$ it is necessary to solve Eq.\ (83) for
$\Lambda_{e\sigma}(i\omega)$ and to insert it into Eq.\ (82). The
next our approximation is the omitting in the right-hand part of
eq.(83) of the polaron function $\Lambda_{p,\sigma}$.The reason
for such approximation is the presence in this term of the phonon
-cloud propagator $ \phi (i\Omega_1)$ which makes the appearance
in the denominator of large quantity $\omega_c$.
 Since $\omega_{c}$ is larger
than other typical energies involved, the term under discussion is at
moderate coupling strength smaller than all other terms in Eq.\ (83)
and can therefore be omitted, which leaves
%
\begin{align}
%
&
\Lambda_{e\sigma}(i\omega)
 = G_\sigma^V(i\omega)
\nonumber \\
& - \frac{1}{\beta N} \sum \limits_{\mathbf{k},\omega_1} \!
   \frac{
   [\tilde{\varepsilon}(\mathbf{k})]^2
   \Lambda_{e}(i\omega_1) \,
   \widetilde{G}_2^{0 \,ir}[\sigma,i\omega;\sigma,i\omega_1|
   \sigma,i\omega_1;\sigma,i\omega]
   }{
   1 - \tilde{\varepsilon}(\mathbf{k}) \,
   \Lambda_{e\sigma}(i\omega )
   }
\nonumber \\
& - \frac{1}{\beta N} \sum \limits_{\mathbf{k},\omega_1} \!
   \frac{
   [\tilde{\varepsilon}(\mathbf{k})]^2
   \Lambda_{e\overline{\sigma}}(i\omega_1) \,
   \widetilde{G}_2^{0 \, ir}[\sigma,i\omega;\overline{\sigma},i\omega_1|
   \overline{\sigma},i\omega_1;\sigma,i\omega]
   }{
   1 - \tilde{\varepsilon}(\mathbf{k}) \,
   \Lambda_{e\overline{\sigma}}(i\omega)
   } ,
\label{(84)}
%
\end{align}
%
where
%
\begin{equation}
%
G_\sigma^V(i\omega) = G_\sigma^0(i\omega)
 - {\textstyle\frac{1}{2}} V_2n^{2c} \,
   \Gamma_{\sigma,\sigma}(i\omega) .
\label{(85)}
%
\end{equation}
%

Equation (84) is identical with the corresponding equation for the
single band Hubbard model without phonons \cite{Vakaru} if we replace
in Eq.\ (84) the renormalized quantities $\mu$, $U$,
$\tilde{\varepsilon}(\mathbf{k})$ and $G_\sigma^V(i\omega)$ by the
initial quantities $\mu_0$, $U_0$ and $\varepsilon(\mathbf{k})$.
This means that we have reduced the investigation of superconductivity
in frame of the Hubbard-Holstein model to the analogical problem
with respect to the single band Hubbard model.

It is instructive to analyze the contributions from the two spin
channels by considering the quantities
%
\begin{subequations} \begin{align}
%
\chi_{+}(i\omega) & =
   \frac{1}{\beta N} \sum_{\omega_1}\sum_{\mathbf{k}}
   \frac{
   [\tilde{\varepsilon}(\mathbf{k})]^2
   \Lambda_{e\sigma}(i\omega_1)
   }{
   1 - \tilde{\varepsilon}(\mathbf{k}) \,
   \Lambda_{e\sigma}(i\omega_1)
   }
\nonumber \\ & \times
   \widetilde{G}_2^{0 \,ir}[\sigma,i\omega;\sigma,i\omega_1|
   \sigma,i\omega_1;\sigma,i\omega] ,
\label{(86a)}
\\*[0.2cm]
\chi_{-}i\omega) & =
   \frac{1}{\beta N} \sum_{\omega_1} \sum_{\mathbf{k}}
   \frac{
   [\tilde{\varepsilon}(\mathbf{k})]^2
   \Lambda_{e\overline{\sigma}}(i\omega_1)
   }{
   1 - \tilde{\varepsilon}(\mathbf{k}) \,
   \Lambda_{e\overline{\sigma}}(i\omega_1)
   }
\nonumber \\ & \times
   \widetilde{G}_2^{0 \, ir}[\sigma,i\omega;\overline{\sigma},i\omega_1|
   \overline{\sigma },i\omega_1;\sigma,i\omega] ,
\label{(86b)}
%
\end{align} \end{subequations}
%
and using the notation
%
\begin{align}
%
\phi^e_\sigma(i\omega) & =
   \frac{1}{N} \sum_{\mathbf{k}}
   \frac{[\tilde{\varepsilon}(\mathbf{k})]^2
   \Lambda_{e\sigma}(i\omega}
   {1 - \tilde{\varepsilon}(\mathbf{k}) \,
   \Lambda_{e\sigma}(i\omega)}
\nonumber \\ & =
   \frac{1}{N} \sum_{\mathbf{k}}
   \frac{\tilde{\varepsilon}(\mathbf{k})}
   {1 - \tilde{\varepsilon}(\mathbf{k)} \,
   \Lambda_{e\sigma}(i\omega)} .
\label{(87)}
%
\end{align}
%
Here it is assumed that
$\varepsilon(\mathbf{k}) =\varepsilon(-\mathbf{k)}$
holds with
$\sum_{\mathbf{k}} \varepsilon(\mathbf{k}) = 0$.
We replace sums by integrals,
%
\begin{align}
%
\frac{1}{N} \sum_{\mathbf{k}} & =
   \int \! d\tilde{\varepsilon} \, \rho_0(\tilde{\varepsilon}) ,
\label{(88)}
\\
\rho_0(\tilde{\varepsilon}) & =
   \frac{4}{\pi\widetilde{W}}
   \sqrt{1 - \left(
   2\tilde{\varepsilon}/{\widetilde{W}} \right)^2 }
   \times \left \lbrace
\begin{tabular}{l}
   $1, \quad |\tilde{\varepsilon}| \leq \widetilde{W}/2$
   \\
   $0, \quad |\tilde{\varepsilon}| > \widetilde{W}/2$
\end{tabular} \right. .
\label{(89)}
%
\end{align}
%
$\widetilde{W}$ is the renormalized band width
$\widetilde{W} = W e^{-W_p}$ and $\rho_0$ the semielliptic model
density of states. Since we do not consider magnetic solutions,
the spin index can be omitted. Making use of (80a-b) for the
irreducible functions we obtain
%
\begin{subequations} \begin{align}
%
\chi_{+}(i\omega) & =
   \frac{
   \beta U^2(1 + e^{\beta\mu})(e^{\beta\mu} + e^{\beta(2\mu - U)})
   }{
   Z_0^2 \, \lambda(i\omega) \, \overline{\lambda}(i\omega)
   }
\nonumber \\ & \times
   \bigg [ \overline{\phi}
 - \frac{\phi^e_{\sigma}(i\omega)}
   {\beta \lambda(i\omega) \, \overline{\lambda}(i\omega)} \bigg ] ,
\label{(90a)}
\\*[0.3cm]
\chi_{-}(i\omega) & =
   \frac{U}{Z_0} \bigg \lbrace
 - \frac{
   2(\mu - U)(e^{\beta(2\mu - U)} - 1) \,
   \phi^e_{\sigma}(-i\omega)
   }{
   \beta(2\mu - U)[\lambda(i\omega) \, \overline{\lambda}(i\omega)]^2
   }
\nonumber \\ & -
   U \frac{
   e^{\beta\mu}\phi^e_{\sigma}(i\omega) + (1 + e^{\beta\mu})
   U \overline{\phi}
   }{
   [\lambda(i\omega) \, \overline{\lambda}(i\omega)]^2
   }
\nonumber \\ & +
   \bigg [
   \frac{\beta Ue^{2\beta\mu} (e^{-\beta U} - 1)\overline{\phi}}{Z_0}
 -  (1 + e^{\beta\mu})(\overline{\phi}_1 + 2\overline{\phi}) \bigg ]
\nonumber \\ & \times
   \frac{1}{\lambda(i\omega) \, \overline{\lambda}(i\omega)}
 - \frac{
   (e^{\beta(2\mu - U)} - 1) \gamma_0(i\omega)
   }{\overline{\lambda}^2(i\omega)}
   \bigg \rbrace ,
\label{(90b)}
%
\end{align} \end{subequations}
%
where
%
\begin{subequations} \begin{align}
%
\overline{\phi} & =
   \frac{1}{\beta} \sum \limits_{\omega_1}
   \frac{\phi^e_{\sigma}(i\omega_1)}
   {\lambda(i\omega_1) \, \overline{\lambda}(i\omega_1)}  ,
\label{(91a)}
\\
\overline{\phi}_1 & =
   \frac{1}{\beta} \sum \limits_{\omega_1}
   \phi^e_{\sigma}(i\omega_1) \left(
   \frac{1}{\lambda(i\omega_1)}
 - \frac{1}{\overline{\lambda}(i\omega_1)} \right)^{\! 2}
\nonumber \\
& = \frac{U^2}{\beta} \sum \limits_{\omega_1}
   \frac{\phi^e_{\sigma}(i\omega_1)}
   {[\lambda(i\omega_1) \, \overline{\lambda}(i\omega_1)]^2},
\label{(91b)}
\\
\gamma_0(i\omega) & =
   \frac{1}{\beta} \sum \limits_{\omega_1}
   \frac{\overline{\lambda}(i\omega_1) + \overline{\lambda}(i\omega)}
   {\overline{\lambda}(i\omega_1) + \lambda(i\omega)}
   \frac{\psi_{\omega + \omega_1,0} \, \phi^e_{\sigma}(i\omega_1)}
   {\overline{\lambda}^2(i\omega_1)} ,
\label{(91c)}
%
\end{align} \end{subequations}
%
and
$\psi_{\omega + \omega_1,0} = 1 - \delta_{\omega + \omega _1,0}$.
In case of half filling when
$\mu = U/2$, $\lambda(i\omega) = i\omega + \mu$,
$\overline{\lambda}(i\omega) = i\omega - \mu$,
we have the antisymmetry property,
$\phi^e(-i\omega) = -\phi^e(i\omega)$,
and therefore $\overline{\phi} = \overline{\phi}_1 = 0$.
Furthermore, we find at half filling,
%
\begin{subequations} \begin{align}
%
\chi_{+}(i\omega) & =
 - \frac{\mu^2 \phi^e(i\omega)}{[(i\omega)^2 - \mu^2]^2} ,
\label{(92a)}
\\
\chi_{-}(i\omega) & =
 - \frac{2\mu^2 \phi^e(i\omega)}{[(i\omega)^2 - \mu^2]^2}
 = 2\chi_{+}(i\omega) ,
\label{(92b)}
%
\end{align} \end{subequations}
%
and therefore Eq.\ (84) is equal to
%
\begin{equation}
%
\chi_{+}(i\omega) + \chi_{-}(i\omega)
 = 3 \chi_{+}(i\omega) .
\label{(93)}
%
\end{equation}
%

Away from half filling we will for simplicity omit the contribution
of the anti-parallel spin channel in Eq.\ (84) and introduce instead
a correction factor $f_s$, which is three at half filling and different
from three at general filling.
The final equation to be investigated is then
%
\begin{align}
%
\Lambda_{e\sigma}(i\omega) & =
   G_{\sigma}^V(i\omega)
\nonumber \\ & -
   \frac{f_s}{\beta N} \sum \limits_{\mathbf{k},\omega_1}
   \frac{[\tilde{\varepsilon}(\mathbf{k})]^2 \Lambda_{e}(i\omega_1)}
   {1 - \tilde{\varepsilon}(\mathbf{k}) \, \Lambda_{e\sigma}(i\omega)}
\nonumber \\ & \times
   \widetilde{G}_2^{0 \, ir}[\sigma,i\omega;\sigma,i\omega_1|
   \sigma,i\omega_1;\sigma,i\omega] .
\label{(94)}
%
\end{align}
%

Further simplifications can be made regarding Eq.\ (82) for the
critical temperature. We note that the main contribution to the last
term results from the minimum values of frequency difference
$\Omega_1 - \Omega_2$. When $\Omega_1 = \Omega_2$ we get after summing over
$\Omega_1$:
%
\begin{equation}
%
\frac{1}{\beta^2} \sum_{\Omega_1}
   [\phi(i\Omega_1)]^2
 = \frac{1}{\beta\omega_c} \coth {\textstyle\frac{1}{2}}
   \beta\omega_c
 + \frac{1}{2 \sinh^2 {\textstyle\frac{1}{2}} \beta\omega_c} ,
\label{(95)}
%
\end{equation}
%
so that
%
\begin{equation}
%
f_c =
   1 + \frac{1}{\beta\omega_c} \coth {\textstyle\frac{1}{2}}
   \beta\omega_c
 + \frac{1}{2 \sinh^2 {\textstyle\frac{1}{2}} \beta\omega_c}
\label{(96)}
%
\end{equation}
%
can be used as a common factor in the remaining function for the
superconducting order parameter, which is
%
\begin{widetext} \begin{equation}
%
Y_{e\sigma,\overline{\sigma}}(i\omega)
 = - \frac{f_c}{\beta N} \sum \limits_{\mathbf{k,}\omega_1}
   \frac{
   \tilde{\varepsilon}(\mathbf{k})\,
   \tilde{\varepsilon}(-\mathbf{k}) \,
   Y_{e\sigma,\overline{\sigma}}(i\omega_1) \,
   \widetilde{G}_2^{0 \, ir}[\sigma,i\omega;\overline{\sigma},-i\omega|
   \sigma,i\omega_1;\overline{\sigma},-i\omega_1]
   }{
   [1 - \tilde{\varepsilon}(\mathbf{k}) \,
   \Lambda_{e\sigma}(i\omega_1)]
   [1 - \tilde{\varepsilon}(-\mathbf{k}) \,
   \Lambda_{e\overline{\sigma}}(-i\omega_1)]
   } .
\label{(97)}
%
\end{equation} \end{widetext}
%
In order to solve Eq.\ (97) for $T_c$, we introduce a new function,
%
\begin{align}
%
\phi^{sc}(i\omega) & =
\frac{1}{N} \sum_{\mathbf{k}}
   \frac{\tilde{\varepsilon}(\mathbf{k}) \,
   \tilde{\varepsilon}(-\mathbf{k})}
   {[1 - \tilde{\varepsilon}(\mathbf{k}) \,
   \Lambda_{e\sigma }(i\omega)]
   [1 - \tilde{\varepsilon}(-\mathbf{k}) \,
   \Lambda_{e\overline{\sigma}}(-i\omega)]}
\nonumber \\ & =
   \frac{\phi^e(i\omega) - \phi^e(-i\omega)}
   {\Lambda_{e\sigma}(i\omega) - \Lambda_{e\overline{\sigma}}
      (-i\omega)} ,
\label{(98)}
%
\end{align}
%
which allows to rewrite Eq.\ (97) in the form
%
\begin{align}
%
Y_{e\sigma,\overline{\sigma}}(i\omega) & =
   - \frac{f_c}{\beta} \sum \limits_{\omega_1}
   \phi^{sc}(i\omega_1) \,
   Y_{e\sigma,\overline{\sigma}}(i\omega_1)
\nonumber \\ & \times
   \widetilde{G}_2^{0 \, ir}[\sigma,i\omega;\overline{\sigma},-i\omega|
   \sigma,i\omega_1;\overline{\sigma},-i\omega_1] .
\label{(99)}
%
\end{align}
%
Using furthermore
%
\begin{subequations} \begin{align}
%
\psi_1 & = \frac{1}{\beta} \sum \limits_\omega
   \frac{\phi^{sc}(i\omega) \, Y_{e\sigma,\overline{\sigma}}(i\omega)}
   {\mu^2 - (i\omega)^2} ,
\label{(100a)}
\\
\psi_2 & = \frac{1}{\beta} \sum \limits_\omega
   \frac{\phi^{sc}(i\omega) \, Y_{e\sigma,\overline{\sigma}}(i\omega)}
   {(\mu - U)^2 - (i\omega)^2} ,
\label{(100b)}
\\
\zeta & = \frac{U^2 f_c}{Z_0} \left \lbrace
   e^{\beta \mu} + \frac{e^{2\beta \mu}(1 - e^{-\beta U})}{Z_0}
   \right \rbrace ,
\label{(100c)}
\\
Q(i\omega) & = 1 + \frac{\zeta \phi^{sc}(i\omega)}
   {[\mu^2 - (i\omega)^2] [(\mu - U)^2 - i\omega)^2]} ,
\label{((100d)}
%
\end{align} \end{subequations}
%
allows to obtain the solution:
%
\begin{align}
%
Y_{e\sigma,\overline{\sigma}}(i\omega ) & =
 - \frac{Uf_c}{Z_0}
   \frac{[1 + U/(2\mu - U)] (1 + e^{\beta\mu}) \, \psi_1}
  {Q(i\omega) \, [ \mu^2 - (i\omega)^2]}
\nonumber \\ & -
 - \frac{Uf_c}{Z_0}
   \frac{[1 - U/(2\mu - U)]
   (e^{\beta\mu} + e^{\beta(2\mu - U)}) \,\psi _2}
   {Q(i\omega) \, [(\mu - U)^2 - (i\omega)^2]} ,
\label{(101)}
%
\end{align}
%
where symmetry properties of $Y_{e\sigma,\overline{\sigma}}(i\omega)$
and $\phi^{sc}(i\omega)$,
%
\begin{equation}
%
\phi^{sc}(-i\omega)  = \phi^{sc}(i\omega), \quad
Y_{\sigma,\overline{\sigma }}(-i\omega) =
   Y_{\sigma,\overline{\sigma}}(i\omega) .
\label{(102)}
%
\end{equation}
%
has been used. The two constants $\psi_1$ and $\psi_2$ are
determined from the following system of equations,
%
\begin{eqnarray}
%
\lefteqn{
   \psi_1 \left[ 1 + \frac{\eta_{11} f_c}{Z_0}
   \left( 1 + \frac{U}{2\mu - U} \right)
   \left( 1 + e^{\beta \mu } \right) \right]
}
\nonumber \\ && {} +
   \psi_2 \frac{\eta_{12 f_c}}{Z_0}
   \left( 1-\frac U{2\mu -U}\right)
   \left( e^{\beta \mu } + e^{\beta(2\mu - U)} \right) = 0 ,
\nonumber  \\*[0.2cm]
\lefteqn{
\psi_2 \left[ 1 + \frac{\eta_{22} f_c}{Z_0}
   \left( 1 - \frac{U}{2\mu - U} \right)
   \left( e^{\beta\mu} + e^{\beta(2\mu - U)} \right) \right]
}
\nonumber \\ && {} +
   \psi_1 \frac{\eta_{12} f_c}{Z_0}
   \left( 1 + \frac{U}{2\mu - U} \right)
   \left( 1 + e^{\beta\mu} \right) = 0 ,
\label{(103)}
%
\end{eqnarray}
%
Where $\eta_{ij}$ are given by the summations of
%
\begin{align}
%
\eta_{11} & =
   \frac{U}{\beta} \sum_{\omega}
   \frac{\phi^{sc}(i\omega)}
   {Q(i\omega) \, [\mu^2 - (i\omega)^2]^2} ,
\nonumber \\
\eta_{12} & =
   \frac{U}{\beta} \sum_\omega
   \frac{\phi^{sc}(i\omega)}
   {Q(i\omega) \, [\mu^2 - (i\omega)^2]
   [(\mu - U)^2 - (i\omega)^2]} ,
\nonumber \\
\eta_{22} & =
   \frac{U}{\beta} \sum_\omega
   \frac{\phi^{sc}(i\omega)}
   {Q(i\omega) \, [(\mu - U)^2 - (i\omega)^2]^2} .
\label{(104)}
%
\end{align}
%

The critical temperature is then obtained by setting $T = T_c$
in the equation:
%
\begin{eqnarray}
%
\lefteqn{
   \left[ 1 + \frac{\eta_{11} f_c}{Z_0}
   \left( 1 + \frac{U}{2\mu - U} \right)
   \left( 1 + e^{\beta\mu} \right) \right]
}
\nonumber \\ && \times
   \left[ 1 + \frac{\eta_{22} f_c}{Z_0}
   \left( 1 - \frac{U}{2\mu - U} \right)
   \left( e^{\beta\mu} + e^{\beta(2\mu - U)} \right)\right]
\nonumber \\ && {} -
   \left( \frac{\eta_{12} f_c^2}{Z_0} \right)^{\! 2}
   \left( 1 - \frac{U^2}{(2\mu - U)^2} \right)
   \left( e^{\beta\mu} + e^{\beta(2\mu - U)} \right)
\nonumber \\ && \times
   \left( 1 + e^{\beta\mu} \right) = 0 .
\label{(105)}
%
\end{eqnarray}
%
Equation (105) is invariant under the particle-hole transformation,
%
\begin{align}
%
n_\sigma & \rightarrow 1 - n_\sigma ,
\nonumber \\
\mu & \rightarrow - \mu + U ,
\nonumber \\
2\mu - U & \rightarrow - (2\mu - U) ,
\nonumber \\
Z_0(\mu) & \rightarrow Z_0(\mu) \exp [\beta( U - 2\mu)] .
\label{(106)}
%
\end{align}
%

\section{Metallic, insulating and superconducting phases}

In order to obtain a better understanding of the properties of the main
equations, we will first check whether the normal state determined by
Eq.\ (94) is metallic or dielectric. This can be done by analyzing
the renormalized density of states, $\rho(E)$, which is given by
%
\begin{subequations} \begin{align}
%
\rho (E) & = -
   \frac{1}{\pi} \, \mathrm{Im} \,
   g(E + i0^{+}) ,
\label{(107a)}
\\
g(i\omega_n) & =
   \frac{1}{N} \sum \limits_{\mathbf{k}}
   \frac{\Lambda(i\omega_n)}
   {1 - \tilde{\varepsilon}(\mathbf{k}) \,
   \Lambda(i\omega_n)} ,
\label{(107b)}
%
\end{align} \end{subequations}
%
where $\Lambda(i\omega_n)$ has to be calculated from Eq.\ (94).
The integration over $\mathbf{k}$ is again done by using the
semielliptical form of model density of states in Eq.\ (89).
This gives for the quantities (87) and (107b) the
following result \cite{Izyumov}:
%
\begin{subequations} \begin{align}
%
\phi^e(i\omega) & =
   {\textstyle\frac{1}{2}} \widetilde{W} \varphi^e(i\omega) ,
\label{(108a)}
\\
\varphi^e(i\omega) & =
   (1 - [1 - \widetilde{\Lambda}^2(i\omega)]^{1/2})^2 /
   \widetilde{\Lambda}^3(i\omega) ,
\label{(108b)}
%
\end{align} \end{subequations}
%
\begin{subequations} \begin{align}
%
g(i\omega) & = (4/\widetilde{W}) \,
   (1 - [1 - \widetilde{\Lambda }^2(i\omega)]^{1/2}) /
   \widetilde{\Lambda}(i\omega) ,
\label{(109a)}
\\
\widetilde{\Lambda}(i\omega) & =
   {\textstyle\frac{1}{2}} \widetilde{W} \Lambda(i\omega) ,
\label{109b}
%
\end{align} \end{subequations}
%
yielding for the renormalized density of states (DOS) in Eq.\ (107a)
%
\begin{equation}
%
\rho(E)\equiv \frac{4 \, r(E)}{\pi \widetilde{W}}
 = - \frac{4}{\pi \widetilde{W}} \, \mathrm{Im} \,
   \frac{1 - \sqrt{1-\widetilde{\Lambda}^2(E)}}
   {\widetilde{\Lambda}(E)} ,
\label{(110)}
%
\end{equation}
%
where $\widetilde{\Lambda}(E)$ is the analytical continuation
of $\widetilde{\Lambda}(i\omega_n)$.

We now address Eq.\ (94) which determines the normal state of the
system. By Using (90a) and (108) we rewrite it in the form:
%
\begin{equation}
%
\widetilde{\Lambda}(i\omega) =
   {\textstyle\frac{1}{2}} \widetilde{W} G^V(i\omega)
 - {\textstyle\frac{1}{2}} f_s \widetilde{W} \chi_{+}(i\omega) ,
\label{(111)}
%
\end{equation}
%
with $\chi_{+}(i\omega)$ given by Eq. (90a). In the limit
$E \to 0$ Eq.\ (111) becomes
%
\begin{equation}
%
\widetilde{\Lambda} \bigg [ 1 - \frac{a^2}{\widetilde{\Lambda}^4}
   \bigg ( 1 - \sqrt{1 - \widetilde{\Lambda}^2} \, \bigg )^{\! 2}
   \bigg ] = b ,
\label{(112)}
%
\end{equation}
%
where $\widetilde{\Lambda} = \widetilde{\Lambda}(E + i\delta)$
for $E = 0$. The two parameters $a$ and $b$ and the function
$G^V(0)$ are given by
%
\begin{subequations} \begin{align}
%
a^2 & = {\textstyle\frac{1}{4}} f_s \widetilde{W}^2 \,
   \frac{U^2(1 + e^{\beta\mu}) (e^{\beta\mu} + e^{\beta(2\mu - U)})}
        {Z_0^2\mu^2(\mu - U)^2} ,
\label{(113a)}
\\*[0.2cm]
b & = {\textstyle\frac{1}{2}} \widetilde{W} G^V(0)
 - {\textstyle\frac{1}{2}} f_s
\nonumber \\ & \times
   \frac{\beta\widetilde{W}U^2( 1 + e^{\beta\mu})
   (e^{\beta\mu} + e^{\beta(2\mu - U)}) \, \overline{\phi}}
   {\mu(\mu - U) Z_0^2} ,
\label{(113b)}
\\*[0.2cm]
G^V(0) & = \frac{1 - \overline{n}_\sigma}{\mu}
 + \frac{\overline{n}_\sigma}{\mu - U}
\nonumber \\ & -
   \frac{n^{2c} \, V_2\beta}{\widetilde{\mu}^2} \left(
   1 + \frac{\beta\mu}{1 + e^{\beta\mu}} \right)
   (1 - \overline{n}) ,
\label{(113c)}
\\*[0.2cm]
\overline{n} & = 2\overline{n}_{\sigma} ,
\label{(113d)}
%
\end{align} \end{subequations}
%
with $\overline{\phi}$ given by Eq.\ (91). The first term
in Eq.\ (113c) is the value of the local one-particle Green's function
of the Hubbard model for $E = 0$ while the second term is the
corresponding contribution from the intersite Coulomb interaction
discussed in Appendix B, see (B16).

Equation (112) has been discussed for the simpler case of half
filling when $b = 0$ in Refs.\ \onlinecite{Entel} and
\onlinecite{Izyumov}. Away from half filling we have
$\overline{n} \neq 1$, $\mu \neq U/2$ and $b \neq 0$.
As Eq.\ (113a-b) show, the parameters $a$ and $b$ depend on
chemical potential $\mu$, strong-coupling parameter $U$ and
mean electron number per lattice site $\overline{n}$.
Moreover, $b$ depends on the intersite Coulomb repulsion.
Although parameters $a$ and $b$ are not independent of each
other, we first try to get information in case that $b = 0$,
i.e., for $\overline{\phi} = G^V(\nu) = 0$, which holds for
$\Lambda(-i\omega) = - \Lambda(i\omega)$. In the half-filled band
case when $\mu = U/2$ and $\overline{n} = 1$, the quantity $a$ is equal
to
%
\begin{equation}
%
a = \sqrt{f_s} \, \widetilde{W}/U, \quad f_s = 3 .
\label{(114)}
%
\end{equation}
%
The equality b=0  allows to determine $\Lambda$ in Eq.\ (112) from a simpler
relation,
%
\begin{equation}
%
\left[ 1 - a \left(
   \frac{1 - \sqrt{1 - \widetilde{\Lambda}^2}}
   {\widetilde{\Lambda}^2} \right) \right]
   \left[ 1 + a \left( \frac{1 - \sqrt{1 - \widetilde{\Lambda}^2}}
   {\widetilde{\Lambda}^2} \right) \right] = 0  .
\label{(115)}
%
\end{equation}
%
For $a > 1$ the first factor gives $\widetilde{\Lambda}^2 = a (2 - a)$
and hence,
%
\begin{subequations} \begin{align}
%
\widetilde{\Lambda} & = \pm \sqrt{a\left( 2-a\right) }, \quad  a < 2 ,
\label{(116a)}
\\
\widetilde{\Lambda} & = \pm i\sqrt{a\left( a-2\right) }, \quad a > 2 .
\label{(116b)}
%
\end{align} \end{subequations}
%

By inserting these solutions in Eq.\ (109) for the renormalized DOS
we obtain results different from zero only for the expression with lower
sign in Eq.\ (116b), i.e., for $a > 2$. Hence we obtain a metallic
state at half filling only if the Coulomb interaction is less
than a critical value \cite{Hubbard},
%
\begin{equation}
%
U < U_c, \quad U_c = {\textstyle\frac{1}{2}} \sqrt{3} \,
   \widetilde{W} .
\label{(117)}
%
\end{equation}
%
In this case there is no gap at the Fermi level and the renormalized
DOS becomes $(a>2)$
%
\begin{equation}
%
\rho(0) = 4 / (\pi\widetilde{W}) \, r(0),
   \quad r(0) = \sqrt{(a-2)/a} .
\label{(118)}
%
\end{equation}
%
The insulating (dielectric) phase exists if the inverse condition holds,
$a < 2$, $U > U_c$, leading to the opening of an energy gap at
the Fermi level.

Away from half filling, for $a>2$ and $b$ sufficiently small,
the solution of Eq.\ (112) can be obtained from a series expansion
in powers of $b$,
%
\begin{equation}
%
\widetilde{\Lambda}_b = \widetilde{\Lambda}
   + \frac{(a - 1)b}{2(a - 2)}
   + \frac{(a^2 - 5a + 3)b^2}
     {8\widetilde{\Lambda} (a - 1)(a - 2)^2} + \ldots \,
\label{(119)}
%
\end{equation}
%
where $\widetilde{\Lambda}$ is given by Eq.\ (116) in zero-order
approximation. This leads to
%
\begin{equation}
%
r(0) = \sqrt{\frac{a - 2}a}
 + \frac{(2a^3 - 8a^2 + 12a - 5) b^2}
   {8a^{3/2} (a - 1)^2 (a - 2)^{5/2}} \ldots
\label{(120)}
%
\end{equation}
%

This result shows that not $b$ but $b/(a - 2)$ should be taken as
expansion parameter. Therefore the expansion is not correct very close
to $a = 2$. Another peculiarity of Eq.\ (120) is its even character in
$b$, which follows from the fact that in the solution of Eq.\ (112),
$\widetilde{\Lambda}$, changes sign when the sign of $b$ is changed.
Therefore, if we have the solutions $\Lambda_1 \pm i\Lambda_2$
of Eq.\ (112), then correspondingly $-\Lambda_1 \mp i\Lambda_2$ are
solutions for $b < 0$. The different signs of the real parts do
not matter since Eq.\ (110) depends only on the absolute value of
 $\Lambda_1$. The numerical investigation shows that for
$a > 2$ the role of $b$ is not decisive, i.e., the system remains
metallic. However, in the range $0 < a <2$, for which the system
is insulating when $b = 0$, the influence of $b$ is decisive
because there exists a critical value, $b_c(a) \approx 1 - a/2$,
such that the system is metallic for $|b| > b_c$ and insulating for
$|b| < b_c$. Note that the parameters $a$ and $b$ are not independent each other
and therefore such extremal parameter values as $a$ small and $b$ large or vice versa are
not admitted by their definition (eq.113).

The metallic state exists for low and high band filling (small and
large values of $\mu$) and near half filling, $(\mu \simeq U/2)$,
provided that $U < U_c$. The physical role of $b$ is to enhance in each
case the tendency towards metalicity. There is also the case
of reappearance of the metallic phase for $a < 2$ and $|b| > b_c(a)$.
The variation of $b_c(a)$ is shown in Fig.\ 9.

\begin{figure}[!h]
%
\centering
\includegraphics[width=0.475\textwidth,clip]{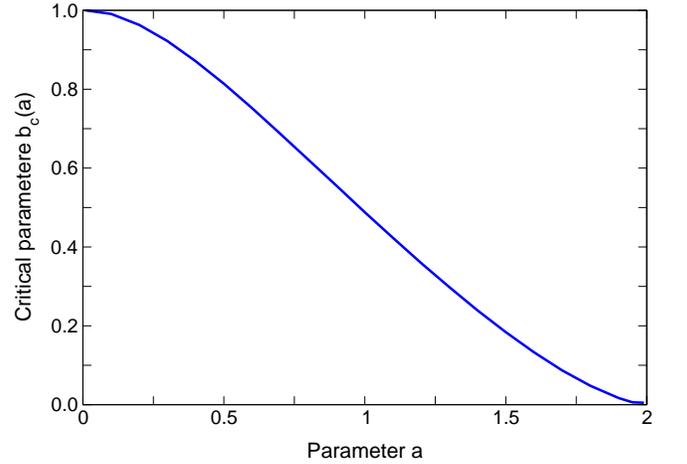}
\caption{
Variation of the critical parmaeter $b_c(a)$ in the range
$0 < a < 2$. For $|b| > b_c$ is the system metallic while it is
insulating for $|b| < b_c$.
For $a > 2$ the system is metallic regardless of the value of $b$.
}
\label{fig-9}
%
\end{figure}


We discuss now the superconducting phase transition with $T_c$
obtained from Eq.\ (105). In order to gain some insight we need to
simplify $\phi^{sc}(i\omega)$ from Eq.\ (98). In the asymptotic limit
$|\omega|\to\infty$ this quantity is equal to
%
\begin{equation}
%
\phi^{sc}(i\omega) \to
   {\textstyle\frac{1}{4} (\widetilde{W}}/2)^2 ,
\label{(121)}
%
\end{equation}
%
whereas in the low-energy limit $E \to 0$ required here, we obtain from Eqs.\
(90a) and (111) the value
%
\begin{equation}
%
\phi^{sc}(E + i\delta) \to \phi^{sc}
 = (1/a)^2 \, (\widetilde{W}/2)^2 .
\label{(122)}
%
\end{equation}
%
The two limits coincide for $a = 2$. Really we are interested
 mainly in the low energy limit (eq.122).
Because of the fast convergence of the sums
in Eq.\ (104) which contain $\phi^{sc}(i\omega)$ and which determine the
parameters $\eta_{ij}$ of Eq.\ (105), we can rewrite $\eta_{ij}$ as
%
\begin{subequations} \begin{align}
%
\eta_{11} & =
   \frac{U\phi^{sc}}{\beta} \sum \limits_{\omega}
   \frac{(d^2 - z^2)}{Q_1(z) (c^2 - z^2)}
\nonumber \\ & =
   U \phi^{sc} [I_1 + I_2(c^2 - d^2)] ,
\label{(123a)}
\\
\eta_{22} & =
   \frac{U\phi^{sc}}{\beta} \sum \limits_{\omega}
   \frac{(c^2 - z^2)}{Q_1(z) (d^2 - z^2)}
\nonumber \\ & =
   U\phi^{sc} [I_1 - I_3(c^2 - d^2)] ,
\label{(123b)}
\\
\eta_{12} & = \frac{U\phi^{sc}}{\beta} \sum \limits_{\omega}
   \frac{1}{Q_1(z)} = U\phi^{sc}I_1 ,
\label{(123c)}
%
\end{align} \end{subequations}
%
where
%
\begin{subequations} \begin{align}
%
I_1 & = \frac{1}{\beta} \sum \limits_{\omega} \frac{1}{Q_1(z)} ,
\label{(124a)}
\\
I_2 & = \frac{1}{\beta} \sum \limits_{\omega} \frac{1}{Q_1(z) ,
   (z^2 - c^2)} ,
\label{(124b)}
\\
I_3 & = \frac{1}{\beta} \sum \limits_{\omega} \frac{1}{Q_1(z)
   (z^2 - d^2)} ,
\label{(124c)}
\\*[0.2cm]
Q_1(z) & = (z^2 - c^2)(z^2 - d^2) + \zeta \phi^{sc} ,
\label{(124d)}
%
\end{align} \end{subequations}
%
and
%
\begin{align}
%
\zeta \phi^{sc} & =
   \frac{f_c\mu^2(\mu - U)^2 [Z_0e^{\beta\mu} + e^{2\beta\mu}
   (1 - e^{-\beta U})]}
   {f_s (e^{\beta\mu} + 1)(e^{\beta\mu} + e^{\beta(2\mu - U)})} ,
\nonumber \\
c^2 & = \mu^2, \quad d^2 = (\mu - U)^2, \quad z = i\omega_n .
\label{(125)}
%
\end{align}
%

In case that $U > 0$ the quantity $\zeta\phi^{sc} \equiv \gamma^4$ is
positive. In the other case of very strong electron-phonon interaction
we have $U < 0$. In this case we will use the notation
$\zeta\phi^{sc} = -\theta^4$ (with negative $\zeta$). Although
$U < 0$ seems not to be the generic case, we believe that in the
metallic phase in case of strong electron-phonon interaction an
effective attractive interaction is possible, because the local
and intersite Coulomb interaction can be largely screened by
the electron-ion interaction. However, calculation of self-consistent
screening in not the aim of this paper. We just require charge
neutrality and take $U$ as a parameter which allows to discuss
the case of effective repulsive as well as attractive interactions.
We first discuss the case $U > 0$ and then
$U < 0$.

The sums in $I_n$ have been evaluated by contour integration with
the help of Poisson's formula, see Eq.\ (C2).
For the case $\gamma^4 \geq (c^2 - d^2)^2/4$ we obtain the
following equation which determines $T_c$:
%
\begin{widetext} \begin{eqnarray}
%
\lefteqn{
0 = 1 + \frac{f_cU\phi^{sc}}{2\gamma^4\sqrt{\gamma^4 + c^2d^2}} \Bigg \lbrace
   \frac{2\gamma^4 - (c^2 - d^2)^2}
        {\sqrt{4\gamma^4 - (c^2 - d^2)^2}}
   \bigg [ 1 + \frac{U(1 - e^{\beta(2\mu - U)})}
                    {Z_0(2\mu - U)} \bigg ]
   \frac{\alpha_2 \sinh \beta\alpha_1 - \alpha_1 \sin \beta\alpha_2}
   {\cosh \beta\alpha_1 + \cos \beta\alpha_2}
}
\nonumber \\ && {} +
   \bigg [ U + \frac{(2\mu - U)(1 - e^{\beta(2\mu - U)})}{Z_0} \bigg ]
   \frac{U[\alpha_1 \sinh \beta\alpha_1 + \alpha_2 \sin \beta\alpha_2]}
   {\cosh \beta\alpha_1 + \cos \beta\alpha_2}
\nonumber \\ && {} +
   \bigg [ \frac{\tanh {\textstyle\frac{1}{2}} \beta d}{dZ_0}
   (\mu - U)(e^{\beta\mu} + e^{\beta(2\mu - U)})
 - \frac{\mu \tanh {\textstyle\frac{1}{2}} \beta c}{cZ_0} (1 + e^{\beta\mu})
   \bigg ] 2U \sqrt{\gamma^4 + c^2d^2} \Bigg \rbrace
\nonumber \\ && {} +
   f_c^2 \, \frac{U^2\mu(\mu - U)(U\phi^{sc})^2 (1 + e^{\beta\mu})
            (e^{\beta\mu} + e^{\beta(2\mu - U)})}
            {\gamma^8 Z_0^2 \sqrt{\gamma^4 + c^2d^2}}
   \Bigg \lbrace \frac{2\gamma^4 - (c^2 - d^2)^2}
                 {\sqrt{4\gamma^4 - (c^2 - d^2)^2}}
   \frac{1}{(c^2 - d^2)}
\nonumber \\ &&\ \times
   \bigg (
   \frac{\tanh {\textstyle\frac{1}{2}} \beta d}{d}
 - \frac{\tanh {\textstyle\frac{1}{2}} \beta d}{c} \bigg )
   \frac{\alpha_2\sinh \beta\alpha_1 - \alpha_1\sin \beta\alpha_2}
        {\cosh \beta\alpha_1 + \cos \beta\alpha_2}
 -  \frac{\sinh^2 \beta\alpha_1 + \sin^2 \beta\alpha_2}
        {[\cosh \beta\alpha_1 + \cos \beta\alpha_2]^2}
\nonumber \\ && {} -
   \sqrt{\gamma^4 + c^2d^2} \,
   \frac{\tanh {\textstyle\frac{1}{2}} \beta d}{d}
   \frac{\tanh {\textstyle\frac{1}{2}} \beta d}{c}
  + \bigg ( \frac{\tanh {\textstyle\frac{1}{2}} \beta d}{d}
         + \frac{\tanh {\textstyle\frac{1}{2}} \beta d}{c} \bigg )
   \frac{\alpha_1 \sinh \beta\alpha_1 + \alpha_2 \sin \beta\alpha_2}
        {\cosh \beta\alpha_1 + \cos \beta\alpha_2}
   \Bigg \rbrace  ,
\label{(126)}
%
\end{eqnarray} \end{widetext}
%
where the parameters $\alpha_1$ and $\alpha_2$ are given by Eq.\ (C3).
For the special case of half filling when $\mu = U/2>0$ and
%
\begin{align}
%
c^2 & = d^2 = \mu^2 = (U/2)^2 , \quad
   \phi^{sc} = {\textstyle\frac{1}{3}} \mu^2 ,
\nonumber \\
\gamma^4 & = f_c\ \frac{\mu^4 (e^{\beta\mu}
 - {\textstyle\frac{1}{3}} )}{e^{\beta\mu} + 1} > 0 ,
\label{(127)}
%
\end{align}
%
Eq.\ (126) is of simpler form,
%
\begin{widetext} \begin{eqnarray}
%
\lefteqn{
0 = 1 + \frac{f_c\mu^5}{3\gamma^4\sqrt{\gamma^4 + \mu^4}}
   \Bigg \lbrace \frac{\gamma^2}{\mu^2}
   \bigg ( 1 - \frac{\beta\mu}{e^{\beta\mu} + 1} \bigg )
   \frac{\alpha_2 \sinh \beta\alpha_1 - \alpha_1\sin \beta\alpha_2}
        {\cosh \beta\alpha_1 + \cos \beta\alpha_2}
 + 4 \,
   \frac{\alpha_1 \sinh \beta\alpha_1 + \alpha_2 \sin \beta\alpha_2}
   {\cosh \beta\alpha_1 + \cos \beta\alpha_2}
}
\nonumber \\ && {} -
   \frac{4}{\mu} \, \sqrt{\gamma^4 + \mu^4} \,
   \tanh {\textstyle\frac{1}{2}} \beta\mu \Bigg \rbrace
 + \frac{4f_c^2\mu^{10}}{9\gamma^8\sqrt{\gamma^4 + \mu^4}} \Bigg \lbrace
   \frac{\sinh^2 \beta\alpha_1 + \sin^2 \beta\alpha_2}
        {[\cosh \beta\alpha_1 + \cos \beta\alpha_2]^2}
 + \frac{\sqrt{\gamma^4 + \mu^4}}{\mu^2} \,
   \tanh^2 {\textstyle\frac{1}{2}} \beta\mu
\nonumber \\ && {} +
   \frac{\gamma^2}{2\mu} \,
   \bigg ( \frac{d}{d\mu} \,
        \frac{\tanh {\textstyle\frac{1}{2}} \beta\mu}{\mu} \bigg )
   \frac{\alpha_2 \sinh \beta\alpha_1 - \alpha_1 \sin \beta\alpha_2}
        {\cosh \beta\alpha_1 + \cos \beta\alpha_2}
 - \frac{2}{\mu} \, \tanh {\textstyle\frac{1}{2}} \beta\mu \,
   \frac{\alpha_1 \sinh \beta\alpha_1 + \alpha_2 \sin \beta\alpha_2}
        {\cosh \beta\alpha_1 + \cos \beta\alpha_2} \Bigg \rbrace ,
\label{(128)}
%
\end{eqnarray} \end{widetext}
%
where
%
\begin{align}
%
\alpha_{1 \atop 2} & = \frac{1}{\sqrt{2}} \left( \sqrt{\gamma^4 \pm \mu^4}
 + \mu^2 \right)^{\! 1/2} ,
%
\label{(129)}
%
\end{align}
%
If we consider at half filling in addition
$\beta\mu \gg 1$ and $\beta\omega _c \gg 1$ we get $\gamma^4 = \mu^4$
and $f_c = 1$ and instead of Eq.\ (129) we obtain
%
\begin{align}
%
\alpha_{1 \atop 2} & = \frac{\mu}{\sqrt{2}}
   \left( \sqrt{2} \pm 1 \right)^{\! 1/2},
\label{(130)}
%
\end{align}
%
and Eq.\ (128) for $T = T_c$ becomes
%
\begin{equation}
%
\beta\mu e^{-\beta\mu} + 1 + 2\sqrt{\sqrt{2}+1}
   \left( 1 + 2\sqrt{2} + 2\sqrt{\sqrt{2}+1} \, \right) = 0 ,
\label{(131)}
%
\end{equation}
%
which can not be fulfilled. This shows that for the half-filled
band case and positive renormalized Coulomb interaction \textit{s}-wave
superconductivity is not favored. This may change for \textit{d}-wave
superconductivity, which would be a technically more demanding
investigation in view of our expansion around the atomic limit,
because of the explicit $\mathbf{k}$ dependence of the
order parameter, which has to be retained in case of
\textit{d}-wave superconductivity.

In the following we will discuss the opposite case, i.e.,
$\zeta\phi^{sc} \leq (c^2 - d^2)^2/4$, which includes the
possibility to consider the negative values of $\zeta$ and hence
$\zeta\phi^{sc} = -\theta^4 < 0$. This case corresponds to
$U < 0$. The corresponding equation for $T_c$ is then
%
\begin{widetext} \begin{eqnarray}
%
\lefteqn{
   0 = 1 - \frac{f_c U\phi^{sc}}{4\theta^4} \Bigg \lbrace
   \frac{(c^2 - d^2)^2 + 2\theta^4}{\sqrt{(c^2 - d^2)^2 + 4\theta^4}}
   \bigg ( 1 - \frac{U(e^{\beta(2\mu - U)} - 1)}{Z_0(2\mu - U)} \bigg )
   \bigg ( \frac{\tanh {\textstyle\frac{1}{2}} \beta z_1}{z_1}
           \frac{\tanh {\textstyle\frac{1}{2}} \beta z_2}{z_2} \bigg )
}
\nonumber \\ & & {} +
   U \bigg ( U + \frac{(2\mu - U)(1 - e^{\beta(2\mu - U)})}{Z_0} \bigg )
   \bigg ( \frac{\tanh {\textstyle\frac{1}{2}} \beta z_1}{z_1}
         + \frac{\tanh {\textstyle\frac{1}{2}} \beta z_2}{z_2} \bigg )
\nonumber \\ && {} +
   4 U \bigg ( (\mu - U)
   \frac{\tanh {\textstyle\frac{1}{2}} \beta d}{d}
   \frac{(e^{\beta\mu} + e^{\beta(2\mu - U)})}{Z_0} 
 - \mu \frac{\tanh {\textstyle{1}{2}} \beta c}{c}
  \frac{(1 + e^{\beta\mu})}{Z_0} \bigg ) \Bigg \rbrace
\nonumber \\ && {} +
   \frac{f_c^2 U^4\mu(\mu - U)(\phi^{sc})^2 (1 + e^{\beta\mu})
   (e^{\beta\mu} + e^{\beta(2\mu - U)})}{2\theta^8Z_0^2}
   \Bigg \lbrace \bigg (
   \frac{\tanh {\textstyle\frac{1}{2}} \beta c}{c}
 + \frac{\tanh {\textstyle\frac{1}{2}} \beta d}{d} \bigg ) \bigg (
   \frac{\tanh {\textstyle\frac{1}{2}} \beta z_1}{z_1}
 + \frac{\tanh {\textstyle\frac{1}{2}} \beta z_2}{z_2} \bigg )
\nonumber \\ && {} -
   2 \, \bigg (
   \frac{\tanh {\textstyle\frac{1}{2}} \beta z_1}{z_1}
   \frac{\tanh {\textstyle\frac{1}{2}} \beta z_2}{z_2}
 + \frac{\tanh  {\textstyle\frac{1}{2}} \beta c}{c}
   \frac{\tanh  {\textstyle\frac{1}{2}} \beta d}{d} \bigg )
 - \frac{1}{(c^2 - d^2)}
\nonumber \\ && \times
   \bigg (
   \frac{\tanh  {\textstyle\frac{1}{2}} \beta c}{c}
 - \frac{\tanh  {\textstyle\frac{1}{2}} \beta d}{d} \bigg)
   \bigg (
   \frac{\tanh {\textstyle\frac{1}{2}} \beta z_1}{z_1}
  -\frac{\tanh {\textstyle\frac{1}{2}} \beta z_2}{z_2} \bigg )
   \frac{(c^2 - d^2)^2 + 2\theta^4}
   {\sqrt{(c^2 - d^2)^2 + 4\theta^4}} \Bigg \rbrace ,
\label{(132)}
%
\end{eqnarray} \end{widetext}
%
where
%
\begin{align}
%
z_{1 \atop 2}  = \frac{1}{\sqrt{2}} \left(
   c^2 + d^2 \pm \sqrt{(c^2 - d^2)^2 + 4\theta^4} \, \right)^{\! 1/2} .
\label{(133)}
%
\end{align}
%

For half filling, $U < 0$ and $\theta^4 < c^2d^2t$ we have
%
\begin{align}
%
\zeta\phi^{sc} & = -\theta^4 = -{\textstyle\frac{1}{3}} f_c \,
   \mu^4 \frac{1 - 3e^{-\beta|U|/2}}{1 + e^{-\beta |U|/2}}
\label{(134)}
%
\end{align}
%
for sufficiently large $|U|$, i.e., $\ln |U\beta| > \ln 9$.

In case of very small $\theta$ values corresponding to the
condition
%
\begin{align}
%
\ln \beta |\mu| = \ln (\beta {\textstyle\frac{1}{2}} |U| )
   \geq \ln 3
\label{(135)}
%
\end{align}
%
Eq.\ (132) simplifies to ($\theta \to 0$) ,
%
\begin{align}
%
1 & - f_c\mu\phi^{sc} \left \lbrace A_1
   \left( 1 - \frac{\beta\mu}{1 + e^{\beta\mu}} \right) +
   4\mu^2 A_2 \right \rbrace
\nonumber \\
  & - 4f_c^2 \mu^6 (\phi^{sc})^2 (A_1A_3 - A_2^2) = 0 ,
\label{(136)}
%
\end{align}
%
where
%
\begin{align}
%
A_n & = \frac{1}{n!} \frac{d^{\, n}}{(d\mu^2)^n} \left(
   \frac{\tanh {\textstyle\frac{1}{2}} \beta\mu}{\mu} \right) .
\label{(137)}
%
\end{align}
%

Because $\beta|\mu|$ is assumed to be larger than 3, the
coefficients in Eq.\ (137) can be approximated by
%
\begin{align}
%
A_1 & \simeq - \frac{1}{2|\mu|^3}, \quad
   A_2 \simeq \frac{3}{8|\mu|^5}, \quad
   A_3 \simeq - \frac{5}{16|\mu|^7}
\label{(138)}
%
\end{align}
%
allowing to replace Eq.\ (136) by
%
\begin{align}
%
1 & + \frac{f_c}{3} - \frac{f_c^2}{144} - \frac{f_c}{6}
   \beta|\mu| = 0 .
\label{(139)}
%
\end{align}
%
For negative chemical potential, $\mu < 0$, and
$f_c = 1 + 1/(\beta\omega_c)$ the equation for the critical
temperature has the simple form:
%
\begin{equation}
%
t^3 - 46t^2 + (24y - 191)t + 24y = 0 ,
\label{(140)}
%
\end{equation}
%
with
%
\begin{equation}
%
t = \frac{k_BT_c}{\omega_c}, \quad
   y = \frac{|\mu|}{\omega_c} ,
\label{(141)}
%
\end{equation}
%
yielding for small $t$ the solution,
%
\begin{equation}
%
t \simeq \frac{\overline{y}}{1 - \overline{y}}
 - \frac{46}{191} \frac{\overline{y}^2}{(1 - \overline{y})^3} ,
\label{(142)}
%
\end{equation}
%
where
%
\begin{align}
%
\overline{y} = \frac{24}{191} y ,
\label{(143)}
%
\end{align}
%
with the requirement that $\overline{y} \ll 1$ is fulfilled.
In the simplest approximation $T_c$ is of the order,
%
\begin{equation}
%
k_B T_c \simeq \frac{24}{191} |\mu| ,
%
\end{equation}
%
showing that $T_c$ is proportional to the renormalized
chemical potential in Eq.\ (14) and increases linearly with increasing
strength of the electron-phonon coupling parameter.

\section{Conclusions}

The interaction of correlated electrons and acoustical
phonons has been discussed by using the canonical transformation
of Lang-Firsov which results in mobile polarons consisting of
electrons surrounded by the acoustical phonon fields (clouds). A kind
of  generalized Wick's theorem is used to handle the strong Coulomb
repulsion between the electrons emerged into the see of phonon fields.

In the strong-coupling limit of the electron-phonon interaction
chronological thermodynamic averages of products of acoustical
phonon-field operators are expressed by averages of one-cloud
operators. For the normal one-cloud propagator the Lorentzian form
in Eq.\ (25) while for anomalous one the Gaussian form in Eq.\ (28)
has been found. Because the latter propagator is considerably smaller
than the first one, we find that the anomalous electronic Green's
functions are more important than the corresponding polaronic
functions. So the superconducting phase transition is determined
as usual by the appearance of electronic Cooper pairs, i.e.,
the pairing of electrons without phonons-clouds is easier to achieve
than the paring of polarons with such clouds.

For the system of renormalized electronic Green's functions
in Eq.\ (64) the diagrammatical structure is analyzed and the Dyson
equations have been derived, see Eqs.\ (66-69).
Besides the full Green's functions the equations contain also three
correlation functions and three mass operators. These quantities
have been calculated by summing infinite series of diagrams after
performing appropriate approximations of Eq.\ (68).
Resulting Eqs.\ (77) and (79) for the superconducting state
are then linearized in terms of the order parameter
$Y_{\sigma,\overline{\sigma}}$ leading to the final Eq.\ (105)
which determines the superconducting transition temperature $T_c$,
which is invariant with respect to particle-hole transformation.
Further analysis shows that the problem of superconductivity
in the frame of the Hubbard-Holstein model is analogous to the
discussion of superconductivity in the frame of the single band
Hubbard model with appropriately renormalized paramaters.

For further discussion the normal state properties given by Eq.\ (94)
are investigated with respect to the metal-insulator transition.
For half filling, i.e., one electron per lattice site and $\mu = U/2$,
the model yields a metallic state provided the renormalized
value of $U$ is smaller than $\sqrt{3}/2 \widetilde{W}$,
where $\widetilde{W}$ is the electron energy band width, which is
narrowed by the effect of the phonon fields, see Eqs.\ (112)
and (115). The second parameter $b$ in Eq.\ (112) determines the
deviation from half-filling. It has been shown that away from
half filling a metallic state is favored for
$U > \sqrt{3}/2 \widetilde{W}$ and $b$ larger than a critical value
$b_c$.

The search for superconductivity has then been performed on the basis
of Eq.\ (105) for \textit{s}-wave superconductivity
for two cases. In the first case, at half filling,
 $(\mu = U/2)$ with positive $U$, Eq.\ (105) has been reduced to Eq.\
(131), which has no solution. In the second case of negative $U$
the non-trivial solution in Eq.\ (144) has been found. As mentioned
before, in a real solid with correlated electrons the quantity $U$
should be replaced by an effective screened parameter.
For an overall attractive interaction mediated by the phonons
the Hubbard-Holstein model can have a superconducting solution
although the bare value $U_0$ can still be substantially large.

\begin{acknowledgments}
%
This work was supported by the Heisenberg-Landau Program. It is a
pleasure to acknowledge discussions with Prof.\ N.\ Plakida and Dr.\
S.\ Cojocaru, to thank Vadim Shulezhko  for asistance in diagram
drawing. V.A.M.\ would like to thank the University of
Duisburg-Essen for financial support. P.E.\ thanks the Bogoliubov
Laboratory of Theoretical Physics, JINR, for the hospitality he received
during his stay in Dubna.
%
\end{acknowledgments}

\appendix

\section{Laplace approximation} \label{AppLapjApr_A}

In this section we provide calculational details of the
Fourier representations $\widetilde{\phi}(i\Omega)$ and
$\widetilde{\varphi}(i\Omega)$ defined in Eqs.\ (24a) and (24b)
by making use of the Laplace method of approximation \cite{Fedoryuk}.
In the strong-coupling limit the integrand is maximal at the end
points $\tau = 0$ and $\tau = \beta$ and is considerably smaller
at other points of the interval of integration $(0,\beta)$.
Therefore, we can replace the initial integral in Eq.\ (24a) by one in
which the region of integration is limited to the two small
intervals $(0,\tau_{0})$ and $(\beta - \tau_{0},\beta)$.
In these intervals insertion of the expansions (19) lead to
%
\begin{align}
%
\widetilde{\phi}(i\Omega) & \simeq
   \int_{0}^{\tau_{0}} \! \! d\tau \,
   e^{i\Omega\tau - \omega_{c}\tau}
 + \int_{\beta - \tau_{0}}^{\beta} \! \! d\tau \,
   e^{i\Omega\tau - \omega_{c}(\beta - \tau)} .
\label{(A1)}
%
\end{align}
%
Then, because in the strong-coupling limit the collective frequency $\omega_c$
is large, we can replace $\tau _{0}$ by infinity, which yields a
Lorentzian for (A1),
%
\begin{align}
%
\widetilde{\phi}(i\Omega) & = \frac{2\omega_c}
   {(i\Omega)^2 - (\omega_c)^2} .
\label{(A2)}
%
\end{align}
%
If we take into account the space dependence and expand
in terms of small distances, $|\mathbf{x}|$, we obtain:
%
\begin{align}
%
\sigma(\mathbf{x}|\tau) & \approx \sigma(0|\tau)
 - {\textstyle\frac{1}{2}} \sigma_1 \mathbf{x}^2 ,
\label{(A3)}
%
\end{align}
%
where
%
\begin{align}
%
\sigma _1 & =
   \frac{1}{6N} \sum_{\mathbf{k}} \, |\overline{g}(\mathbf{k})|^2 \,
   \frac{\mathbf{k}^2
   \coth {\textstyle\frac{1}{2}} \beta\omega_{\mathbf{k}}}
  {\sinh {\textstyle\frac{1}{2}} \beta\omega_{\mathbf{k}}} .
\label{(A4)}
%
\end{align}
%
For small values of $|\mathbf{x}|$ the function in Eq.\ (21) can then
be written in factorized form,
%
\begin{align}
%
\phi(\mathbf{x}|\tau) & \simeq \phi(\mathbf{x)} \phi(0|\tau)  ,
\label{(a5)}
%
\end{align}
%
where
%
\begin{align}
%
\phi(0|\tau) & = e^{-\sigma(0|0) + \sigma(0|\tau)}
 = \frac{1}{\beta} \sum_{\Omega} \widetilde{\phi}(i\Omega) \,
   e^{-i\Omega\tau} ,
\label{(A6)}
\\
\phi(\mathbf{x}) & = e^{-\sigma_1\mathbf{x}^2/2}
 = \frac{1}{N} \sum_{\Omega} \widetilde{\phi}(q) \, e^{-i\mathbf{qx}} ,
\label{(A7)}
\\
\widetilde{\phi}(q) & = (2\pi/\sigma_1)^{3/2} \,
   e^{-q^2/\sigma_1} .
\label{(A8)}
%
\end{align}
%

For the Fourier representation of the anomalous phonon-cloud
propagator $\varphi(\mathbf{x}|\tau)$ we consider first
$|\mathbf{x}| = 0$. In this case we need the $\tau$ expansion of
$\sigma\left(0|\tau\right)$ near the midpoint of the interval
$\tau = \beta/2$ where this function is minimal. Near this minimum we
use the following expression:
%
\begin{align}
%
\sigma(0|\tau) & \simeq \sigma(0|\beta) +
   {\textstyle\frac{1}{2}}
   \sigma^{\prime\prime}(0|\beta/2) \, (\tau - \beta/2)^{2} .
\nonumber
%
\end{align}
%
This approximation can be used in the integral (24b), which allows
to rewrite it as
%
\begin{align}
%
\widetilde{\varphi}(i\Omega_n) & \simeq \! \!
   \int_{\beta/2 - \tau_0}^{\beta/2 + \tau_0} \! \! d\tau \,
   e^{-\sigma(0|0) - \sigma(0|\beta/2)}
\nonumber \\ & \times
   e^{i\Omega\tau - {\textstyle\frac{1}{2}} (\tau - \beta/2)^2 \,
   \sigma^{\prime\prime}(0|\beta/2)} .
\label{A9}
%
\end{align}
%
The width of the small interval, $2\tau_0$, is now extended to
infinity because the second derivative, $\sigma^{\prime\prime}(0|\beta/)$,
is large in the strong-coupling limit, which yields a Gaussian
distribution,
%
\begin{align}
%
\widetilde{\varphi}(i\Omega_n) & = \sqrt{2\pi/\sigma_2} \,
   e^{-\sigma(0|0) - \sigma(0|\beta/2)
 +  i\beta\Omega/2 - \Omega^2/(2\sigma_2)} ,
\label{(A10)}
\\
\sigma_2 & = \sigma^{\prime\prime}(0|\beta/2) .
\nonumber
%
\end{align}
%
This allows to get an approximate expression for Eq.\ (38).
With Eqs.\ (30) and (32) we have, for example, to evaluate
%
\begin{eqnarray}
%
\lefteqn{
\phi_2(\mathbf{x}_1,i\Omega_1;\mathbf{x}_2,i\Omega_2|
   \mathbf{x}_3,i\Omega_3;\mathbf{x}_4,i\Omega_4)
}
\nonumber \\ &&
  = \int_0^\beta \! \! d\tau_1 \ldots d\tau_4 \,
   e^{i\Omega_1\tau_1 + i\Omega_2\tau_2 - i\Omega_3\tau_3
    - i\Omega_4\tau_4}
\nonumber \\ && \times
   \exp \big \lbrace \sigma(\mathbf{x}_1 - \mathbf{x}_3||
   \tau_1 - \tau_3|)
 + \sigma(\mathbf{x}_1 - \mathbf{x}_4||\tau_1 - \tau_4|)
\nonumber \\ && {}
 + \sigma(\mathbf{x}_2 - \mathbf{x}_3||\tau_2 - \tau_3|)
 + \sigma(\mathbf{x}_2 - \mathbf{x}_4||\tau_2-\tau_4|)
\nonumber \\ && {}
 - \sigma(\mathbf{x}_1 - \mathbf{x}_2||\tau_1 - \tau_2|)
 - \sigma(\mathbf{x}_3 - \mathbf{x}_4||\tau_3 - \tau_4|)
\nonumber \\ && {}
 - 2\sigma(0|0) \big \rbrace .
\label{(A11)}
%
\end{eqnarray}
%
Equation (A11) leads to a sum of 24 fourfold integrals with different
chronological order of $\tau_n$, $n=1 \ldots 4$  (for example,
$\phi_2^{\tau_1 > \tau_2 > \tau_3 > \tau_4}$ is defined by
from $\beta > \tau_1 > \tau_2 > \tau_3 > \tau _4 > 0$), of which
only 16 make an essential contribution in the strong-coupling
limit. It is convenient to combine the 16 terms pairwise like
%
\begin{align}
%
\big [
     \tau_1 > \tau_3 > \tau_2 > \tau_4 & \quad
     \& \quad \tau_2 > \tau_4 > \tau_1 > \tau_3 \big ],
\nonumber \\
\big [
     \tau_1 > \tau_3 > \tau_4 > \tau_2 & \quad
     \& \quad \tau_4 > \tau_2 > \tau_1 > \tau_3 \big ],
\nonumber \\
\big [
     \tau_1 > \tau_4 > \tau_2 > \tau_3 & \quad
     \& \quad \tau_2 > \tau_3 > \tau_1 > \tau_4 \big ],
\nonumber \\
 \big [
      \tau_1 > \tau_4 > \tau_3 > \tau_2 & \quad
      \& \quad \tau_3 > \tau_2 > \tau_1 > \tau_4 \big ].
\nonumber
%
\end{align}
%
Further pairwise terms are obtained from the chronological orders
by changing in the first two groups the order of $\tau_1$ and $\tau_3$,
and in the last two groups the order of $\tau_1$ and $\tau_4$.
All integrals are then calculated by using the maximum possible
value of $\Sigma$ in Eq.\ (32) in the Laplace approximation, i.e.,
maximal contributions arise from positive terms in Eq.\ (32) and
coinciding arguments in each $\sigma$ function.
In addition the $\tau$ space, for which the integrand is maximal,
should be large enough. For example, in the integration over the
first pairwise terms we take
$|\tau_1 - \tau_3|$ and $|\tau_2 - \tau_4|$ as well as
$|\mathbf{x}_1 - \mathbf{x}_3|$ and $|\mathbf{x}_2-\mathbf{x}_4|$
as small quantities. This leads with $\tau_1 - \tau_3 = t_1$
and $\tau_2 - \tau_4 = t_2$ and, assuming for simplicity,
$\mathbf{x}_1 = \mathbf{x}_3$ and $\mathbf{x}_2=\mathbf{x}_4$,
to the following argument of the exponential function in
Eq.\ (A11):
%
\begin{eqnarray}
%
\lefteqn{
   \sigma(0|t_1) +\sigma(0|t_2)
}
\nonumber \\ && {}
 + \sigma(\mathbf{x}_1 - \mathbf{x}_2|\tau_1 - \tau_2 + t_2)
 + \sigma(\mathbf{x}_2 - \mathbf{x}_1|\tau_1 - \tau_2 - t_1)
\nonumber \\ && {}
 - \sigma(\mathbf{x}_1 - \mathbf{x}_2|\tau_1 - \tau_2)
 - \sigma(\mathbf{x}_1 - \mathbf{x}_2|\tau_1 - \tau_2 - t_1 + t_2)
\nonumber \\ && {}
 - 2\sigma(0|0) ,
 \label{(A12)}
%
\end{eqnarray}
%
which simply reduces to $-\omega_c (t_1 + t_2)$
when expanding in $t_1$ and $t_2$. The corresponding integrations
over $t_1$  and $t_2$ in the interval $(0,\tau_0) \to (0,\infty)$ lead
to
%
%
\begin{eqnarray}
%
\lefteqn{
\phi_2^{\tau_1 > \tau_3 > \tau_2> \tau_4}
}
\nonumber \\ &&
 = \int_0^\beta \! \! \! d\tau_1 \! \int_0^{\tau_1} \! \! \! d\tau_2 \,
   e^{i\tau_1(\Omega_1 - \Omega_3) + i\tau_2(\Omega_2 - \Omega_4)}
\nonumber \\ && \times \!
   \int_0^{\tau_0} \! \! \! dt _1 \! \int_0^{\tau_0} \! \! \! dt_2 \,
   \delta_{\mathbf{x}_1,\mathbf{x}_3} \,
   \delta_{\mathbf{x}_2,\mathbf{x}_4} \,
   e^{t_1(-\omega_c + i\Omega_3) + t_2(-\omega_c + i\Omega_4)}
\nonumber \\ && {}
 = \frac{\delta_{\mathbf{x}_1,\mathbf{x}_3} \,
         \delta_{\mathbf{x}_2,\mathbf{x}_4}}
        {(-\omega_c + i\Omega_3) (-\omega_c + i\Omega_4)}
\nonumber \\ && \times \!
   \int_0^{\beta} \! \! \! d\tau_1 \! \int_0^{\tau_1} \! \! \! d\tau_2 \,
   e^{i\tau_1(\Omega_1 - \Omega_3) + i\tau_2(\Omega_2 - \Omega_4)} .
\label{(A13)}
%
\end{eqnarray}
%
$\phi_2^{\tau_2 > \tau_4 > \tau_1 > \tau_3}$ differs from
$\phi_2^{\tau_1 > \tau_3 > \tau_2 > \tau_4}$ by the last twofold
integrals, which can be written as
%
\begin{eqnarray}
%
\lefteqn{
\int_{0}^{\beta} \! \! d\tau_2 \! \int_{0}^{\tau_2} \! \! d\tau_1 \,
   e^{i\tau_1(\Omega_1 - \Omega_3) + i\tau_2(\Omega_2 - \Omega_4)}
}
\nonumber \\ && {}
  = \int_{0}^{\beta} \! \! d\tau_1 \! \int_{\tau_1}^{\beta} \! \!
    d\tau_2 \,
   e^{ i\tau _1(\Omega_1 - \Omega_3) + i\tau_2(\Omega_2 - \Omega_4)} .
\label{(A14)}
%
\end{eqnarray}
%
By combining the last two integrals, (A13)and (A14), we obtain the
law of conservation for the frequencies,
%
\begin{eqnarray}
%
\lefteqn{
   \phi_2^{\tau_1 > \tau_3 > \tau_2 > \tau_4}
 + \phi_2^{\tau_2 > \tau_4 > \tau_1 > \tau_3}
}
\nonumber \\ && {}
 = \frac{\delta_{\mathbf{x}_1,\mathbf{x}_3} \,
   \delta_{\mathbf{x}_2,\mathbf{x}_4} \, \beta^2
   \delta_{\Omega_1,\Omega_3} \,
   \delta_{\Omega_2,\Omega_4}}
   {(-\omega_c + i\Omega_3) (-\omega_c + i\Omega_4)} .
\label{(A15)}
%
\end{eqnarray}
%
The same procedure can be used for the other 7 groups of integrals,
which finally leads to
%
\begin{eqnarray}
%
\lefteqn{
\phi_2(\mathbf{x}_1,i\Omega_1;\mathbf{x}_2,i\Omega_2|
   \mathbf{x}_3,i\Omega_3;\mathbf{x}_4,i\Omega_4)
}
\nonumber \\ && {}
 = \big \lbrace
   \delta_{\mathbf{x}_1,\mathbf{x}_3} \,
   \delta_{\mathbf{x}_2,\mathbf{x}_4} \, \beta^2 \,
   \delta_{\Omega_1,\Omega_3} \,
   \delta_{\Omega_{2,\Omega_4}}
\nonumber \\ && {}
 + \delta_{\mathbf{x}_1,\mathbf{x}4} \,
   \delta_{\mathbf{x}_2,\mathbf{x}_3} \, \beta^2 \,
   \delta_{\Omega_1,\Omega_4} \,
   \delta_{\Omega_{2,\Omega3}} \big \rbrace
\nonumber \\ && \times
   \frac{(2\omega_c)^2}
   {[(i\Omega_1)^2 - (\omega_c)^2]
    [(i\Omega_2)^2 - (\omega_c)^2]} ,
\label{(A16)}
%
\end{eqnarray}
%
This equation can be rewritten in the form,
%
\begin{eqnarray}
%
\lefteqn{
\phi_2(\mathbf{x}_1,i\Omega_1;\mathbf{x}_2,i\Omega_2|
   \mathbf{x}_3,i\Omega_3;\mathbf{x}_4,i\Omega_4)
}
\nonumber \\ && {}
 = \phi(\mathbf{x}_1,i\Omega_1|\mathbf{x}_3,i\Omega_3) \,
   \phi(\mathbf{x}_2,i\Omega_2|\mathbf{x}_4,i\Omega_4)
\nonumber \\ && {}
 + \phi(\mathbf{x}_1,i\Omega_1|\mathbf{x}_4,i\Omega_4) \,
   \phi(\mathbf{x}_2,i\Omega_2|\mathbf{x}_3,i\Omega_3) ,
\label{(A17)}
\\*[0.2cm]
\lefteqn{
\phi(\mathbf{x}_1,i\Omega_1|\mathbf{x}_3,i\Omega_3)
 = \widetilde{\phi}(i\Omega_1) \beta \,
   \delta_{\Omega_1,\Omega_3} \, \delta_{\mathbf{x}_1,\mathbf{x}_3} .
}
\label{(A18)}
%
\end{eqnarray}
%
From these equations we finally obtain Eq.\ (36). In similar manner we
calculated the Fourier representation of the two-cloud function
$\varphi_2$, which leads to Eq.\ (37).

\section{Determination of the Green's function
\boldmath
$\Gamma_{\sigma,\sigma^{\prime}}(i\omega)$}
\unboldmath
\label{AppLapjApr_B}

The evaluation of $\Gamma_{\sigma,\sigma^{\prime}}(i\omega)$
requires knowledge of the following two- and three-particle
Green's functions:
%
\begin{subequations} \begin{align}
%
G_2^0[\sigma,\tau;\sigma^{\prime},\tau^{\prime}|\tau_1] & =
   \langle \mathrm{T} \, a_\sigma(\tau) \,
   \overline{a}_{\sigma^{\prime}}(\tau^{\prime}) \, n(\tau_1)
   \rangle_0 ,
\label{(B1a)}
\\
G_3^0[\sigma,\tau;\sigma {\prime},\tau^{\prime}|\tau_1,\tau_2]  & =
   \langle \mathrm{T} \, a_\sigma (\tau) \,
   \overline{a}_{\sigma^{\prime}}(\tau^{\prime}) \, n(\tau_1)n(\tau_2)
   \rangle_0 ,
\label{(B1b)}
%
\end{align} \end{subequations}
%
where the statistical averaging is determined by the local part of
the Hubbard Hamiltonian. In the calculation it is necessary to
switch over to Hubbard transfer operators, $X^{mn}$, where the indices m and n denote
ionic quantum states, $m = 0, \, \pm 1, \, 2$:
%
\begin{align}
%
a_\sigma & = X^{0\sigma} + \sigma X^{\overline{\sigma}2}, \quad
   a_\sigma^{\dag} = X^{\sigma 0} + \sigma X^{2\overline{\sigma}} .
\label{(B2)}
%
\end{align}
%
We omit calculational details and list only the resulting expressions,
%
\begin{widetext}
\begin{eqnarray}
%
\lefteqn{
G_2^0[\sigma,\tau;\sigma^{\prime},\tau^{\prime}|\tau_1] =
   \frac{\delta_{\sigma,\sigma^{\prime}}}{Z_0} \Big \lbrace
   \theta(\tau - \tau^{\prime}) \theta(\tau^{\prime} - \tau_1) \,
   e^{-\beta E_{\overline{\sigma}} + (\tau - \tau^{\prime})
   \overline{\Delta}}
}
\nonumber \\*[-0.15cm] && {}
 + \theta(\tau - \tau_1)\theta(\tau_1 - \tau^{\prime}) \,
   \big [ e^{-\beta E_0 + (\tau - \tau^{\prime})\Delta}
 + 2 e^{-\beta E_{\overline{\sigma}} +(\tau - \tau^{\prime})
   \overline{\Delta}} \big ]
\nonumber \\ && {}
 - \theta(\tau^{\prime} - \tau)\theta(\tau - \tau_1)
   \big [ e^{-\beta E_\sigma + (\tau - \tau^{\prime})\Delta}
 + 2 e^{-\beta E_2 + (\tau - \tau^{\prime})\overline{\Delta}} \big ]
 - \theta(\tau^{\prime} - \tau_1)\theta(\tau_1 -\tau) \,
   e^{-\beta E_2 + (\tau - \tau^{\prime})\overline{\Delta}}
\nonumber \\ && {}
 + \theta(\tau_1 - \tau) \theta(\tau - \tau^{\prime}) \,
   e^{-\beta E_{\overline{\sigma}} + (\tau -\tau^{\prime})
   \overline{\Delta}}
 - \theta(\tau_1 - \tau^{\prime})\theta(\tau^{\prime} - \tau) \,
  \big [ e^{-\beta E_{\sigma} + (\tau - \tau^{\prime})\Delta}
 + 2 e^{-\beta E_2 + (\tau - \tau^{\prime})\overline{\Delta}}
 \big ] \Big \rbrace ,
\label{(B3)}
\\*[0.2cm]
\lefteqn{
G_{3}^{0}[\sigma,\tau;\sigma^{\prime},\tau^{\prime}|\tau_{1},\tau_{2}]
   = \frac{\delta_{\sigma,\sigma^{\prime}}}{Z_0} \Big \lbrace
   \big [ \theta(\tau - \tau^{\prime}) \theta(\tau^{\prime} - \tau_1)
   \theta (\tau_1 - \tau _2) + \theta(\tau - \tau^{\prime})
   \theta (\tau^{\prime} - \tau_2) \theta (\tau_2 - \tau_1) \big ]
   e^{-\beta E_{\overline{\sigma}} + (\tau - \tau^{\prime})
   \overline{\Delta}}
}
\nonumber \\*[-0.15cm] && {}
 + \big [ \theta(\tau - \tau_1)\theta(\tau_1 - \tau^{\prime})
   \theta (\tau^{\prime} - \tau_2) + \theta (\tau  - \tau_2)
   \theta (\tau_2 - \tau^{\prime})\theta (\tau^{\prime} - \tau_1)
   \big  ] \,
   e^{-\beta E_{\overline{\sigma}} + (\tau - \tau^{\prime})
  \overline{\Delta}}
\nonumber \\ && {}
+ \big [ \theta(\tau - \tau_1)\theta (\tau_1 - \tau_2)
  \theta(\tau_2 - \tau^{\prime}) + \theta (\tau - \tau_2)
  \theta(\tau_2 - \tau_1) \theta(\tau_1 - \tau^{\prime}) \big]
   \big [ e^{-\beta E_0 + (\tau - \tau^{\prime})\Delta}
+  4 e^{-\beta E_{\overline{\sigma }} + (\tau - \tau^{\prime})
   \overline{\Delta}} \big ]
\nonumber \\ && {}
-  \big [ \theta(\tau^{\prime} - \tau)\theta(\tau - \tau_1)
   \theta(\tau_1 - \tau_2)
 + \theta(\tau^{\prime} - \tau)\theta(\tau - \tau_2)
   \theta(\tau_2 - \tau_1) \big ]
  \big [ e^{-\beta E_{\sigma} + (\tau - \tau^{\prime})\Delta}
 + 4 e^{-\beta E_{2} + (\tau - \tau^{\prime})\overline{\Delta}} \big ]
\nonumber \\ && {}
- \big [ \theta(\tau^{\prime} - \tau_ 1 ) \theta (\tau_1 - \tau_2)
  \theta(\tau_2 - \tau) + \theta(\tau^{\prime} - \tau_2)
  \theta(\tau_2 - \tau_1)\theta(\tau_1 - \tau) \big ] \,
  e^{-\beta E_2 + (\tau - \tau^{\prime})\overline{\Delta}}
\nonumber \\ && {}
 - \big [ \theta(\tau^{\prime} - \tau_1)\theta(\tau_1 - \tau)
   \theta(\tau - \tau_2) + \theta(\tau^{\prime} - \tau_2)
   \theta(\tau_2 - \tau)\theta(\tau - \tau_1) \big ] \,
   2 e^{-\beta E_2 + (\tau - \tau^{\prime})\overline{\Delta}}
\nonumber \\ && {}
+ \big [ \theta(\tau_1 - \tau)\theta(\tau - \tau^{\prime})
  \theta(\tau^{\prime} - \tau_2) + \theta(\tau_2 - \tau)
  \theta (\tau - \tau^{\prime})\theta(\tau^{\prime} - \tau_1) \big ]
  e^{-\beta E_{\overline{\sigma}} + (\tau - \tau^{\prime})
  \overline{\Delta}}
\nonumber \\ && {}
 + \big [ \theta(\tau_1 - \tau)\theta(\tau - \tau_2)
   \theta(\tau_2 - \tau^{\prime}) + \theta(\tau_2 -\tau)
   \theta(\tau - \tau_1)\theta(\tau_1 - \tau^{\prime}) \big ] \,
   2 e^{-\beta E_{\overline{\sigma}} + (\tau - \tau^{\prime})
   \overline{\Delta}}
\nonumber \\ && {}
 + \big[ \theta(\tau_1 - \tau_2)\theta(\tau_2 - \tau)
   \theta(\tau - \tau^{\prime}) + \theta(\tau_2 - \tau_1)
   \theta(\tau_1 - \tau)\theta(\tau - \tau^{\prime}) \big ] ß,
   e^{-\beta E_{\overline{\sigma}} + (\tau - \tau^{\prime})
   \overline{\Delta}}
\nonumber \\ && {}
 - \big [ \theta(\tau_1 - \tau^{\prime})\theta(\tau^{\prime} - \tau)
   \theta(\tau - \tau_2)
 + \theta(\tau_2 - \tau^{\prime})\theta(\tau^{\prime} - \tau)
   \theta(\tau - \tau_1) \big]
   \big [ e^{-\beta E_{\sigma} + (\tau - \tau^{\prime}) \Delta}
+ 4 e^{-\beta E_{2} + (\tau - \tau^{\prime })\overline{\Delta}} \big ]
\nonumber \\ && {}
 - \big [ \theta(\tau_{1} - \tau^{\prime})\theta(\tau^{\prime} - \tau_2)
   \theta(\tau_2 - \tau) + \theta(\tau_2 - \tau^{\prime})
   \theta (\tau^{\prime} - \tau_1)\theta(\tau_1 - \tau) \,
   \big ]
   2 e^{-\beta E_{2} + (\tau - \tau^{\prime}) \overline{\Delta}}
\nonumber \\ && {}
- \big [ \theta(\tau_1 - \tau_2)\theta(\tau_2 - \tau^{\prime})
  \theta(\tau^{\prime} - \tau) + \theta(\tau_2 - \tau_1)
  \theta(\tau_1 - \tau^{\prime})\theta (\tau^{\prime} - \tau)
  \big ] \big [
  e^{-\beta E_{\sigma} + (\tau - \tau^{\prime}) \Delta}
 + 4 e^{-\beta E_2 + (\tau - \tau^{\prime})\overline{\Delta}} \big ]
 \Big \rbrace ,
\label{b4}
%
\end{eqnarray}
\end{widetext}

%
where
%
\begin{align}
%
\Delta & = E_0 - E_{\sigma} = \mu, \quad
   \overline{\Delta} = E_{\overline{\sigma}} - E_2 =\mu - U,
\label{(B5)}
%
\end{align}
%
and $\theta(x)$ is here the step function. Integration over $\tau_1$ and
$\tau_2$ yields:
%
\begin{eqnarray}
%
\lefteqn{
\psi_2(\tau - \tau^{\prime})
 = \int_0^{\beta} \! \! d\tau_1 \,
   \langle \mathrm{T} \, a_\sigma(\tau) \,
   \overline{a}_{\sigma^{\prime}}(\tau^{\prime}) \, n(\tau_1)
    \rangle_0
}
\nonumber \\*[-0.15cm] && {}
 = \frac{\delta_{\sigma,\sigma^{\prime}}}{Z_0} \Big \lbrace
  \theta(\tau - \tau^{\prime}) \, (\tau - \tau^{\prime}) \,
  e^{-\beta E_{\sigma} + (\tau - \tau^{\prime})\Delta}
\nonumber \\ && {}
 + \theta(\tau - \tau^{\prime}) \, (\beta + \tau - \tau^{\prime}) \,
    e^{-\beta E_{\overline{\sigma }} + (\tau - \tau^{\prime})
    \overline{\Delta}}
\nonumber \\ && {}
 - \theta(\tau^{\prime} - \tau) \, (\beta + \tau - \tau^{\prime}) \,
   e^{- \beta E_\sigma + (\tau - \tau ^{\prime}) \Delta}
\nonumber \\ && {}
 - \theta(\tau - \tau^{\prime}) \, (2\beta + \tau - \tau^{\prime}) \,
   e^{-\beta E_2 + (\tau -\tau^{\prime})\overline{\Delta }}
   \Big \rbrace ,
\label{(B6)}
%
\end{eqnarray}
%
%
\begin{eqnarray}
%
\lefteqn{\hspace*{-0.75cm}
\psi_3(\tau - \tau^{\prime}) = \int_0^{\beta} \! \! d\tau_1 \!
   \int_0^{\beta} \! \! d\tau_2 \,
   \langle \mathrm{T} \, a_\sigma(\tau) \,
   \overline{a}_{\sigma^{\prime}}(\tau^{\prime}) \,
   n(\tau_1)n(\tau_2) \rangle_0
}
\nonumber \\ && {}
 = \frac{\delta_{\sigma,\sigma^{\prime}}}{Z_0} \Big \lbrace
   \theta(\tau - \tau^{\prime}) \, (\tau - \tau^{\prime})^2 \,
   e^{-\beta E_0 + (\tau - \tau^{\prime})\Delta}
\nonumber \\ && {}
 + \theta(\tau - \tau^{\prime}) \, (\beta + \tau - \tau^{\prime})^2 \,
   e^{-\beta E_{\overline{\sigma}} + (\tau - \tau^{\prime})
   \overline{\Delta}}
\nonumber \\ && {}
- \theta(\tau^{\prime} - \tau) \,
   (\beta + \tau - \tau^{\prime})^2 \,
   e^{-\beta E_\sigma + (\tau - \tau^{\prime})\Delta}
\nonumber \\ && {}
  - \theta(\tau^{\prime} - \tau) \, (2\beta + \tau - \tau^{\prime})^2 \,
   e^{-\beta E_2 + (\tau - \tau^{\prime})\overline{\Delta}}
   \Big \rbrace .
\label{(B7)}
%
\end{eqnarray}
%
The Fourier representation of these functions,
%
\begin{align}
%
\psi_n(\tau - \tau^{\prime}) & = \frac{1}{\beta} \sum_{\omega_n}
   e^{i\omega_n(\tau^{\prime} - \tau)} \,
   \widetilde{\psi}_n(i\omega_n) ,
\label{(B8)}
%
\end{align}
%
are given by
%
\begin{align}
%
\psi_2(i\omega) & = \frac{\delta_{\sigma,\sigma^{\prime}}}{Z_0}
   \bigg \lbrace
   \left( e^{-\beta E_0} + e^{-\beta E_\sigma} \right)
   \left( \frac{1}{\lambda^2(i\omega)}
 - \frac{\beta}{\lambda(i\omega)} \right)
\nonumber \\ &
 + \left( e^{-\beta E_{\overline{\sigma}}} + e^{-\beta E_2} \right)
   \left( \frac{1}{\overline{\lambda }^2(i\omega)}
 - \frac{\beta}{\overline{\lambda}(i\omega)} \right)
\nonumber \\ &
 + \beta \left( \frac{e^{-\beta E_0}}{\lambda(i\omega)}
 - \frac{e^{-\beta E_2}}{\overline{\lambda}(i\omega) } \right)
   \bigg \rbrace ,
\label{(B9)}
%
\end{align}
%
%
\begin{align}
%
\psi_3(i\omega) & = \frac{\delta_{\sigma,\sigma^{\prime}}}{Z_0}
   \bigg \lbrace
 - \frac{2(e^{-\beta E_0} + e^{-\beta E_\sigma})}
   {\lambda^3(i\omega)}
\nonumber \\ &
 - \frac{2(e^{-\beta E_{\overline{\sigma}}} + e^{-\beta E_2})}
   {\overline{\lambda}^3(i\omega)}
\nonumber \\ &
 + \frac{2\beta \, e^{-\beta E_\sigma}}{\lambda^2(i\omega)}
 + \frac{2\beta \, e^{-\beta E_{\overline{\sigma}}}}
   {\overline{\lambda}^2(i\omega)}
 + \frac{4\beta \, e^{-\beta E_2}}{\overline{\lambda}^2(i\omega)}
\nonumber \\ &
 - \frac{\beta^2 \, e^{-\beta E_\sigma}}{\lambda(i\omega)}
 - \frac{\beta^2 \, e^{-\beta E_{\overline{\sigma}}}}
   {\overline{\lambda}(i\omega)}
 - \frac{4\beta^2 \, e^{-\beta E_2}}{\overline{\lambda}(i\omega)}
   \bigg \rbrace .
\label{(B10)}
%
\end{align}
%
This allows to evaluate the
$\Gamma_{\sigma,\sigma^{\prime}}(\tau - \tau^{\prime})$
function,
%
\begin{align}
%
\Gamma_{\sigma,\sigma^{\prime}}(\tau - \tau^{\prime}) &
 = \psi_3(\tau - \tau^{\prime}) - 2\langle n \rangle_0 \beta \,
   \psi_2(\tau - \tau^{\prime})
\nonumber \\ &
 + \beta^2 G_{\sigma,\sigma^{\prime}}^0(\tau - \tau^{\prime}) \,
   ( \langle n^2 \rangle_0 - 2 \langle n \rangle_0^2) ,
%
\end{align}
%
and to obtain its Fourier representation,
%
%
\begin{align}
%
\Gamma_{\sigma,\sigma^{\prime}}(i\omega) & =
   \frac{\delta_{\sigma,\sigma^{\prime}}}{Z_0} \bigg \lbrace
 - \frac{2(e^{-\beta E_0} + e^{-\beta E_\sigma})}
   {\lambda^3(i\omega)}
\nonumber \\ &
 - \frac{2(e^{-\beta E_{\overline{\sigma}}} + e^{-\beta E_2})}
   {\overline{\lambda}^3(i\omega)}
\nonumber \\ &
 + \frac{2\beta}{\lambda^2(i\omega)} \big [
   e^{-\beta E_\sigma } - \langle n \rangle_0 (e^{-\beta E_0} +
   e^{-\beta E_\sigma}) \big]
\nonumber \\ &
 + \frac{2\beta}{\overline{\lambda}^2(i\omega)} \big [
   e^{-\beta E_{\overline{\sigma}}} + 2e^{-\beta E_2}
\nonumber \\ &
 - \langle n \rangle_0 (e^{-\beta E_{\overline{\sigma}}}
 + e^{-\beta E_2}) \big]
\nonumber \\ &
 + \frac{\beta^2}{\lambda(i\omega)} \big [
   -(1 - 2 \langle n \rangle_0 ) \, e^{-\beta E_\sigma}
\nonumber \\ &
 + ( \langle n^2 \rangle_0 - 2\langle n \rangle_0^2)
   (e^{-\beta E_0} + e^{-\beta E_\sigma})
   \big ]
\nonumber \\ &
 + \frac{\beta^2}{\overline{\lambda}(i\omega)}
   \big [ -(1 - 2 \langle n \rangle_0) \, e^{-\beta E_{\overline{\sigma}}}
\nonumber \\ &
 + -(1 - \langle n \rangle_0) \, 4e^{-\beta E_2}
\nonumber \\ &
 + ( \langle n^2 \rangle_0 - 2 \langle 2n \rangle_0^2) \,
   (e^{-\beta E_{\overline{\sigma}}} + e^{-\beta E_2})
   \big] \bigg \rbrace .
\label{(B11)}
%
\end{align}
%
For the case of half filling we have,
%
\begin{align}
%
Z_0 & = 2 (1 + e^{\beta\mu}), \quad \lambda(i\omega) = i\omega + \mu,
\nonumber \\
\overline{\lambda}(i\omega) &
 = i\omega -\mu, \quad G_\sigma^0(i\omega)
 = \frac{i\omega}{(i\omega)^2 - \mu^2} ,
\label{(B12)}
%
\\*[0.2cm]
%
\psi_2(i\omega) & = \delta_{\sigma,\sigma^{\prime}} \bigg \lbrace
 - \frac{\beta i\omega}{(i\omega)^2 - \mu^2}
 + \frac{(i\omega)^2 + \mu^2}{[(i\omega)^2 - \mu^2]^2}
\nonumber \\ &
 - \frac{\beta\mu}{(1 + e^{\beta\mu})[(i\omega)^2 - \mu^2]} \bigg
 \rbrace ,
\label{B(13)}
%
\\*[0.2cm]
%
\psi_3(i\omega) & = \frac{\delta_{\sigma,\sigma^{\prime}}}
                    {2(1 + e^{\beta\mu})} \bigg \lbrace
 - \frac{4i\omega[(i\omega)^2 + 3\mu^2] \, (1 + e^{\beta\mu})}
   {[(i\omega)^2 - \mu^2]^3}
\nonumber \\ &
 + \frac{4\beta[(i\omega)^2 + \mu^2] \, e^{\beta\mu}}
   {[(i\omega)^2 - \mu^2]^2}
 + \frac{4\beta[(i\omega)^2 + \mu^2 + 2i\omega\mu]}
   {[(i\omega)^2 - \mu^2]^2}
\nonumber \\ &
 - \frac{2i\omega\beta^2e^{\beta\overline{\mu}}}
   {(i\omega)^2 - \mu^2}
 - \frac{4\beta^2(i\omega + \mu)}
   {(i\omega)^2 - \mu^2} \bigg \rbrace .
\label{(B14)}
%
\end{align}
%

The corresponding analytical continuation $i\omega \to E+i0^{+}$
for $E = 0$ gives to,
%
\begin{subequations} \begin{align}
%
\psi _2(0) & = \frac{\delta_{\sigma,\sigma^{\prime }}}
                    {\mu^2}
   \Big ( 1 + \frac{\beta\mu}{1 + e^{\beta\mu}} \Big ) ,
\\
\psi_3(0) & = \frac{\delta_{\sigma,\sigma^{\prime}}2\beta}
                     {\mu^2}
   \Big ( 1 + \frac{\beta\mu}{1 + e^{\beta\mu}} \Big ) ,
\\
G_{\sigma}^0(0 & = 0.
\label{(B15)}
%
\end{align} \end{subequations}
%
In this case the $\Gamma$ function is of particular simple
form,
%
\begin{align}
%
\Gamma_{\sigma}(0) & = \frac{\delta_{\sigma,\sigma^{\prime}}2\beta}
   {\mu^2} \Big ( 1 + \frac{\beta\mu}{1 + e^{\beta\mu}} \Big )
   [1 - \langle n \rangle_0] .
\label{(B16)}
%
\end{align}
%

\section{The critical temperature}

With help of the Poisson summation formula,
%
\begin{align}
%
\frac{1}{\beta} \sum\limits_{\omega_n}f(i\omega_n) & =
   \frac{1}{4\pi i} \int_C \! \! dz \,
   \tanh {(\textstyle\frac{1}{2} \beta z )} f(z) ,
\label{(C1)}
%
\end{align}
%

where $C$ is the usual counterclockwise contour of the imaginary
axis, we obtain for case that the parameters in Eq.\ 123 obey
$\gamma^4 > (c^2 - d^2)^2/4$,
%
\begin{subequations} \begin{align}
%
I_1 & = \frac{1}{\beta} \sum_{\omega } \frac{1}{Q_1(i\omega)}
 = - \frac{1}{\sqrt{4\gamma^4 - (c^2 - d^2)^2}}
\nonumber \\ & \times
   \mathrm{Im} \,
   \bigg ( \frac{\tanh {\textstyle\frac{1}{2}}
          \beta(\alpha_1 + i\alpha_2)}{\alpha_1 + i\alpha_2}
   \bigg ) ,
\label{(C2a)}
\\
I_2 & = - \frac{c^2 - d^2}{2\gamma^4} I_1
 - \frac{\tanh {\textstyle\frac{1}{2}} \beta c}{2c\gamma^4}
 + \frac{1}{2c\gamma^4}
\nonumber \\ & \times
   \mathrm{Re} \,
   \frac{\tanh {\textstyle\frac{1}{2}}
         \beta(\alpha_1 + i\alpha_2)}{\alpha_1 + i\alpha_2} ,
\\
I_3 & = - \frac{c^2 - d^2}{2\gamma^4} I_1
 - \frac{\tanh {\textstyle\frac{1}{2}} \beta d}{2d\gamma^4}
 + \frac{1}{2d\gamma^4}
\nonumber \\ & \times
   \mathrm{Re} \,
   \frac{\tanh {\textstyle\frac{1}{2}}
         \beta(\alpha_1 + i\alpha_2)}{\alpha_1 + i\alpha_2} ,
%
\end{align} \end{subequations}
%
with
%
\begin{align}
%
\alpha_{1 \atop 2} & =
   \frac{1}{\sqrt{2}} \left( \sqrt{\gamma^4 + c^2d^2}
   \pm \frac{c^2 + d^2}{2} \right)^{\! 1/2} ,
%
\end{align}
%
and
%
\begin{eqnarray}
%
\lefteqn{
   \mathrm{Re} \,
   \frac{\tanh {\textstyle\frac{1}{2}}\beta(\alpha_1 + i\alpha_2)}
   {\alpha_1 + i\alpha_2}
}
\nonumber \\ && {} =
   \frac{\alpha_1\sinh \beta\alpha_1 + \alpha_2 \sin \beta\alpha_2}
   {(\alpha_1^2 + \alpha_2^2)
   [\cosh \beta\alpha_1 + \cos \beta\alpha_2} ,
\\*[0.2cm]
\lefteqn{
   \mathrm{Im} \,
   \frac{\tanh {\textstyle\frac{1}{2}}\beta(\alpha_1 + i\alpha_2)}
   {\alpha_1 + i\alpha_2}
}
\nonumber \\ && {} = -
   \frac{\alpha_2\sinh \beta\alpha_1 + \alpha_1 \sin \beta\alpha_2}
   {(\alpha_1^2 + \alpha_2^2)
   [\cosh \beta\alpha_1 + \cos \beta\alpha_2} .
\label{(C4)}
%
\end{eqnarray}

Inserting the results in Eq.\ (C2) into Eq.\ (123) allows us
to obtain expressions for the parameters
$\eta_{11}$, $\eta_{22}$ and $\eta_{12}$.
Furthermore, this allows to decompose the $T_c$ equation with the
help of the following abbreviations,
%
\begin{align}
%
A & = \bigg(
   1 - \frac{U^2}{(2\mu - U)^2}
   \frac{(1 + e^{\beta\mu}) (e^{\beta\mu}
 + e^{\beta(2\mu - U)})}{Z_0^2} \bigg)
\nonumber \\ & \times
   (\eta_{11}\eta_{22} - \eta_{12}^2)
\nonumber \\ & =
   \bigg( 1 - \frac{U^2}{(2\mu - U)^2} \bigg)
   \frac{(1 + e^{\beta\mu}) (e^{\beta\mu} + e^{\beta(2\mu - U)})}
   {Z_0^2}
\nonumber \\ & \times
   (U\phi^{sc})^2 \frac{(c^2 - d^2)^2}{4\gamma^8}
   \bigg \lbrace \bigg(
   \frac{\tanh {\textstyle\frac{1}{2}} \beta c}{c}
 + \frac{\tanh {\textstyle\frac{1}{2}} \beta d}{d}
   \bigg)
\nonumber \\ & \times
   \mathrm{Re} \, \bigg(
   \frac{\tanh {\textstyle\frac{1}{2}} \beta(\alpha_1 + i\alpha_2)}
   {\alpha_1 + i\alpha_2} \bigg)
 - \frac{\tanh {\textstyle\frac{1}{2}} \beta c}{c}
   \frac{\tanh {\textstyle\frac{1}{2}} \beta c}{c}
\nonumber \\ &
 - \bigg(
   \frac{\tanh {\textstyle\frac{1}{2}} \beta(\alpha_1 + i\alpha_2)}
   {\alpha_1 + i\alpha_2} \bigg)^{\! \! 2}
 + \frac{2\gamma^4 - (c^2 - d^2)^2}
   {\sqrt{4\gamma^4 - (c^2 - d^2)^2}}
\nonumber \\ & \times
   \mathrm{Im} \,
   \bigg(
   \frac{\tanh {\textstyle\frac{1}{2}} \beta(\alpha_1 + i\alpha_2)}
   {\alpha_1 + i\alpha_2} \bigg)
   \bigg(
   \frac{\tanh {\textstyle\frac{1}{2}} \beta d}{d}
 - \frac{\tanh {\textstyle\frac{1}{2}} \beta c}{c}
   \bigg)
\nonumber \\ & \times
   \frac{1}{c^2 - d^2}
   \bigg \rbrace ,
\label{(C5)}
%
\end{align}
%
and
%
\begin{align}
%
B & =  \bigg( 1 +\frac{U}{2\mu - U} \bigg)
   \frac{1 + e^{\beta\mu}}{Z_0} \, \eta_{11}
\nonumber \\ &
 + \bigg( 1 - \frac{U}{2\mu - U} \bigg)
   \frac{e^{\beta\mu} + e^{\beta(2\mu - U)}}{Z_0} \, \eta_{22}
\nonumber \\ &
 = \frac{U\phi^{sc}}{2\gamma^4} \bigg \lbrace
   [2\gamma^4 - (c^2 - d^2)^2] \bigg(
   1 + \frac{U(1 - e^{\beta(2\mu - U)})}{Z_0(2\mu - U)} \bigg)
   I_1
\nonumber \\ &
 + \mathrm{Im} \, \bigg(
   \frac{\tanh {\textstyle\frac{1}{2}} \beta(\alpha_1 + i\alpha_2)}
   {\alpha_1 + i\alpha_2} \bigg)
\nonumber \\ & \times
   \bigg( U + \frac{(2\mu - U) (1 - e^{\beta(2\mu - U)})}{Z_0}
   \bigg)
\nonumber \\ &
 + 2U \bigg( (\mu - U)
   \frac{\tanh {\textstyle\frac{1}{2}} \beta d}{d}
   \frac{e^{\beta\mu} + e^{\beta(2\mu - U)}}{Z_0}
\nonumber \\ &
 - \mu \frac{\tanh {\textstyle\frac{1}{2}} \beta c}{c}
   \frac{(1 + e^{\beta\mu})}{Z_0} \bigg)
   \bigg \rbrace ,
\label{(C6)}
%
\end{align}
%
where the value of $I_1$ is determined by Eq.\ (C2a).
Equation (105), which determines $T_c$, can then be written as
%
\begin{align}
%
1 + Bf_c + Af_c^2 & = 0 ,
 \label{C7}
%
\end{align}
%
where the quantity $f_c$ is given by Eq.\ (96).
Equation (C7) would allow to investigate in detail the interplay
of the different parameters and renormalized quantities
obtained after eliminating the electron-phonon interaction
by the Lang-Firsov transformation. The influence of these parameters
on $T_c$ can however only be obtained by
numerical work, which has yet to be undertaken. The simplified
discussion in the main text shows that $T_c$ is proportional
to the typical energy scale involved, which is the renormalized
chemical potential and not the bare quantities of the
original model. This means that due to the Lang-Firsov
transformation the proportionality of $T_c$ to a typical
phonon frequency as in BCS theory or in the Eliashberg formulation
is lost. This is an interesting observation but must be tested
numerically by analyzing the more complex expressions of this
paper. In order to find out whether the proportionality
$T_c \propto |\mu|$, i.e., $T_c$ proportional to a
renomalized electronic energy scale and not proportional to an
average phononic frequency,
$\langle \omega_{\mathbf{k}} \rangle$, is an intrinsic
property of the model, we will perform investigations in infinite dimensions.
This is nowadays much at debate by using other approaches like the
dynamical mean field theory, which allows to evaluate more or less
accurate materials properties if combined with density functional
theory \cite{Georges,Lichtenstein,Chioncel,Kotliar}.
We believe that our approach allows for similar
accurate description of materials including superconducting
properties. This is left for future work.

Finally we would like to stress that the present approach, to start from
the exact local (atomic) description and to take into account the
properties of the correlated tight-binding electrons on a
lattice by perturbation theory in the transfer integral, has
conceptually some advantages compared to other theories.



\end{document}